\documentclass[12pt]{article}
\usepackage{amsmath,amsthm,amsfonts,amssymb,amscd,graphicx,array,dsfont,booktabs,multirow,mathtools}
\usepackage[nosort]{cite}

\newlength{\xtrawidth}
\newlength{\xtraheight}
\numberwithin{equation}{section}
\numberwithin{table}{section}
\numberwithin{figure}{section}

\newcolumntype{R}{>{\raggedleft\arraybackslash}p{0.47\textwidth}}
\newcolumntype{L}{>{\raggedleft\arraybackslash}p{0.47\textwidth}}

\begin{document}
%%%%%%%%%
\begin{titlepage}
\begin{center}
\hfill BONN--TH--2016--02\\
\vskip 1in
{\Large\bf{Quantum periods of Calabi--Yau fourfolds}}
\vskip 0.6in
{\large{Andreas Gerhardus and Hans Jockers}}\\
\vskip 0.4in
{\it Bethe Center for Theoretical Physics, Physikalisches Institut\\
der Universit\"at Bonn, Nussallee 12, D-53115 Bonn, Germany}
\vskip 0.2in
{\tt gerhardus@th.physik.uni-bonn.de}\\
{\tt jockers@uni-bonn.de}
\end{center}
\vskip 0.1in
\begin{center} {\bf Abstract} \end{center}
In this work we study the quantum periods together with their Picard--Fuchs differential equations of Calabi--Yau fourfolds. In contrast to Calabi--Yau threefolds, we argue that the large volume points of Calabi--Yau fourfolds generically are regular singular points of the Picard--Fuchs operators of non-maximally unipotent monodromy. We demonstrate this property in explicit examples of Calabi--Yau fourfolds with a single K\"ahler modulus. For these examples we construct  integral quantum periods and study their global properties in the quantum K\"ahler moduli space with the help of numerical analytic continuation techniques. Furthermore, we determine their genus zero Gromov--Witten invariants, their Klemm--Pandharipande meeting invariants, and their genus one BPS invariants. In our computations we emphasize the features attributed to the non-maximally unipotent monodromy property. For instance, it implies the existence of integral quantum periods that at large volume are purely worldsheet instanton generated. To verify our results, we also present intersection theory techniques to enumerate lines with a marked point on complete intersection Calabi--Yau fourfolds in Grassmannian varieties.

\vfill
\noindent April, 2016

\end{titlepage}
%%%%%%%%%%%%%%%%%%%%%%%%%%%%%%%%%%%%%%%%%%%%%%%%%%%%%%%%%%%%%%%%%%%%
\tableofcontents
\newpage
%%%%%%%%%%%%%%%%%%%%%%%%%%%%%%%%%%%%%%%%%%%%%%%%%%%%%%%%%%%%%%%%%%%%

%%%%%%%%%%%%%%%%%%%%%%%%%%%%%%
\section{Introduction} \label{sec:Intro}
%%%%%%%%%%%%%%%%%%%%%%%%%%%%%%
It is well-established that non-perturbative worldsheet instanton corrections of string compactifications on Calabi--Yau manifolds are captured in terms of the quantum cohomology ring \cite{Witten:1988xj,MR1366548,MR1286255,MR1604364}, which arises from a deformation of the classical intersection product. From the string worldsheet point of view the quantum cohomology ring is identified with the chiral--anti-chiral ring of the two-dimensional $N=(2,2)$ conformal field theory \cite{Lerche:1989uy,Cecotti:1989gv}. In this work we study the quantum cohomology of Calabi--Yau fourfolds \cite{Greene:1993vm,Mayr:1996sh} --- in particular with a single K\"ahler modulus. Due to $N=2$ special geometry \cite{Strominger:1990pd,Ceresole:1992su,Bershadsky:1993cx} for Calabi--Yau threefolds the number of generators of the quantum cohomology ring is essentially determined by the dimension of the K\"ahler moduli space, which corresponds to the number of marginal chiral--anti-chiral operators of the two-dimensional worldsheet theory. For Calabi--Yau fourfolds, however, the ring structure of the quantum cohomology ring is less constrained by target space symmetries. As a consequence, the number of generators of their quantum cohomology ring is generically only given by the number of both marginal and certain irrelevant chiral--anti-chiral operators. That is to say the number of generators cannot simply be deduced from the dimensionality of the K\"ahler moduli space. 

This basic observation has an interesting immediate consequence on the level of quantum periods, which describe quantum corrected volumes of even-dimensional cycles in Calabi--Yau manifolds. Namely, we find that while the classical K\"ahler volume of certain quantum cycles vanishes their respective quantum volume can nevertheless be non-zero. As a consequence, in the large volume regime there are non-vanishing integral quantum periods of the form
\begin{equation} \label{eq:InstPer}
  \Pi(J) \,=\, \mathcal{O}(e^{2\pi \int J} ) \,\ne\,0 \ ,
\end{equation}
in terms of the K\"ahler form $J$ in flat coordinates. Such quantum periods can never occur in Calabi--Yau threefolds as all even-dimensional quantum cycles are governed by the generators of their K\"ahler moduli spaces. Similarly, as a consequence of the Jurkiewicz--Danilov theorem and the quantum Lefschetz hyperplane theorem~\cite{MR1719555,MR1839288,MR2276766}, this phenomenon seems difficult to realize in smooth complete intersection Calabi--Yau fourfolds in compact toric varieties \cite{MR1328251,Hosono:1994ax,Hosono:1995bm,Klemm:1996ts} --- at least not within the toric part of the moduli space and not for the quantum periods describable in terms of the ambient compact toric varieties. However, for generic Calabi--Yau fourfolds the structure of the even-degree cohomology is not entirely determined by the dimensionality of the K\"ahler moduli space anymore. Therefore, the appearance of integral quantum periods purely generated by instanton numbers may not come as a surprise. Indeed, such examples have already appeared for complete intersection Calabi--Yau fourfolds in ambient complex Grassmannians \cite{Honma:2013hma},\footnote{The Pl\"ucker map embeds complex Grassmannians into projective spaces as non-complete intersections. Thus these Calabi--Yau fourfolds are projective varieties of the non-complete intersection type. As consequence the Jurkiewicz--Danilov theorem and the quantum Lefschetz hyperplane theorems are not applicable.} and are in general expected for non-complete intersections Calabi--Yau fourfolds in toric varieties, as recently also observed in ref.~\cite{Gerhardus:2015sla}.

We determine the quantum periods of Calabi--Yau fourfolds as solutions to Picard--Fuchs differential equations. With the help of gauge theory techniques \cite{Benini:2012ui,Doroud:2012xw,Jockers:2012dk,Gerchkovitz:2014gta,Gomis:2012wy}, we extract these Picard--Fuchs differential equations of non-complete intersection Calabi--Yau fourfolds with a single K\"ahler modulus for examples with a purely instanton-generated quantum period~\eqref{eq:InstPer}. A characteristic feature of such Calabi--Yau fourfolds are non-factorizable Picard--Fuchs operators of order six (or higher). Furthermore, due to the additional quantum period the regular singular point of the large volume phase does not have maximally unipotent monodromy with respect to the Picard--Fuchs operator. Hence, computing the integral quantum periods becomes more challenging, because the integration constants are not entirely determined by the perturbative asymptotic behavior, as --- for instance --- computed by the Gamma class of the Calabi--Yau fourfold \cite{Libgober:1999aaa,Iritani:2007aaa,MR2553377,MR2483750,Halverson:2013qca,Hori:2013ika}. In addition, we use the regular singular point in K\"ahler moduli space, where the quantum volume of the 8-brane vanishes.\footnote{The existence of such a singularity is predicted by the Strominger--Yau--Zaslow mirror symmetry conjecture \cite{Strominger:1996it}.} Here, the monodromy behavior of the integral quantum periods is determined by a Thomas--Seidel twist \cite{MR1831820}. We demonstrate that for the analyzed examples the knowledge of these two monodromies combined with numerical analytic continuation techniques is actually sufficient to unambiguously calculate the integral quantum periods. As a non-trivial check we establish --- again with numerical analytic continuation techniques --- that the monodromy matrices at the remaining regular singular points in K\"ahler moduli space are indeed integral as well.\footnote{Combining numerical analytic continuation techniques with the requirement of integral monodromy matrices has for instance been used extensively before in the context of the moduli spaces of Calabi--Yau threefolds \cite{Hofmann:2013PhDThesis}. For Calabi--Yau geometries associated to hypergeometric functions a systematic treatment towards analytic continuation has recently been given in refs.~\cite{Knapp:2016rec,Scheidegger:2016ysn}. Generalizing further the methods of ref.~\cite{Puhlfuerst:2015zqw} to resonant periods arising in Calabi--Yau geometries would offer a powerful framework to study analytic continuations systematically.}

With the integral quantum periods at hand, we explicitly extract the instanton corrections entering the quantum cohomology rings, which geometrically amounts to extracting genus zero Gromov--Witten invariants. Using global properties of the quantum periods in the vicinity of singular points in quantum K\"ahler moduli space, we determine the generalized topological index of the $N=(2,2)$ superconformal worldsheet theory \cite{Bershadsky:1993ta,Bershadsky:1993cx}. The genus zero Gromov--Witten invariants also define recursively the Klemm--Pandharipande meeting invariants of Calabi--Yau fourfolds, which then allow us to enumerate genus one BPS invariants of the examined Calabi--Yau fourfolds \cite{Klemm:2007in}. The intricate integrality property of these genus one invariants furnishes yet another non-trivial check on the proposed integral quantum periods.

To further check our enumerative results, we present intersection theory techniques that allow us to directly enumerate lines with a marked point on complete intersection Calabi--Yau fourfolds embedded in Grassmannians. While these intersection calculations are developed for complete intersection Calabi--Yau fourfolds in Grassmannians, our results easily generalize to enumerate lines on other complete intersection varieties embedded in Grassmannians.

Finally, let us briefly remark that our findings may have phenomenological applications as well. The study of global properties of quantum periods --- in particular the analysis of their monodromy behavior around singular divisors in moduli space --- exhibits many characteristic features of monodromy inflation in string cosmology \cite{Silverstein:2008sg}. In the context of Calabi--Yau fourfold compactifications of type~IIA strings to two dimensions, the quantum periods~\eqref{eq:InstPer} give rise to flux-induced superpotentials of the form
\begin{equation} \label{eq:FluxW}
   W_\text{flux}(t) \,=\, \sum_i a_i t^i + b + W_{inst}(t) \ , \qquad W_{inst}(t)\,=\, \mathcal{O}(e^{2\pi t^i \int \omega_i}) \,\ne\, 0 \ .
\end{equation}
Here the K\"ahler form $J=\sum_i t^i \omega_i$ is expanded in a basis of harmonic two forms $\omega_i$. Depending on the details of the chosen background fluxes all of the constants $a_i$ and $b$ can either be chosen to vanish or some of them not to vanish. Assuming further that the mirror Calabi--Yau fourfold of the analyzed fourfold has a suitable elliptic fibration, the superpotentials~\eqref{eq:FluxW} can also be interpreted in four space-time dimensions. Then the superpotential arises from four-form fluxes in F-theory on the elliptically-fibered mirror Calabi--Yau fourfold, where the chiral fields $t^i$ parametrize the mirror complex structure moduli space in the vicinity of a large complex structure point. Such large complex structure points in F-theory have been considered recently in the context of string cosmology in refs.~\cite{Hebecker:2014kva},\footnote{More generally, inflationary models in string cosmology arising from F-term axion monodromies have been introduced in refs.~\cite{Marchesano:2014mla,Hebecker:2014eua}.} where the hierarchy between polynomial and exponential suppressed terms is explored.

The outline of this work is as follows: In Section~\ref{sec:metho} we introduce the necessary ingredients and establish the computational techniques to derive the integral quantum periods for the class of studied Calabi--Yau fourfolds. Moreover, we recall some properties of enumerative invariants in Calabi--Yau fourfolds relevant for this work. In Section~\ref{sec:Examples} we exemplify in detail how to compute integral quantum periods and how to extract Gromov--Witten invariants. We tabulate our results for all the studied Calabi--Yau fourfold examples in Appendix~\ref{app:tables}. To further confirm our results, in Appendix~\ref{app:InterTheory} we calculate genus zero Gromov--Witten invariants for Calabi--Yau fourfolds directly using intersection theory methods. Our conclusions are presented in Section~\ref{sec:con}.

%%%%%%%%%%%%%%%%%%%%%%%%%%%%%%
\section{Methodology} \label{sec:metho}
%%%%%%%%%%%%%%%%%%%%%%%%%%%%%%
The aim of this section is to establish the computational tools that are necessary to analyze the quantum periods of the Calabi--Yau fourfolds  studied in Section~\ref{sec:Examples}. We review certain aspects of the quantum cohomology ring of Calabi--Yau fourfolds. Then we recall gauged linear sigma model techniques to determine the Picard--Fuchs differential equations for the quantum periods. Next we construct the asymptotic behavior of the quantum integral periods --- corresponding to B-brane central charges ---  in the vicinity of the large volume point and the singular locus, where the 8-brane becomes massless. Finally, we describe the numerical analytic continuation techniques that allow us to determine integral quantum periods from their global structure  and their asymptotic behavior at certain singular points in the quantum K\"ahler moduli space.

%%%%%%%%
\subsection{Quantum cohomology of Calabi--Yau fourfolds}
\label{sec:QuantumCohomology}
%%%%%%%%
The chiral--anti-chiral ring of $N=(2,2)$ worldsheet theories of the Calabi--Yau manifold~$X$ is given by its quantum cohomology ring, i.e., the even-dimensional cohomology group $\bigoplus_k H^{k,k}(X)$ together with the cup product deformed by genus zero worldsheet instanton corrections \cite{Witten:1988xj,Lerche:1989uy,Cecotti:1989gv,MR1366548,MR1286255,MR1604364}. 

Marginal operators of the chiral--anti-chiral ring correspond to cohomology elements of $H^{1,1}(X)$. For worldsheet theories associated to Calabi--Yau threefolds all chiral--anti-chiral ring elements are generated from such marginal deformations. This is a consequence of the underlying $N=2$ special geometry \cite{Strominger:1990pd,Ceresole:1992su,Bershadsky:1993cx}. However, for Calabi--Yau manifolds of complex dimension four or greater the chiral--anti-chiral ring need not be generated just by marginal chiral--anti-chiral ring elements anymore, but may require additional generators from the higher dimensional cohomology groups $H^{k,k}(X)$ for $k>1$ \cite{Greene:1993vm,Mayr:1996sh}. We study this phenomenon of quantum cohomology rings in the context of Calabi--Yau fourfolds.

A standard technique to study the quantum cohomology rings of a compact Calabi--Yau manifold~$X$ uses a quantum version of the Lefschetz hyperplane theorem~\cite{MR1719555,MR1839288,MR2276766}. That is to say, the information about the quantum cohomology ring of the Calabi--Yau manifold~$X$ is inferred from the quantum cohomology of some ambient space. 

Mirror symmetry furnishes another very powerful --- but yet indirect method --- to deduce the quantum cohomology~\cite{Candelas:1990rm,Witten:1991zz,Aspinwall:1993rj,Morrison:1994fr,Hori:2000kt}. For compact complete intersection Calabi--Yau manifolds in toric varieties the Batyrev--Borisov mirror construction relates K\"ahler moduli induced from the ambient space to polynomial complex structure deformations given in terms of the defining complete intersection equations \cite{MR1328251,Batyrev:1994pg,Morrison:1995yh,Hori:2000kt}. That is to say, the structure of the quantum cohomology ring is again inferred via mirror symmetry from the cohomology elements induced from some ambient toric variety.

As a consequence, for Calabi--Yau manifolds~$X$ embedded in toric ambient spaces $X_\Sigma$ of complete fans~$\Sigma$, one typically studies the quantum cohomlogy ring of those cohomology elements in $\bigoplus_k H^{k,k}(X)$ that are induced via pullback from the cohomology ring~$H^*(X_\Sigma)$ of the toric ambient space $X_\Sigma$. The Jurkiewicz--Danilov theorem for complete compact toric varieties $X_\Sigma$ guarantees that the entire cohomology ring~$H^*(X_\Sigma)$ is generated by $H^{1,1}(X_\Sigma)$. As a result (the part of) the quantum cohomology ring of $\bigoplus_k H^{k,k}(X)$ induced from the embedding of $X$ into $X_\Sigma$ is also generated by $H^{1,1}(X)$. Hence for compact smooth Calabi--Yau fourfolds $X$ embedded as complete intersections in toric varieties the part of the quantum cohomology induced from the toric ambient space is always generated by marginal operators of the chiral--anti-chiral ring. 

To study the more general --- and actually generic --- structure of the quantum cohomology ring with additional generators apart from marginal operators, we focus on Calabi--Yau fourfolds~$X$ embedded as complete intersections in compact complex ambient spaces~$Y$, whose even-dimensional cohomology ring is not just generated by $H^{1,1}(Y)$. This happens for instance for non-toric GIT quotients $Y$, which in the physics literature arise from two-dimensional $N=(2,2)$ non-Abelian gauged linear sigma models \cite{Witten:1993yc,Witten:1993xi,Lerche:2001vj,Hori:2006dk,Donagi:2007hi,Hori:2011pd,Jockers:2012zr,Sharpe:2012ji,Hori:2013gga,Gerhardus:2015sla,Sharpe:2015vza}. The simplest examples of this kind arise from complete intersection Calabi--Yau fourfolds $X$ embedded in complex Grassmannians $Y$ \cite{Honma:2013hma}. Namely, for Grassmannians $Y=\operatorname{Gr}(k,n)$ with $k>2$, the cohomology group $H^{1,1}(Y)$ is generated by the Schubert cycle $\sigma_1$, while the cohomology group $H^{2,2}(Y)$ is generated by the two Schubert cycles $\sigma_{1,1}$ and $\sigma_2$, related to $\sigma_1$ via the relation $\sigma_1^2 = \sigma_{1,1} +\sigma_2$. Thus $\sigma_1$ alone does not generate $H^{2,2}(Y)$; an additional generator is required. From the gauged linear sigma model point of view, such GIT quotients are obtained from two-dimensional non-Abelian gauge theories based on the gauge group $U(k)$ \cite{Witten:1993xi,Lerche:2001vj,Hori:2006dk,Hori:2011pd,Sharpe:2015vza}. 

In this note we focus on Calabi--Yau fourfolds $X$ with $\dim H^{1,1}(X) = 1$ --- that is to say with a single K\"ahler modulus --- and with one additional non-trivial generator in $H^{2,2}(X)$ induced from the embedding ambient space~$Y$, i.e., $\dim H^{2,2}(Y) = 2$. As mentioned before such scenarios occur for instance for Calabi--Yau fourfolds embedded as complete intersections in Grassmannian ambient spaces or flag manifolds. From a gauged linear sigma model point of view, such examples can be constructed from gauge groups $U(1) \times G$ (or discrete quotients thereof) with the semi-simple Lie group factor~$G$. Here, the Fayet--Iliopoulos term of the Abelian gauge group factor~$U(1)$ realizes the single K\"ahler modulus \cite{Witten:1993yc}, while the non-Abelian gauge group factor~$G$ can give rise to additional operators, geometrically corresponding to elements of the ambient space cohomology group $H^{2,2}(Y)$ \cite{Closset:2015rna}. An example of a Calabi--Yau fourfold~$X$ of this more general kind has been constructed in ref.~\cite{Gerhardus:2015sla}.

Thus we determine a chiral--anti-chiral ring of a Calabi--Yau fourfold $X$ with the ring elements $\phi_1$ generating $H^{1,1}(X)$ and the ring elements $\phi_{2,(1)}$ and $\phi_{2,(2)}$ furnishing two independent generators of $H^{2,2}(X)$.\footnote{Strictly speaking, we are considering a subring of the entire chiral--anti-chiral ring. This subring is generated by the ring elements induced from the embedding space $Y$.} The general structure of the quantum product then yields
\begin{equation} \label{eq:QuantumProduct}
  \phi_1 * \phi_1 \,=\, C^{(1)}(q)\, \phi_{2,(1)} + C^{(2)}(q) \,\phi_{2,(2)} \ ,
\end{equation}
where the coefficient functions are given in terms of the worldsheet instanton action $q=e^{2\pi i t}$ with the flat coordinate $t$ as
\begin{equation} \label{eq:MCg0}
  C^{(a)}(q) \,=\, c^{(a)} + \sum_{d=1}^{\infty} n_{0,d}^{(a)} \frac{d^2\,q^d}{1-q^d} \ , \qquad a=1,2 \ .
\end{equation}  
Here the classical ring structure constants are defined by the cup product $\phi_1 \cup \phi_1 = \sum_a c^{(a)} \phi_{2,(a)}$. The integral genus zero worldsheet instanton numbers of degree $d$ are denote by $n_{0,d}^{(a)}$, where the superscript refers to a single marked point constrained to lie on the algebraic cycle class $\phi_{2,(a)}$.

These genus zero worldsheet instanton numbers recursively define the symmetric Klemm--Pandharipande meeting invariants $m_{d_1,d_2}\equiv m_{d_2,d_1}$ according to \cite{Klemm:2007in} \footnote{Note the genus zero invariants $n_{0,d}(\phi_{2,(a)})$ of ref.~\cite{Klemm:2007in} relate to the genus zero invariants $n_{0,d}^{(a)}$ defined here with the identity $n_{0,d}(\phi_{2,(a)}) = \int_X\phi_{2,(a)}\cup \left(\sum_b n_{0,d}^{(b)} \phi_{2,(b)}\right)$.}
\begin{equation}
\label{eq:MeetingInvariants}
\begin{aligned}
     m_{d_1,d_2} \,&=\,0 \  \text{ for } d_1\le 0 \text{ or } d_2 \le 0 \ , \\
     m_{d_1,d_2} \,&=\,\sum_{a,b}g_{ab} n_{0,d}^{(a)} n_{0,d}^{(b)}  + m_{d_1,d_2-d_1} + m_{d_1-d_2,d_2} \  \text{  for } d_1 \ne d_2 \ , \\
     m_{d,d} \,&=\, \sum_{a,b} g_{ab} c_2^{(a)} n_{0,d}^{(b)} + \sum_{a,b}  g_{ab} n_{0,d}^{(a)} n_{0,d}^{(b)} - \sum_{k=1}^{d-1} m_{k,d-k}\ .
\end{aligned}      
\end{equation}
Here $g_{ab}$ is the intersection pairing $g_{ab} \,=\, \int_X \phi_{2,(a)} \cup \phi_{2,(b)}$, and $c_{2}^{(a)}$ are the expansion coefficient of the second Chern class of the Calabi--Yau fourfold $X$, i.e., $c_2(X) = \sum_a c_2^{(a)} \phi_{2,(a)}$ viewed as a cohomology element of $H^{2,2}(X)$. The genus zero invariants $n_{0,d}^{(a)}$ together with the meeting invariants $m_{d_1,d_2}$ are essential to extract the integral genus one invariants $n_{1,d}$ of the Calabi--Yau fourfold $X$, as all of them appear non-trivially in the multi-covering formula for the rational genus one invariants $N_{1,d}$ given by \cite{Klemm:2007in}
\begin{equation} \label{eq:MCg1}
   \sum_d N_{1,d} q^d \,=\, \sum_{d,\ell} n_{1,d} \frac{\sigma_1(\ell)}\ell q^{d\ell} 
   +\frac1{24}\left(  \sum_{d,a,b} g_{ab}c_{2}^{(a)} n_{0,d}^{(b)} - \sum_{d,k} m_{k,d-k} \right) \log(1-q^d) \ .
\end{equation}
Here $\sigma_1(\ell) = \sum_{i|\ell} i$ is the divisor function such that the integers $n_{1,d}$ enumerate elliptic curves rather than BPS states; cf., with the discussion in refs.~\cite{Bershadsky:1993ta,Katz:1999xq,Klemm:2007in}. 

The genus one invariants $n_{1,d}$ appear in the topological limit $F_1^\text{top}$ of the generalized topological index of the $N=(2,2)$ superconformal worldsheet theory \cite{Bershadsky:1993ta,Bershadsky:1993cx}, which for Calabi--Yau fourfolds with $h^{2,1}=0$ takes the form \cite{Cecotti:1992vy,Bershadsky:1993ta,Bershadsky:1993cx,Klemm:2007in}
\begin{equation}
\label{eq:F1}
  F_1^\text{top} \,=\, \left( \frac{\chi}{24} - h^{1,1} - 2 \right) \log \Pi_{\mathcal{O}_\text{pt}}
  +\log\det \left(\frac1{2\pi i} \frac{\partial z}{\partial t} \right) + \sum_\alpha b_\alpha \log \Delta_\alpha \ .
\end{equation}
Here, $\chi$ is the Euler characteristic and $\Pi_{\mathcal{O}_\text{pt}}(z)$ denotes the fundamental quantum period with respect to the large volume point of the Calabi--Yau fourfold~$X$. Furthermore, the (vector-valued) function $z(t)$ is the mirror map of the algebraic coordinates $z$ to the flat coordinates $t$. It is the inverse of the (vector-valued) function
\begin{equation}
   t(z)= \frac{1}{2\pi i} \frac{\Pi_{\vec{\mathcal{C}}}}{\Pi_{\mathcal{O}_\text{pt}}} \ ,
\end{equation}
with a basis of 2-branes $\vec{\mathcal{C}}$ representing the Mori cone of the Calabi--Yau fourfold~$X$. Finally, $\Delta_\alpha$ are the factors of the discriminant locus of the quantum K\"ahler moduli space with rational coefficients (including the large volume divisor). The coefficients $b_\alpha$ reflect the holomorphic ambiguity of $F_1^\text{top}$ \cite{Bershadsky:1993ta,Bershadsky:1993cx}, and they need to be determined by the boundary conditions in the quantum K\"ahler moduli space. Namely, in the vicinity of the large volume point $t\to\infty$ the index $F_1^\text{top}$ for Calabi--Yau fourfolds takes the asymptotic form \cite{Bershadsky:1993ta,Bershadsky:1993cx}
\begin{equation}
\label{eq:F1LV}
  F_1^\text{top} \,=\, - \frac1{24} \int_X c_3(X) \cup J + \text{(regular)} \ .
\end{equation}
Here, $c_3(X)$ and $J$ are the third Chern class and the K\"ahler form of the Calabi--Yau fourfold $X$, respectively. Another boundary condition yields the vicinity of the divisor in the quantum K\"ahler moduli space, where the volume of the 8-brane $\mathcal{O}_X$ vanishes. There one expects the universal asymptotic behavior \cite{Klemm:2007in}
\begin{equation}
\label{eq:F1Coni}
   F_1^\text{top} \,=\, -\frac{1}{24} \log \Delta_{\mathcal{O}_X} + \text{(regular)} \ ,
\end{equation}
where $\Delta_{\mathcal{O}_X}$ is the factor of the discriminant locus that vanishes at this divisor.

If the above boundary conditions are not sufficient to fix all coefficients $b_\alpha$, we employ the additional boundary conditions from further singular loci in the quantum K\"ahler moduli space characterized by vanishing quantum volumes of other branes, which exhibit the same universal asymptotic property~\eqref{eq:F1Coni}. Employing these boundary conditions, we observe for all our examined examples that the genus one invariants at degree one and two vanish, i.e., $n_{1,1}=n_{1,2}=0$.

For the examples studied in this note, we will explicitly extract the above described integral invariants for low degrees. Due to the intricate multicovering formulas \eqref{eq:MCg0} and \eqref{eq:MCg1} the confirmed integrality of the invariants $n_{0,d}^{(a)}$ and $n_{1,d}$ yields non-trivial consistency checks on our findings. Note that for Calabi--Yau fourfolds with a single K\"ahler modulus there are at each degree as many genus zero Gromov--Witten invariants as there are non-trivial chiral--anti-chiral quantum cohomology ring elements in $H^{2,2}(X)$. However, independently of the quantum cohomology ring structure there is just a single genus one BPS invariant $n_{1,d}$ because these invariants do not depend on a marked point.

%%%%%%%%
\subsection{Picard--Fuchs operators via gauged linear sigma models}
\label{sec:GaugeTheories}
%%%%%%%%
Our approach to extract the quantum cohomology ring is to first determine the Picard--Fuchs differential equation that governs the quantum periods of the examined Calabi--Yau fourfold~$X$. 

Using variation of Hodge structure techniques of the holomorphic four-form $\Omega$ of the mirror Calabi--Yau geometry furnishes a standard technique to derive the Picard--Fuchs operators for the quantum periods. However, this approach requires a construction of the mirror Calabi--Yau fourfold, which for non-toric ambient spaces or non-complete intersection Calabi--Yau fourfolds in toric ambient spaces --- as studied in this note --- can be rather cumbersome or is even unknown.\footnote{Using the Pl\"ucker embedding of Grassmannians in projective spaces, the work of ref.~\cite{MR1619529} reduces the problem of constructing a mirror Calabi--Yau fourfold to the Baytrev--Borisov mirror recipe for complete intersections in toric varieties, which is further generalized to complete intersections in flag manifolds in ref.~\cite{MR1756568}. A mirror proposal has also been presented for certain non-complete intersection Calabi--Yau manifolds in toric varieties in ref.~\cite{Boehm:2011rr}.}

Here we follow a different approach that allows us to determine the Picard--Fuchs operators directly from the sphere or hemisphere partition function of the gauged linear sigma models, which describe the Calabi--Yau fourfolds under consideration as a geometric target space phase. The sphere partition function $Z_{S^2}$ computes the exponentiated sign-reversed K\"ahler potential of the Calabi--Yau variety \cite{Jockers:2012dk}, while the hemisphere partition function $Z_{D^2,\partial D^2}$ directly gives rise to quantum periods for appropriate boundary conditions of the gauged linear sigma model at $\partial D^2$ \cite{Hori:2013ika}. Both quantities are annihilated by the Picard--Fuchs operators $\mathcal{L}_i$, i.e., 
\begin{equation}
     \mathcal{L}_i(z_a,\theta_a) Z_{S^2}(z_a) \,=\,0 \ , \quad
     \mathcal{L}_i(z_a,\theta_a) Z_{D^2,\partial D^2}(z_a) \,=\,0 \ , \qquad
     \theta_a\,=\,z_a  \frac{\partial}{\partial z_a} \ ,
\end{equation}
in terms of the algebraic coordinates $z_a$ with $a=1,\ldots,h^{1,1}(X)$. Thus the Picard--Fuchs operators $\mathcal{L}_i$ can be determined by the requirement to annihilate $Z_{S^2}(z_a)$ and $Z_{D^2,\partial D^2}(z_a)$. This approach has also been employed for instance in refs.~\cite{Honma:2013hma,Gerhardus:2015sla}.

As we focus on Calabi--Yau geometries with a single K\"ahler modulus, there is just a single Picard--Fuchs operators $\mathcal{L}(z,\theta)$ depending on a single algebraic coordinate~$z$. While for Calabi--Yau threefolds such a Picard--Fuchs operator is always of order four (due to the aforementioned ring structure of the quantum cohomology ring), for Calabi--Yau fourfolds the order of the Picard--Fuchs operator is given by\footnote{The highest power of the logarithmic derivative $\theta$ is the order of the differential operator $\mathcal{L}(\theta,z)$.} 
\begin{equation}
  \operatorname{ord} \mathcal{L}(z,\theta) \,=\, 4 + \#(\phi_2) \ .
\end{equation}  
Here $\#(\phi_2)$ denotes the number of chiral--anti-chiral ring generators associated to $H^{2,2}(X)$ that non-trivially participate in the quantum product $\phi_1 * \phi_1$. Thus the order of the Picard--Fuchs operator $\mathcal{L}(z,\theta)$ is at least five or higher. For the particular quantum products~\eqref{eq:QuantumProduct} studied in this work we obtain Picard--Fuchs operators of order six.

Note that --- from the A-variation of Hodge structure point of view (see for instance refs.~\cite{MR1442525,Iritani:2007aaa}) --- a large volume point in quantum K\"ahler moduli space of a Calabi--Yau $n$-fold $X_n$ is always of unipotent monodromy of index $n$. For one-dimensional quantum K\"ahler moduli spaces and from a mirror symmetry perspective this is a consequence of the Landman monodromy theorem \cite{MR0344248} applied to the middle dimensional cohomology of the mirror Calabi--Yau $n$-fold $\widehat X$. It states that the monodromy transformation $M$ acting on $H^n(\widehat X)$ about a singular point in the mirror complex structure moduli space is quasi-unipotent with index of at most $n$, i.e., $(M^k - \operatorname{id})^{n+1} = 0$ for some integer $k$. In particular, a large complex structure point in the complex structure moduli space of $\widehat X_n$ --- which is mirror to a large volume point in quantum K\"ahler moduli space of $X_n$ --- is unipotent with the maximal index $n$, i.e, $(M - \operatorname{id})^{n+1}=0$ but $(M -\operatorname{id})^{n} \ne 0$. This implies that the Picard--Fuchs differential equation associated to $X_n$ is always of unipotency of index $n$ at large volume --- independently of the order of the Picard--Fuchs operator. 

In particular, large volume points of Calabi--Yau fourfolds are always points of unipotency of index four. Hence they furnish regular singular points of maximally unipotent monodromy of the differential equation only for Picard--Fuchs operators of order five. This is, for instance, the case for those complete intersection Calabi--Yau fourfolds in toric varieties with a single K\"ahler modulus studied in refs.~\cite{MR1328251,Klemm:1996ts,MR1442525,Klemm:2007in,Bizet:2014uua}. For the examples of order six Picard--Fuchs operators appearing in ref.~\cite{Honma:2013hma} and studied here, the large volume points in the quantum K\"ahler moduli space are not of maximally unipotent monodromy anymore. It is this property of the order six Picard--Fuchs operators, which yields the interesting structure of the quantum cohomology ring discussed in Section~\ref{sec:QuantumCohomology}. In the following we also refer to such examples as Calabi--Yau fourfolds with Picard--Fuchs operators of non-minimal order.

%%%%%%%%
\subsection{B-branes, quantum periods and monodromies}
\label{sec:Bbranes}
%%%%%%%%
In a compact Calabi--Yau manifold~$X$ of complex dimension $d$ the topological B-branes $\mathcal{E}^\bullet$ on $X$ are represented by the objects in the derived category of bounded complexes of coherent sheaves $D^b(X)$ \cite{Sharpe:1999qz,Diaconescu:2001ze,Douglas:2000gi}. Furthermore, to each B-brane $\mathcal{E}^\bullet$ we assign a quantum period $\Pi_{\mathcal{E}^\bullet}(J)$, which is a function of the (complexified) K\"ahler class $J = \sum_a t^a  D_a$ in terms of the K\"ahler moduli $t^a$ with $a=1,\ldots, h^{1,1}(X)$ and the generators of the K\"ahler cone given in terms of divisors $D_a$.\footnote{For simplicity we assume here that the K\"ahler cone is generated by $h^{1,1}(X)$ divisors.} For stable BPS branes~$\mathcal{E}^\bullet$ the quantum periods enjoy the interpretation of a (K\"ahler moduli dependent) central charge, whose magnitude is its BPS mass that enjoys also the interpretation of a calibrated quantum volume. For further details on B-branes and their notion of stability, we refer the reader for instance to the review~\cite{Aspinwall:2004jr}.

The quantum periods depend only on the B-brane charges, which are captured by elements of the algebraic K-group $K^0_\text{alg}(X)$ \cite{Witten:1998cd}. In this note we want to construct a basis of (torsion free) integral quantum periods for B-branes, which corresponds to integral generators of the torsion-free part of the algebraic K-theory group $K^0_\text{alg}(X)$. The asymptotic behavior of quantum periods~$\Pi_{\mathcal{E}^\bullet}^{\text{asy}}$ in the large volume regime of the Calabi--Yau manifold~$X$ constrains --- and for large volume points with maximally unipotent mondromy unambiguously determines --- the integration constants of the integral quantum periods $\Pi_{\mathcal{E}^\bullet}(J)$ as solutions to the associated system of Picard--Fuchs differential equations. In terms of the flat K\"ahler coordinates the large volume asymptotics reads \cite{Halverson:2013qca}
\begin{equation}
    \Pi_{\mathcal{E}^\bullet}^{\text{asy}}(J) \,=\, \int_X e^J \, \Gamma_\mathbb{C}(X) \, \operatorname{ch} \mathcal{E}^{\bullet\,\lor} \ .
\end{equation}    
Here $\Gamma_\mathbb{C}(X)$ is the (multiplicative) characteristic Gamma class, which for Calabi--Yau manifolds with $c_1=0$ enjoys the expansion\footnote{The gamma class~$\Gamma_\mathbb{C}(X)$ is based upon the series $\Gamma_\mathbb{C}(z) \,=\, e^\frac{z}{4} \, \Gamma(1-\tfrac{z}{2\pi i} )$.}
\begin{equation}
    \Gamma_\mathbb{C}(X)\,=\,1+\frac{1}{24}c_2+\frac{i  \zeta (3)}{8 \pi ^3}c_3+\frac{1}{5\,760}(7 c_2^2-4 c_4) + \ldots \ ,
\end{equation}
where $c_k\equiv c_k(X)$ are the Chern classes of $X$.

For the Calabi--Yau manifold  $X$ (of real dimension $2d$) there are some universal B-branes that always correspond to integral generators of the K-theory group $K^0_\text{alg}(X)$:
\begin{itemize}
\item The $2d$-brane of the structure sheaf~$\mathcal{O}_X$ --- with the trivial Chern character $\operatorname{ch} \mathcal{O}_X=1$ --- readily yields the asymptotic quantum period
\begin{equation}
    \Pi_{\mathcal{O}_X}^{\text{asy}}(J) \,=\, \int_X e^J \, \Gamma_\mathbb{C}(X) \ .
\end{equation}    
\item  A collection of $2(d-1)$-branes $\mathcal{E}_a^\bullet$ associated to the K\"ahler cone divisors $D_a$ are given by the complexes
\begin{equation}
  \mathcal{E}_a^\bullet: \quad 0 \longrightarrow \mathcal{O}_X(-D_a) \longrightarrow \mathcal{O}_X \longrightarrow 0 \ .
\end{equation}
Their asymptotic periods read
\begin{equation}
    \Pi_{\mathcal{E}_\alpha^\bullet}^{\text{asy}}(J) \,=\, \int_X e^J \, \Gamma_\mathbb{C}(X) \, \left(1 - \operatorname{ch}\,\mathcal{O}_X(D_\alpha) \right) \ .
\end{equation}
\item We construct a collection of $2$-branes $\mathcal{C}^\bullet_a$ as follows: Given the embedded Mori cone curves $\iota: \mathcal{C}_a\hookrightarrow X$ dual to the K\"ahler cone divisors $D_a$, we consider their structure sheaf $\mathcal{O}_{\mathcal{C}_a}(K_{\mathcal{C}_a}^{1/2})$ twisted by a spin structure $K_{\mathcal{C}_a}^{1/2}$ of $\mathcal{C}_a$. Then the $2$-branes $\mathcal{C}^\bullet_\alpha$ are given by
\begin{equation}
   \mathcal{C}^\bullet_a = \iota_! \mathcal{O}_{\mathcal{C}_a}(K_{\mathcal{C}_a}^{1/2}) \ ,
\end{equation}
in terms of the K-theoretic push-forwards $\iota_!: K^0(\mathcal{C}_a) \to K^0(X)$. The Chern character of $\mathcal{C}^\bullet_a$ is computed by the Grothendieck--Riemann--Roch formula
\begin{equation}
     \operatorname{ch}\,\mathcal{C}^\bullet_a \,=\, 
     \frac{\iota_* \left(\operatorname{ch}K_{\mathcal{C}_a}^{1/2}\,\operatorname{Td}(\mathcal{C}_a)\right)}{\operatorname{Td}(X)}
     \,=\, [ \mathcal{C}_a ] \ ,
\end{equation}
because $\operatorname{ch}K_{\mathcal{C}_a}^{1/2}\operatorname{Td}(\mathcal{C}_a)=(1+\frac12 K_{\mathcal{C}_a})(1 - \frac12 K_{\mathcal{C}_a})=1$ and for Calabi--Yau manifolds $\operatorname{Td}_1(X) = \frac{1}{2} c_1 = 0$. Here $ [ \mathcal{C}_a ] $ denotes the Poincar\'e dual cohomology class of the curve $\mathcal{C}_a$, such that its asymptotic quantum period becomes 
\begin{equation}
    \Pi_{\mathcal{C}_a^\bullet}^{\text{asy}}(J) \,=\, (-1)^{d-1}  \int_X e^J \, [ \mathcal{C}_a ]  \,=\, (-1)^{d-1} t^a \ .
\end{equation}
\item Finally, we consider the skyscraper sheaf $\mathcal{O}_\text{pt}$ for $0$-brane located at a point $\iota: \text{pt} \hookrightarrow X$ in the Calabi--Yau manifold $X$. Employing again the Grothendieck--Riemann--Roch theorem for the Chern character of the K-theoretic push-forward $\operatorname{ch}\iota_! \text{pt}$ we find the asymptotic period
\begin{equation}
  \Pi_{\mathcal{O}_\text{pt}}^{\text{asy}}(J) \,=\, (-1)^d \int_X e^J \, [ \text{pt} ] \,=\, (-1)^d \ .
\end{equation}
\end{itemize}
For Calabi--Yau threefolds the above described integral quantum periods generate all central charges associated to the torsion-free elements in $K^0_\text{alg}(X)$. However, for higher-dimensional Calabi--Yau manifolds we also need to construct algebraic cycles representing $p$-branes of even dimension $p=4,\ldots, 2(d-2)$. In particular, for Calabi--Yau fourfolds we determine the quantum periods of algebraic cycles of 4-branes for cohomology elements in $H^{2,2}(X) \cap H^4(X,\mathbb{Z})$. As such algebraic cycles depend on the details of the Calabi--Yau manifold $X$, we construct them for the explicit examples studied in Section~\ref{sec:Examples}.

For Calabi--Yau fourfolds with Picard--Fuchs operators of non-minimal order the large volume asymptotics of integral quantum periods does not determine all integration constants of their solutions to the Picard--Fuchs differential equations. As a consequence there are (integral linear combination) of quantum periods with vanishing classical terms in the large volume regime. Such quantum periods are purely instanton generated as described in formula~\eqref{eq:InstPer}. Then the integration constants must be further constrained by monodromies around other singularities in moduli space. They are deduced from the monodromies of the associated B-branes about singularities in moduli space, which are described by Fourier--Mukai transformations acting upon the derived category of bounded complexes of coherent sheaves $D^b(X)$ \cite{Aspinwall:2001dz,Distler:2002ym,Jockers:2006sm,Brunner:2008fa}. This allows us to derive the monodromy behavior of the integral quantum periods.

The Strominger--Yau--Zaslow picture of mirror symmetry for Calabi--Yau $d$-folds~$X$ conjectures a singular point $t_{\mathcal{O}_X}$ in the quantum K\"ahler moduli space, where the $2d$-brane --- represented by the structure sheaf $\mathcal{O}_X$ --- becomes massless \cite{Strominger:1996it}.\footnote{In order for the $2d$-brane to become massless a suitable path from the large volume point to the singularity $t_{\mathcal{O}_X}$ must be specified.}  The Seidel--Thomas twist captures the monodromy at the singular point  $t_{\mathcal{O}_X}$, which is represented by the Fourier--Mukai kernel \cite{Kontsevich:1994dn,MR1831820}
\begin{equation}
  \mathcal{K}_{\mathcal{O}_X} \,=\, \operatorname{Cone}\left( \eta: \mathcal{E}^{\bullet\,\lor} \boxtimes \mathcal{E}^\bullet \to \mathcal{O}_\Delta \right) \ .
\end{equation}
The Seidel--Thomas twist is interpreted as the formation of bound states between the brane $\mathcal{E}^\bullet$ --- adiabatically encircling the singularity $t_{\mathcal{O}_X}$ --- and the (massless) brane~$\mathcal{O}_X$, while the index $\chi(\mathcal{E}^\bullet,\mathcal{O}_X)$  of the open strings stretching between the branes $\mathcal{E}^\bullet$ and $\mathcal{O}_X$ becomes the index for the (relative) number of formed bound states \cite{Brunner:2001eg,Jockers:2006sm}. Therefore, on the level of quantum periods the Seidel--Thomas twist induces the monodromy transformation
\begin{equation} \label{eq:STtwist}
   M_{t_{\mathcal{O}_X}}: \  
   \Pi_{\mathcal{E}^\bullet} \mapsto \Pi_{\mathcal{E}^\bullet} - \chi(\mathcal{E}^\bullet,\mathcal{O}_X) \, \Pi_{\mathcal{O}_X} \ .
\end{equation}
Here $\Pi_{\mathcal{E}^\bullet}$ and $ \Pi_{\mathcal{O}_X}$ are the quantum periods of the branes $\mathcal{E}^\bullet$ and $\mathcal{O}_X$, respectively, whereas the index of open strings is computed by the Hirzebruch--Riemann--Roch pairing \cite{Witten:1998cd,Brunner:1999jq}
\begin{equation}
    \chi(\mathcal{E}^\bullet,\mathcal{F}^\bullet) \,=\, \int_X \operatorname{Td}(X) \operatorname{ch}(\mathcal{E}^{\bullet\,\lor}) \operatorname{ch}(\mathcal{F}^{\bullet}) \ .
\end{equation}

We observe that the open-string index $\chi(\mathcal{O}_X,\mathcal{O}_X)$ simplifies to the arithmetic genus of the Calabi--Yau manifold~$X$, i.e., 
\begin{equation}
    \chi(\mathcal{O}_X,\mathcal{O}_X) \,=\, \int_X \operatorname{Td}(X) \,=\, \sum_{p} (-1)^p\, h^{0,p}(X) \ .
\end{equation}    
Thus for Calabi--Yau threefolds with $SU(3)$ holonomy and not a subgroup thereof, we have $\chi(\mathcal{O}_X,\mathcal{O}_X)=0$ for the open-string index between two 6-branes $\mathcal{O}_X$. Furthermore, the open-string index between a 0-brane $\mathcal{O}_\text{pt}$ and a 6-brane $\mathcal{O}_X$ computes to $\chi(\mathcal{O}_\text{pt},\mathcal{O}_X)=1$. Hence, for the dual pair of quantum periods $(\Pi_\text{pt},\Pi_{\mathcal{O}_X})$ of Calabi--Yau threefolds, the Seidel--Thomas twist yields the characteristic monodromy
\begin{equation} \label{eq:STCY3}
  M_{t_{\mathcal{O}_X}}: \ 
  \begin{pmatrix} \Pi_\text{pt}\\ \Pi_{\mathcal{O}_X} \end{pmatrix} \mapsto
  \begin{pmatrix} 1 & -1 \\ 0 & 1\end{pmatrix}  \begin{pmatrix} \Pi_\text{pt}\\ \Pi_{\mathcal{O}_X} \end{pmatrix} \ ,
\end{equation}
which --- in the four-dimensional $N=2$ effective theory of type~II strings on Calabi--Yau threefolds --- is due to additional massless BPS blackhole states at the singularity~$t_{\mathcal{O}_X}$ \cite{Strominger:1995cz,Greene:1995hu}.

For Calabi--Yau fourfolds with $SU(4)$ holonomy and not a subgroup thereof, the arithmetic genus yields the open-string index~
$\chi(\mathcal{O}_X,\mathcal{O}_X)=2$. Hence the monodromy~\eqref{eq:STtwist} of $t_{\mathcal{O}_X}$ maps the quantum period of the 8-brane $\Pi_{\mathcal{O}_X}$ to $-\Pi_{\mathcal{O}_X}$. As result we find that, applying the monodromy transformation~\eqref{eq:STtwist} twice, maps any quantum period back to itself. That is to say we find that\footnote{More generally, $1+(-1)^d$ is the arithmetic genus of any Calabi--Yau $d$-fold with $SU(d)$ holonomy of dimension $d>0$. Hence, at the singularity $t_{\mathcal{O}_X}$ we find the monodromy behavior~\eqref{eq:STCY3} for odd and \eqref{eq:STCY4} for even dimensional Calabi--Yau manifolds, respectively.} 
\begin{equation} \label{eq:STCY4}
  M_{t_{\mathcal{O}_X}}^2 \,=\, \operatorname{id} \ ,
\end{equation}  
where $M_{t_{\mathcal{O}_X}}$ generates a $\mathbb{Z}_2$~group action on the set of all quantum periods. The $\mathbb{Z}_2$~monodromy around the singularity $t_{\mathcal{O}_X}$ in Calabi--Yau fourfolds has previously been studied in ref.~\cite{Grimm:2009ef,Bizet:2014uua}.

%%%%%%%%
\subsection{Numerical analytical continuation}
\label{sec:Continuation}
%%%%%%%%
Starting from the Picard--Fuchs operator $\mathcal{L}(z,\theta)$ for the periods in the quantum K\"ahler moduli space --- for instance to be determined by gauged linear sigma model methods described in Section~\ref{sec:GaugeTheories} --- we now describe the use of numerical analytic continuation techniques to establish the global structure of quantum periods. In particular this allows us to determine linear combinations of solutions to the Picard--Fuchs differential equations corresponding to integral quantum periods.

In a local patch $U_\alpha$ on the quantum K\"ahler moduli space in the vicinity of the origin of the algebraic coordinate $z_\alpha$ the Picard--Fuchs operator takes the form
\begin{equation} \label{eq:PFGeneral}
  \mathcal{L}_\alpha(z_\alpha,\theta_\alpha)\,=\, \sum_{k=0}^n h^{(k)}_\alpha(z_\alpha)\, \theta_{\alpha}^k\ ,
  \qquad \theta_{\alpha}\,=\,z_\alpha \frac{\partial}{\partial z_\alpha}\ ,
\end{equation}
in terms of some polynomials $h^{(k)}_\alpha(z_\alpha)$. The integer $n$ is the order of the Picard--Fuchs operator, which --- as discussed --- for Calabi--Yau fourfolds is at least five but can be greater. Note that the operator $\mathcal{L}_\alpha(z_\alpha,\theta_\alpha)$ is also well-defined in the vicinity of the origin of the algebraic coordinate $z_\beta = z_\alpha- z'$ with $z'\neq \infty$, which allows us to rewrite the Picard--Fuchs operator in the local patch $U_\beta$ associated to the new algebraic coordinate $z_\beta$ according to
\begin{equation}
\mathcal{L}_\beta(z_\beta,\theta_\beta)\,=\, z_\beta^n \cdot \mathcal{L}_\alpha\left(z_\beta+z',(1+\tfrac{z'}{z_\beta})\theta_\beta\right) \ .
\end{equation}
Note that the prefactor $z_\beta^n$ renders the new coefficient functions $h^{(k)}_\beta(z_\beta)$ to be polynomial. Similarly, for $z'= \infty$ we set $z_\beta= z_\alpha^{-1}$ and have
\begin{equation}
  \mathcal{L}_\beta(z_\beta,\theta_\beta)\,=\, z_\beta^m \cdot \mathcal{L}_\alpha\left(z_\beta^{-1},-\theta_\beta\right)\ ,
\end{equation}
where $m$ is the maximal degree of the polynomials $h^{(k)}_\alpha$ in eq.~\eqref{eq:PFGeneral}.

\begin{figure}[tb]
\centering \includegraphics[width=0.6\textwidth]{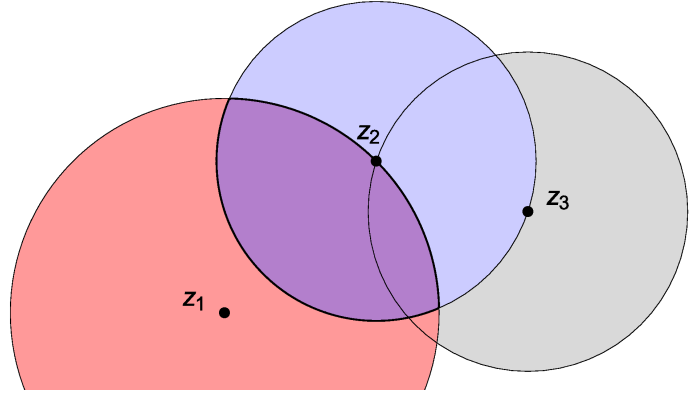}
\caption{\label{fig:ConvAreas} Partwise illustration of a K\"ahler moduli space, which shows three regular singular points, $z_1$, $z_2$ and $z_3$. The solutions $\Pi^{(k)}_{\alpha}$ for $\alpha=1,2,3$ converge within circles around $z_\alpha$ whose radii are given by the distance to the closest other $z_\alpha$. On the overlaps of convergence areas --- such as the intersecting region of the circles around $z_1$ and $z_2$ --- there is a $\operatorname{GL}(n,\mathbb{C})$ transformation relating the respective solutions $\Pi^{(k)}_{\alpha}$.}
\end{figure}

For any operator~$\mathcal{L}_\alpha(z_\alpha,\theta_\alpha)$ there are $n$ linearly independent solutions $\Pi^{(k)}_\alpha(z_\alpha)$ to the Picard--Fuchs differential equation, i.e.,
\begin{equation}
\mathcal{L}_\alpha(z_\alpha,\theta_\alpha)\, \Pi^{(k)}_\alpha(z_\alpha) = 0 \ ,
\end{equation}
which can be determined by the Frobenius method as an infinite series expansion in the local coordinates~$z_\alpha$. As illustrated in Figure~\ref{fig:ConvAreas}, these solutions are valid within a radius of convergence around the origin of the coordinate $z_\alpha$ that is given by the distance to the closest regular singular point. To be precise, in particular we determine solutions in the vicinity of regular singular points. This means that we allow for a pole at or a branch cute emanating from the origin within the radius of convergence. In the following we denote by the patch $U_\alpha$ the disk of convergence for the solutions to the Picard--Fuchs operator around the origin of the local coordinate $z_\alpha$.

Regular singular points of the Picard--Fuchs differential equations are points in the quantum K\"ahler moduli space that exhibit non-trivial monodromy behavior. Let $z_\alpha$ be a regular singular point. In terms of the period vector $\vec\Pi_\alpha \,=\, ( \Pi^{(1)}_\alpha,\ldots,\Pi^{(n)}_\alpha)^T$ the monodromy matrix $M_{\alpha}$ is given by
\begin{equation} \label{eq:Mono}
  \vec{\Pi}_\alpha\big(z_\alpha \,e^{2\pi i}\big) = M^T_{\alpha} \cdot \vec{\Pi}_\alpha(z_\alpha) \ , 
\end{equation}
deviating from the identity matrix. A necessary condition for $z_\alpha$ to be a regular singular point is that
\begin{equation}
  z_\alpha = 0 \quad \text{or} \quad z_\alpha = \infty  \quad \text{or} \quad h^{(n)}_\alpha(z_\alpha) = 0 \ .
\end{equation}
Note, however, that the converse is not true in general.\footnote{For the general theory of ordinary differential equations with regular singular points, see for instance ref.~\cite{Forsyth:1959}.}

Since the Picard--Fuchs operator is defined globally, the quantum periods as their solutions can be analytically continued over the entire quantum K\"ahler moduli space. Therefore, as long as the disks $U_\alpha$ and $U_\beta$ overlap, there exists a transformation matrix~$A_{\alpha\beta}$ in $\operatorname{GL}(n,\mathbb{C})$ that relates their solutions on the overlap $U_\alpha\cap U_\beta$ as
\begin{equation} \label{eq:MatchingPeriods}
  \vec{\Pi}_\alpha(z_\alpha)\,=\, A_{\alpha\beta} \cdot \vec{\Pi}_\beta(z_\beta(z_\alpha)) \ , 
\end{equation}
where we express the local coordinate $z_\beta$ in terms of $z_\alpha$ in the overlap $U_\alpha\cap U_\beta$.

By repeating this analytic continuation successively from patch to patch, we see that a set of quantum periods $\vec\Pi$ can be analytically continued along any path (avoiding the regular singular points) in the quantum K\"ahler moduli space. From the analytic continuation along suitable paths we then deduce the monodromy behavior of a basis of quantum periods around regular singular points $z_\alpha$ according to eq.~\eqref{eq:Mono}. As the quantum periods describe central charges of B-branes, the monodromy matrix $M_{z_\alpha}$ (and its inverse) for the regular singular point $z_\alpha=0$ must actually be integral for a generating basis of integral quantum periods \cite{Brunner:1999jq,Diaconescu:1999vp,Scheidegger:1999ed,Diaconescu:2000ec,Mayr:2000as}, i.e.,\footnote{In the context of Calabi--Yau threefolds, $N=2$ special geometry restricts the monodromy action on integral quantum periods to integral symplectic transformations \cite{Strominger:1990pd,Ceresole:1992su,Bershadsky:1993cx}. For Calabi--Yau fourfolds algebraic relations among quantum periods put similar but yet less restrictive constraints on the possible integral monodromy transformation matrices $M_{z_\alpha}$ \cite{Hosono:1993qy,Alim:2011rp,Bizet:2014uua}. It would be interesting to study the properties of these algebraic constraints systematically, so as to further develop the notion of $N=1$ special geometry \cite{Lerche:2002yw,Lerche:2002ck}.}
\begin{equation}  \label{eq:IntM}
  M_{z_\alpha} \,\in\, \operatorname{GL}(n,\mathbb{Z}) \quad \text{for $z_\alpha=0$ a regular singular point} \ .
\end{equation}  
Our goal is now to find a generating set of integral quantum periods together with their integral monodromy matrices $M_{z_\alpha}$ around regular singular points $z_\alpha$ by combining the methods of Section~\ref{sec:Bbranes} with the strong integrality constraint \eqref{eq:IntM}. Then the integral quantum periods in turn allow us to extract the quantum cohomology rings and Gromov--Witten invariants discussed in Section~\ref{sec:QuantumCohomology}.

In practice we perform the analytic continuations numerically. This is done by inserting $n^2$ different points for $z$ --- chosen from the overlap and according to the prescription to be given in the next paragraph --- in eq.~\eqref{eq:MatchingPeriods}, which gives a set of $n^2$ linear equations for the $n^2$ entries of $A_{\alpha\beta}$. If the period vectors $\vec{\Pi}_\alpha$ and $\vec{\Pi}_\beta$ could be evaluated at a given point exactly, the results would not depend on the particular choice of the points~$z$. However, since we approximate their value up to a certain fixed expansion order in the respective variables $z_\alpha$ and $z_\beta$ only, the resulting values of the periods are approximations themselves. In order to get an estimate of the error, we choose the $n^2$ values for $z$ randomly several times and check, how much the results fluctuate. Moreover, we perform the continuation in both directions and check, to what precision the products $A_{\alpha \beta} \cdot A_{\beta \alpha}$ and  $A_{\beta \alpha} \cdot A_{\alpha \beta}$ agree with the unit matrix. A final check of the numerical precision is, whether the appropriately ordered product of all monodromy matrices --- representing a contractible path of analytic continutation with respect to a fixed basis of periods --- indeed equals unity.

As mentioned before the periods at a regular singular point necessarily involve functions with a branch cut, such as roots and logarithms. When choosing values for $z$ as described in the previous paragraph, one has to ensure that they are always on a definite side of these branch cuts with respect to a chosen path of analytic continuation. In the vicinity of a particular regular singular point we work with implementations of $\sqrt[k]{z}$ and $\operatorname{ln}z$, which have their branch cuts on the negative real axis. Since all regular singular points turn out to be located on the real axis (in terms of the algebraic coordinate $z_{\text{LV}}$ of the regular singular point associated to the large volume limit), all branch cuts are then located on the real axis.\footnote{Note that the negative real axis is mapped to itself under $f:z \mapsto z^{-1}$. Hence, the branch cuts of periods in the vicinity of $z_{\text{LV}}=\infty$ are also located on the negative real axis.} Our convention is to choose all values for $z$ above the real axis of with respect to the coordinate $z_{\text{LV}}$.

Let us close this section with a practical remark: The area of convergence associated to a regular singular point always intersects with that of another regular singular point. It is thus in principle possible to analytically continue the periods at these two points directly to each other. If, however, the overlap of convergence areas is close to the border of converge for one of the points, the corresponding periods will convergence very slowly. For a high numerical precision one would hence have to expand these periods to very high orders, which is computationally expensive. In these situations it can be better, to perform the continuation in several steps via appropriately chosen regular points in between the two singular points.

%%%%%%%%%%%%%%%%%%%%%%%%%%%%%%
\section{Examples} \label{sec:Examples}
%%%%%%%%%%%%%%%%%%%%%%%%%%%%%%
In this section we discuss in detail two examples of Calabi--Yau fourfolds with a single K\"ahler modulus, whose Picard--Fuchs operators are of order six. We explicitly construct a basis of integral periods on the entire quantum K\"ahler moduli space and determine the monodromy matrices in this basis. Furthermore, for these examples we work out the quantum cohomology ring and determine the genus zero Gromov--Witten invariants. Due to the non-maximally unipotent monodromy property at large volume arising from the Picard--Fuchs operators of order six there are two independent genus zero Gromov--Witten invariants at each degree. Using the recursive definition of the Klemm--Pandharipande meeting invariants we deduce the genus one BPS invariants as well. Their non-trivial integrality properties furnish a consistency check on our calculations. Our results for these and further Calabi--Yau fourfold examples are tabulated in Appendix~\ref{app:tables}.

%%%%%%%%%%%%%%%%%%%%%%%%%%%%%%
\subsection{Calabi--Yau fourfold $X_{1,4}\subset\operatorname{Gr}(2,5)$}
\label{sec:Example1}
%%%%%%%%%%%%%%%%%%%%%%%%%%%%%%
We describe the Calabi--Yau fourfold $X_{1,4}$ as a complete intersection of codimension two in the complex six-dimensional Grassmannian $\operatorname{Gr}(2,5)$. The Schubert classes $\sigma_k$ and $\sigma_{k-a,a}$ with $1\le a \le \lfloor\frac{k}2\rfloor$ generate the individual cohomology groups $H^{2k}(\operatorname{Gr}(2,5),\mathbb{Z})$ (while the cohomology ring $H^*(\operatorname{Gr}(2,5),\mathbb{Q})$ is generated by $\sigma_1$ and $\sigma_2$). We realize the family of Calabi--Yau fourfolds $\iota: X_{1,4} \hookrightarrow \operatorname{Gr}(2,5)$ as the zero locus of sections of the rank two bundle $\mathcal{O}(\sigma_1)\oplus\mathcal{O}(4\sigma_1)$, such that $[X_{1,4}]=4\sigma_1^2$ is the class of the Calabi--Yau fourfold in $\operatorname{Gr}(2,5)$. Using standard Schubert calculus techniques --- see, e.g., ref.~\cite{MR1288523} --- together with intersection formula
\begin{equation}
  \int_{X_{1,4}} \iota^* \alpha \,=\, 4\int_{\operatorname{Gr}(2,5)} \sigma_1^2 \cup\alpha \ ,
\end{equation}
we determine the intersection numbers of the Schubert cycles on the Calabi--Yau fourfold~$X_{1,4}$ to be\footnote{For ease of notation we denote the pullbacks $\iota^*\sigma_k$ and $\iota^*\sigma_{k_1,k_2}$ also by $\sigma_k$ and $\sigma_{k_1,k_2}$, respectively.}
\begin{equation} \label{eq:SomeIntegralsX14}
\begin{aligned}
  &\int\limits_{X_{1,4}} \sigma_1^4 = 20 \ ,\  &&\int\limits_{X_{1,4}} \sigma_1^2\cup \sigma_{1,1} = 8 \ , \ 
  &&\int\limits_{X_{1,4}} \sigma_1^2\cup \sigma_2 = 12 \ , \  &&\int\limits_{X_{1,4}} \sigma_1 \cup \sigma_3 = 4\ ,\\
  &\int\limits_{X_{1,4}} \sigma_{1,1}^2 = 4 \ , \quad &&\int\limits_{X_{1,4}} \sigma_{1,1}\cup \sigma_2 = 4 \ ,\ 
  &&\int\limits_{X_{1,4}} \sigma_{2}^2 = 8 \ , \  &&\int\limits_{X_{1,4}} \sigma_{2,2} = 4 \ .
\end{aligned}
\end{equation}  
Combining the Lefschetz hyperplane theorem together with Poincar\'e duality we further deduce the relations $2\sigma_3 \sim \sigma_{2,1}$ and $\sigma_{3,1}\sim\sigma_{2,2}$ among Schubert classes on the Calabi--Yau fourfold $X_{1,4}$ as well as the cohomology generators
\begin{equation}
\begin{aligned}
  H^0(X_{1,4},\mathbb{Z}) &=\langle\!\langle 1 \rangle\!\rangle \ , \   
  &H^2(X_{1,4},\mathbb{Z})&=\langle\!\langle \sigma_{1} \rangle\!\rangle \ , \
  &H^4(X_{1,4},\mathbb{Z})&\supset \langle\!\langle \sigma_{1,1},\sigma_{2} \rangle\!\rangle\ , \\[0.2em]
  H^6(X_{1,4},\mathbb{Z}) &=\langle\!\langle  \tfrac14\sigma_{3}  \rangle\!\rangle \ , \
  &H^8(X_{1,4},\mathbb{Z})&=\langle\!\langle \tfrac14\sigma_{2,2} \rangle \!\rangle \ .
\end{aligned}
\end{equation}
For the middle dimensional cohomology group $H^4(X_{1,4},\mathbb{Z})$ the Lefschetz hyperplane theorem only states that the pullback $\iota^*$ acts injectively. This implies that the classes $\sigma_2$ and $\sigma_{1,1}$ are linearly independent in $H^4(X_{1,4},\mathbb{Z})$. However, these classes are not necessarily integral generators of $H^4(X_{1,4},\mathbb{Z})$. Finally, by adjunction the total Chern class of $X_{1,4}$ reads
\begin{equation}
\begin{aligned}
   c(X_{1,4}) \,&=\, \frac{c(\operatorname{Gr}(2,5))}{(1+\sigma_1)(1+4\sigma_1)} \,=\, 1 + (8 \sigma_{1,1} + 7 \sigma_2) -440 \frac{\sigma _3}{4} + 1\,848 \frac{\sigma _{2,2}}4 \ ,
\end{aligned}   
\end{equation}
which in particular shows that the first Chern class vanishes and determines the Euler characteristic $\chi=1\,848$ of the Calabi--Yau fourfold $X_{1,4}$.

The K\"ahler class of the ambient Grassmannian space reads $J = t \sigma_1$, and it canonically induces the K\"ahler class $J$ on the Calabi--Yau complete intersection $X_{1,4}$. This allows us to determine the asymptotic periods according to
\begin{equation} \label{eq:PerX14}
     \Pi_{\mathcal{E}^\bullet}^{\text{asy}}(t) \,=\, \int_{X_{1,4}} e^{t \sigma_1} \, \Gamma_\mathbb{C}(X_{1,4}) \, \operatorname{ch} \mathcal{E}^{\bullet\,\lor} \ .
\end{equation}     
In addition to the described canonical B-branes we find additional B-branes arising from algebraic four cycles. There is the algebraic four cycle $\mathcal{S}_1$ of the zero section of $\mathcal{O}(\sigma_1)\oplus\mathcal{O}(\sigma_1)$ intersected with $X_{1,4}$ and there is the algberaic four cycle $\mathcal{S}_2$ of the zero section of the rank two universal subbundle $\mathcal{U}$ of $\operatorname{Gr}(2,5)$ intersected with $X_{1,4}$. The associated $4$-brane $\mathcal{S}^\bullet_\ell$ are the push-forwards $\iota_! \mathcal{S}_\ell$ for $\ell=1,2$. Their Chern characters are computed by the Grothendieck--Riemann--Roch theorem and read
\begin{equation}
\label{eq:ChSiX14}
    \operatorname{ch}\,\mathcal{S}^\bullet_1 \,=\, (\sigma_{1,1} + \sigma_2) - 5 \sigma_3 + \frac{35}{12} \sigma_{2,2} \ , \quad
    \operatorname{ch}\,\mathcal{S}^\bullet_2 \,=\, \sigma_{1,1} + \frac12 \sigma_{2,1} + \frac14 \sigma_{2,2} \ .
\end{equation}

For the tuple of B-branes
\begin{equation}
  {\vec{\mathcal{E}}}^\bullet \,=\, \left( \mathcal{E}^\bullet_k \right)_{k=0,\ldots,5} \,=\, 
  \left(  
    \mathcal{O}_\text{pt},\, 
    \mathcal{C}^\bullet[1],\, 
    \mathcal{S}^\bullet_1,\, 
    \mathcal{S}^\bullet_2,\,
    \mathcal{E}^\bullet,\,
    \mathcal{O}_X \right) \ ,
\end{equation}
given in terms of the canonical B-branes together with the $4$-branes $\mathcal{S}_\ell^\bullet$, we now determine with eq.~\eqref{eq:PerX14} their asymptotic integral period vector $\vec\Pi^\text{asy} = \left( \Pi_{\mathcal{E}_k^\bullet}^\text{asy}\right)_{k=0,\ldots,5}$ to be
\begin{equation} \label{eq:IntPeriods}
    \vec\Pi^{\text{asy}}(t) \,=\,
    \begin{pmatrix} 1 \\
     t  \\
     10 t^2+20 t+\frac{107}{6}  \\
    4 t^2-4 t+\frac{7}{2} \\
     -\frac{10}{3} t^3-5 t^2-\frac{19}{2} t-\frac{47}{12}+\frac{55 i \zeta (3)}{\pi^3}  \\
    \frac{5}{6} t^4+\frac{37}{12}t^2 - \frac{55 i  \zeta (3)}{\pi ^3}t  + \frac{7}{144}
    \end{pmatrix} \ .
\end{equation}
The symmetric intersection pairing is readily computed to be
\begin{equation}
\label{eq:SymIntersect}
  \chi({\vec{\mathcal{E}}}^\bullet,{\vec{\mathcal{E}}}^\bullet) \,=\,
  \begin{pmatrix}
0 & 0 & 0 & 0 & 0 & 1 \\
 0 & 0 & 0 & 0 & 1 & 0 \\
 0 & 0 & 20 & 8 & 10 & 24 \\
 0 & 0 & 8 & 4 & -8 & 6 \\
 0 & 1 & 10 & -8 & -14 & -7 \\
 1 & 0 & 24 & 6 & -7 & 2
  \end{pmatrix} \ ,
\end{equation}
which determines the monodromy matrix $M_{t_{\mathcal{O}_X}}$ with the help of eq.~\eqref{eq:STtwist} to be
\begin{equation}
\label{eq:MTox}
  M_{t_{\mathcal{O}_X}} \,=\, 
  \begin{pmatrix}
    1 & 0 & 0 & 0 & 0 & 0 \\
    0 & 1 & 0 & 0 & 0 & 0 \\
    0 & 0 & 1 & 0 & 0 & 0 \\
    0 & 0 & 0 & 1 & 0 & 0 \\
    0 & 0 & 0 & 0 & 1 & 0 \\
   -1 & 0 & -24 & -6 & 7 & -1
  \end{pmatrix} \ .
\end{equation}

%%%%%%%%%%%%%%%%%%%%%%%%%%%%%%
\subsubsection{Picard--Fuchs system}
%%%%%%%%%%%%%%%%%%%%%%%%%%%%%%
In ref. \cite{Honma:2013hma} Honma and Manabe analyze the Calabi--Yau fourfold~$X_{1,4}$ with gauge theory techniques as described in Section~\ref{sec:GaugeTheories}. For the quantum periods they find the order six Picard--Fuchs operator
\begin{equation}
\label{eq:PFX14}
\begin{aligned}
\mathcal{L}(z,\theta)\,=\,&(\theta -1) \theta ^5-8 z(2 \theta +1) (4 \theta +1) (4 \theta +3) \left(11\theta ^2+11 \theta +3\right) \theta\\
&-64 z^2(2 \theta +1) (2 \theta +3) (4 \theta +1) (4 \theta +3) (4 \theta +5) (4 \theta +7)\ .
\end{aligned}
\end{equation}
Here, $z$ is the local algebraic coordinate in the large volume regime. In addition to the large volume limit at $z=0$ there are three additional regular singular points at $z=\infty$, $z=z_1$ and $z_2$, where the latter two points arise from the zero locus of the discriminant factor
\begin{equation} \label{eq:DisX14}
   \Delta(z)=1-2\,816z -65\,536z^2 \ ,
\end{equation}   
i.e., $z_1\approx-0.043$ and $z_2\approx 3.5\cdot 10^{-4}$.

Note that the same discriminant locus~\eqref{eq:DisX14} arises directly in the gauged linear sigma model description of the Calabi--Yau fourfold~$X_{1,4}$. In this context the discriminant locus $\Delta(z)$ describes the locus in the quantum-corrected Fayet--Iliopoulos parameter space with emerging non-compact strata in the gauge theory moduli space \cite{Morrison:1994fr}. Analogously as in refs.~\cite{Hori:2006dk,Jockers:2012zr,Hori:2011pd} --- comparing to the expression~\eqref{eq:DisX14} of the discriminant --- we find that all singularities arise from non-compact strata attributed to the pure Coulomb branch with no contributions from mixed Higgs--Coulomb branches. This observation carries over to all our other examples collected in Appendix~\ref{app:tables} as well.

As described in Section~\ref{sec:Continuation} we are eventually interested in the monodromy matrices expressed in terms of integral periods. To this end, we first have to find a basis of solutions to the Picard--Fuchs equation at all singular points. The structure of these solutions is conveniently summarized by the Riemann P-symbol, which for the present example reads
\begin{equation}\label{eq:PSymbolX14}
\left\{
\begin{array}{cccc}
0 & \infty & z_1 & z_2\\[0.1em]
\hline
0 & \frac14 & 0 & 0 \\[0.1em]
0 & \frac12 & 1 & 1 \\[0.1em]
0 & \frac34 & 2 & 2 \\[0.1em]
0 & \frac54 & 3 & 3 \\[0.1em]
0 & \frac32 & 4 & 4 \\[0.1em]
1 & \frac74 & \frac{3}{2} & \frac{3}{2} \\
\end{array}
\right\}\ .
\end{equation}
Let us briefly recall its meaning: The first row lists the positions of the regular singular points, here given in terms of the algebraic coordinate $z$. To each such point $\tilde{z}$ the symbol associates the six --- i.e., the order of the operator --- rational numbers that are written in the corresponding column below the horizontal line. For example, the symbol associates the numbers $0$, $1$, $2$, $3$, $4$ and $3/2$ to $z_1$. These so called characteristic exponents are the rational roots of the indicial equation
\begin{equation}\label{eq:indicial}
\mathcal{L}_{\tilde{z}}(u_{\tilde{z}},\theta_{\tilde{z}})\, u_{\tilde{z}}^\alpha = \mathcal{O}\left(u_{\tilde{z}}^\alpha\right)\ , \quad \alpha\in\mathbb{Q}
\end{equation}
for the exponent $\alpha$, where $u_{\tilde{z}}$ is a local coordinate on a patch around $\tilde{z}$.\footnote{Explicitly: For $\tilde{z}\neq \infty$ we have $u_{\tilde{z}}= z - \tilde{z}$, otherwise $u_{\tilde{z}}=z^{-1}$. How $\mathcal{L}_{\tilde{z}}(u_{\tilde{z}},\theta_{\tilde{z}})$ can be deduced from $\mathcal{L}(z,\theta)$ has been explained in section~\ref{sec:Continuation}.} The number of times that a particular solution $\alpha_0$ is listed in the corresponding column of the Riemann P-symbol precisely is the order to which it is a root of eq.~\eqref{eq:indicial}. Now let $\alpha_1<\ldots<\alpha_p$ for $1\leq p \leq 6$ be the distinct roots of the indicial equation at $\tilde{z}$, whose respective orders are $m_1,\ldots,m_p$ such that $\sum m_k \alpha_k=6$. A set of linearly independent solutions to the Picard--Fuchs equation on a disk $U_{\tilde{z}}$ around $\tilde{z}$ is then given by
\begin{equation}
\begin{aligned}
\Pi_{\tilde{z}}^{(k,0)}(u_{\tilde{z}}) &=  u_{\tilde{z}}^{\alpha_k}\, \big(1 + \mathcal{O}(u_{\tilde{z}})\big)\ , \\
\Pi_{\tilde{z}}^{(k,l)}(u_{\tilde{z}})&= \Pi_{\tilde{z}}^{(k,0)}\cdot\frac{(\operatorname{ln}u_{\tilde{z}})^l}{(2\pi i)^l} + \mathcal{O}\left((\operatorname{ln}u_{\tilde{z}})^{l-1}\right) \quad \text{with} \quad 1\leq l \leq m_k-1
\end{aligned}
\end{equation}
for all $1\leq k \leq p$. In the vicinity of the regular singular point $\tilde z$ the $m_{k_0}$ solutions $\Pi_{\tilde{z}}^{(k,0)}, \ldots ,\Pi_{\tilde{z}}^{(k,m_k-1)}$ thus transform irreducibly amongst each other when transported around $\tilde{z}$ by $u_{\tilde{z}}\to u_{\tilde{z}}\cdot e^{2\pi i}$, which leads to a non-trivial monodromy due to the branch cut of the logarithm or of a root $\sqrt[k]{z}$. Consequently, the Jordan normal form $J_{M_{\tilde{z}}}$ of the monodromy matrix $M_{\tilde{z}}$ is the block matrix
\begin{equation}
J_{M_{\tilde{z}}} = \left(
\begin{array}{ccc}
J_1 & & \\
 & \ddots & \\
& & J_p
\end{array}
\right)\quad \text{with} \quad J_q=
\left(\begin{array}{cccc}
e^{2\pi i \alpha_q} & 1 & & \\
 & \ddots & \ddots & \\
 & &  e^{2\pi i \alpha_q} & 1 \\
 & & &  e^{2\pi i\alpha_q}
\end{array}\right) \ ,
\end{equation}
where the Jordan block $J_q$ is a matrix of dimension $m_q\times m_q$. 

It would be interesting to see how the information encoded in the Riemann P-symbol relates to the associators for systems of differential equations recently presented in ref.~\cite{Puhlfuerst:2015zqw}. Developing such a relationship promises to shed light on the global analytic structure of solutions to the Picard--Fuchs differential equations.

For the present example of $X_{1,4}$ the Riemann P-symbol in eq.~\eqref{eq:PSymbolX14} shows that the large volume point at $z=0$ does not have maximially unipotent monodromy due to the additional solution
\begin{equation}
\Pi_0^{(2,0)}(z) \,=\,z\, \big(1 + \mathcal{O}(z) \big) \ .
\end{equation}
As a result, the monodromy matrix consists of two Jordan blocks rather than only one. Note that for Calabi--Yau threefolds in general and for those Calabi--Yau fourfolds with order five Picard--Fuchs operators the large volume point is always a regular singular point of maximally unipotent monodromy.

Let us now focus on the large volume point in more detail. As seen from the Riemann P-symbol~\eqref{eq:PSymbolX14} there are two regular solutions with the expansions
\begin{equation} \label{eq:FPex1}
\begin{aligned}
\Pi_0^{(1,0)}(z) &=\Pi_0(z) = 1 + 72 z + 47\,880 z^2 + 54\,331\,200 z^3 + \ldots \ , \\[0.2em]
\Pi_0^{(2,0)}(z) &=  z \left(1+\frac{2\,625 z}{4}+\frac{6\,702\,850 z^2}{9}+\frac{17\,302\,910\,625 z^3}{16}+ \ldots\right)\ , 
\end{aligned}
\end{equation}
while there are four logarithmic solutions $\Pi_l \equiv \Pi_0^{(1,l)}$, $l=1,\ldots,4$, based upon the period $\Pi_0\equiv \Pi_0^{(1,0)}$. The logarithmic period $\Pi_1$ determines the flat coordinate $t$ in the large volume regime according to 
\begin{equation}\label{eq:FlatCo}
t(z) = \frac{\Pi_1(z)}{\Pi_0(z)} = \frac{\operatorname{ln}z + \mathcal{O}(z)}{2\pi i} \ .
\end{equation}
With these ingredients a period vector $\vec{\Pi}=(\Pi_0,\ldots,\Pi_5)^T$ with asymptotic limit $\vec\Pi^{\text{asy}}$ as given by eq.~\eqref{eq:IntPeriods} in general reads
\begin{equation}\label{eq:IntPeriodVector}
\vec{\Pi}(z) = \Pi_0(z)\, \left[\vec\Pi^{\text{asy}}(t(z)) + \mathcal{O}(z)\right] + \frac{1}{\pi^2}\Pi_0^{(2,0)}(z)\, (0,0,\alpha_2,\alpha_3,\alpha_4, \alpha_5)^T\ .
\end{equation}
Note that by the second term on the right hand side of this equation we have added a multiple of $\Pi_0^{(2,0)}(z)$ to the at least doubly logarithmic solutions. This is possible --- in fact it is necessary to make $\vec{\Pi}$ integral --- since $\Pi_0^{(2,0)}(z)$ vanishes in the asymptotic limit $z\to 0$. As the additional period $\Pi^{(2,0)}_0$ relates to the existence of B-branes on the two non-trivial algebraic cycles associated to the described cohomology classes in $H^4(X_{1,4},\mathbb{Z})$ there are no such ambiguities for the quantum periods $\Pi_0$ and $\Pi_1$ for B-branes in higher codimension. Since the values for the integration constants $\alpha_2, \ldots, \alpha_5$, cannot be fixed by large volume asymptotics, we momentarily determine them by analyzing their global properties in the quantum K\"ahler moduli space.

In a next step we analytically continue the period vector $\vec{\Pi}$ to the other three singular points by the method described in Section~\ref{sec:Continuation}. As a result we obtain numerical expressions for the monodromy matrices $M_0$, $M_\infty$, $M_{z_1}$ and $M_{z_2}$ in the large volume basis $\{\Pi_0,\ldots,\Pi_5\}$, which still depend on the parameters $\alpha_2,\ldots,\alpha_5$. As discussed in Section~\ref{sec:Bbranes} we know, however, that one of the monodromy matrices should take the form $M_{t_{\mathcal{O}_X}}$ given in eq.~\eqref{eq:MTox}. For the given example, this match can only be achieved for the monodromy matrix~$M_{z_2}$, whose last row reads\footnote{With the calculated numerical precision we are able to identify the exact numerical rational values. While strictly speaking this is an educated guess, the integrality of the genus zero Gromov--Witten invariants and the determined number of lines --- computed independently in Appendix~\ref{app:InterTheory} via intersection theory --- confirms these rational numbers.}
\begin{equation}
\left(-1\,,\,0\,,\,-24-\frac{\alpha_2}{720}\,,\,-\frac{179}{30}-\frac{\alpha_3}{720}\,,\,7-\frac{\alpha_4}{720}\,,\,-\frac{719}{720}-\frac{\alpha_5}{720}\right)\ .
\end{equation}
By matching this to the last row of the monodromy matrix~$M_{t_{\mathcal{O}_X}}$ in eq.~\eqref{eq:MTox} we identify the parameters as
\begin{equation}
\alpha_2=\alpha_4=0 \ , \quad \alpha_3 =24\ , \quad \alpha_5=1 \ .
\end{equation}
With these values all four monodromy matrices are indeed integral and become
\begin{equation}
\begin{aligned}
M_0&=\left(
\begin{array}{*{6}{>{\scriptstyle}c}}
 1 & 1 & 30 & 0 & 0 & 0 \\
 0 & 1 & 20 & 8 & 0 & 0 \\
 0 & 0 & 1 & 0 & -1 & 0 \\
 0 & 0 & 0 & 1 & 0 & 0 \\
 0 & 0 & 0 & 0 & 1 & -1 \\
 0 & 0 & 0 & 0 & 0 & 1 \\
\end{array}
\right)\ , \quad 
M_\infty =\left(
\begin{array}{*{6}{>{\scriptstyle}c}}
 -19 & -11 & -670 & -72 & 270 & -40 \\
 -40 & -19 & -1340 & -168 & 540 & -80 \\
 2 & 1 & 67 & 8 & -27 & 4 \\
 -3 & -1 & -90 & -13 & 35 & -5 \\
 2 & 1 & 66 & 8 & -27 & 4 \\
 5 & 2 & 156 & 22 & -63 & 9 \\
\end{array}
\right)\ , \\[0.2em]
M_{z_1} &=\left(
\begin{array}{*{6}{>{\scriptstyle}c}}
 21 & 10 & 700 & 80 & -300 & 50 \\
 40 & 21 & 1400 & 160 & -600 & 100 \\
 -2 & -1 & -69 & -8 & 30 & -5 \\
 2 & 1 & 70 & 9 & -30 & 5 \\
 -2 & -1 & -70 & -8 & 31 & -5 \\
 -4 & -2 & -140 & -16 & 60 & -9 \\
\end{array}
\right)\ ,\quad
M_{z_2} = \left(
\begin{array}{*{6}{>{\scriptstyle}c}}
 1 & 0 & 0 & 0 & 0 & 0 \\
 0 & 1 & 0 & 0 & 0 & 0 \\
 0 & 0 & 1 & 0 & 0 & 0 \\
 0 & 0 & 0 & 1 & 0 & 0 \\
 0 & 0 & 0 & 0 & 1 & 0 \\
 -1 & 0 & -24 & -6 & 7 & -1 \\
\end{array}
\right)\ .
\end{aligned}
\end{equation}

Note that these results are in accord with the consistency condition $M_\infty M_{z_2} M_0 M_{z_1} = \mathds{1}$. Moreover, the monodromy matrix $M_{z_2}$ indeed agrees with the expected matrix $M_{t_{\mathcal{O}_X}}$ given in eq.~\eqref{eq:MTox}. According to Section~\ref{sec:Bbranes} this shows that the $8$-brane $\mathcal{O}_X$ (described by the integral period $\Pi_5$) becomes massless at the point $z_2$. We also observe that at $z_1$ --- which is a second point of $\mathbb{Z}_2$-monodromy --- the brane $\mathcal{B}_{z_1}$ associated to the integral period
\begin{equation}
\Pi_{\mathcal{B}_{z_1}}=10 \Pi_0 + 20 \Pi_1 - \Pi_2 + \Pi_3 - \Pi_4 - 2\Pi_5
\end{equation}
becomes massless. The monodromy $M_{z_1}$ is thus described by a Seidel--Thomas twist as in eq.~\eqref{eq:STtwist} with the 8-brane $\mathcal{O}_X$ being replaced by the brane $\mathcal{B}_{z_1}$, with a spherical open-string index $\chi(\mathcal{B}_{z_1},\mathcal{B}_{z_1})=2$. This observation in fact carries over to all examples analzyed in this paper: At every point of $\mathbb{Z}_2$-monodromy there is a vanishing integral period and the monodromy is described by a Seidel--Thomas twist.

As anticipated in the introduction --- due to the non-maximally unipotent monodromy property with respect to the large volume regular singular point of the Picard--Fuchs operator --- the structure of the integral quantum periods of the Calabi--Yau fourfold $X_{1,4}$ indeed admits integral linear combinations, which give rise to flux-induced superpotentials of the form~\eqref{eq:FluxW}. Namely, in terms of the flat coordinate $t$ we find for instance the superpotentials
\begin{equation}
\begin{aligned}
  W^{(1)}_\text{flux}(t) \,&=\, \frac{1}{\Pi_0} \left( 109 \Pi_0 + 360 \Pi_1 -12 \Pi_2 + 30 \Pi_3  \right) = \frac{2\,880}{4\pi^2} e^{2\pi i t} + \mathcal{O}(e^{4\pi i t})  \ , \\
  W^{(2)}_\text{flux}(t) \,&=\, \frac{1}{\Pi_0} \left(60 \Pi_1 -2 \Pi_2 + 5 \Pi_3  \right) = -\frac{109}{6} + \frac{480}{4\pi^2} e^{2\pi i t} + \mathcal{O}(e^{4\pi i t})  \ , \\
  W^{(3)}_\text{flux}(t) \,&=\, \frac{1}{\Pi_0} \left( 109 \Pi_0  -12 \Pi_2 + 30 \Pi_3  \right) = -360\,t + \frac{2\,880}{4\pi^2} e^{2\pi i t} + \mathcal{O}(e^{4\pi i t})  \ .
\end{aligned}
\end{equation}
Here, the integral coefficients in the presented linear combinations should be interpreted as flux quantum numbers. The leading non-perturbative terms arise from genus zero worldsheet instantons, which we study in the next subsection in the context of the quantum cohomology ring of the Calabi--Yau fourfold~$X_{1,4}$.

%%%%%%%%%%%%%%%%%%%%%%%%
\subsubsection{Gromov--Witten invariants and quantum cohomology ring}
In section~\ref{sec:QuantumCohomology} we have introduced the quantum cohomology ring of Calabi--Yau fourfolds with Picard--Fuchs operators of non-minimal order. We now explicitly determine the quantum cohomology ring and calculate the genus zero Gromov--Witten invariants $n^{(a)}_{0,d}$ of the Calabi--Yau fourfold~$X_{1,4}$. Furthermore, with the help of the Klemm--Pandharipande meeting invariants we also infer the genus one invariants $n_{1,d}$ defined in eq.~\eqref{eq:MCg1}.

%%%%%%%%%%%%%%%%%%%%%%%%
\begin{table}[t]\centering
\footnotesize{
\begin{tabular}{|r|r|r|}
\hline $d$ & $n^{(1)}_{0,d}$ & $n^{(2)}_{0,d}$ \\
\hline  1 & 400 & 520 \\
 2 & 208\,240 & 226\,480 \\
 3 & 175\,466\,480 & 191\,464\,760 \\
 4 & 196\,084\,534\,160 & 213\,155\,450\,240 \\
 5 & 255\,402\,582\,828\,400 & 277\,092\,686\,601\,400 \\
 6 & 367\,048\,595\,782\,193\,680 & 397\,700\,706\,634\,553\,680 \\
 7 & 564\,810\,585\,071\,858\,496\,880 & 611\,416\,342\,763\,726\,567\,800 \\
 8 & 913\,929\,133\,261\,543\,393\,001\,760 & 988\,670\,017\,271\,687\,389\,572\,480 \\
 9 & 1\,536\,929\,129\,164\,031\,410\,293\,358\,720 & 1\,661\,748\,145\,541\,449\,358\,296\,013\,440 \\
 10 & 2\,664\,576\,223\,763\,330\,924\,317\,069\,072\,400 & 2\,879\,777\,881\,450\,393\,936\,532\,565\,976\,400 \\ \hline
\end{tabular}}
\caption{\label{tab:Genus0X14}Genus zero Gromov--Witten invariants $n^{(1)}_{0,d}$ and $n^{(2)}_{0,d}$ of the Calabi--Yau fourfold~$X_{1,4}$ associated to $\phi_{2,(1)} = \sigma_{1,1}$ and $\phi_{2,(2)} = \sigma_{2}$ up to degree $d=10$.}
\end{table}
%%%%%%%%%%%%%%%%%%%%%%%%%

First of all, with the classical ring structure encoded in the intersections~\eqref{eq:SomeIntegralsX14}, we determine the genus zero Gromov--Witten invariants from the identity
\begin{equation}\label{eq:Genus0Formula}
\frac{\partial^2}{\partial t^2}\frac{\Pi_{\mathcal{S}^\bullet_i}(z(t))}{\Pi_{\mathcal{O}_\text{pt}}(z(t))} = \int\limits_{X_{1,4}} \left(\sigma_1 * \sigma_1 \right)\cup \operatorname{ch} \mathcal{S}^\bullet_i\ , \qquad i=1,2 \ ,
\end{equation}
in terms of the mirror map $z(t)$ for the flat coordinate $t$. Note that this formula holds because in Gromov--Witten theory the metric for the chiral--anti-chiral operators is identified with the classical intersection pairing. Since we have previously determined the integral periods, the left hand side of this equation is known. Using the intersection numbers~\eqref{eq:SomeIntegralsX14}, the explicit Chern characters~\eqref{eq:ChSiX14} as well as the identification $\phi_{2,(1)} = \sigma_{1,1}$ and $\phi_{2,(2)} = \sigma_{2}$ in the quantum product~\eqref{eq:QuantumProduct}, we arrive at 
\begin{equation}
\begin{aligned}
\frac{\partial^2}{\partial t^2}\frac{\Pi_{\mathcal{S}^\bullet_1}}{\Pi_{\mathcal{O}_\text{pt}}} &= 20 + \sum_{d=1}^\infty d^2\frac{q^d}{1-q^d} \left(8n^{(1)}_{0,d}+12n^{(2)}_{0,d}\right)\ , \\
\frac{\partial^2}{\partial t^2}\frac{\Pi_{\mathcal{S}^\bullet_2}}{\Pi_{\mathcal{O}_\text{pt}}} &= 8 + \sum_{d=1}^\infty d^2\frac{q^d}{1-q^d} \left(4n^{(1)}_{0,d}+4n^{(2)}_{0,d}\right)\ .
\end{aligned}
\end{equation}
By expanding these equations in $q$ we obtain two independent equations for each degree $d$ and are thus able to identify the unknowns $n^{(1)}_{0,d}$ and $n^{(2)}_{0,d}$. We have checked integrality up to degree $50$ and list the numbers up to degree $10$ in Table~\ref{tab:Genus0X14}. With the help of the recursive definition~\eqref{eq:MeetingInvariants} we further deduce the associated Klemm--Pandharipande meeting invariants listed in Table~\ref{tab:MeetingInvariantsX14}.

In Appendix~\ref{app:InterTheory} we employ intersection theory techniques to directly compute the number of lines with a marked point restricted to the codimension two Schubert classes $\sigma_{1,1}$ and $\sigma_2$. As further explained there, these results are in agreement with the genus zero Gromov--Witten invariants $n_{0,1}^{(1)}=400$ and $n_{0,1}^{(2)}=520$ at degree one. This provides for yet another independent consistency check on the linear combinations of the obtained integral quantum periods.

%%%%%%%%%%%%%%%%%%%%
\begin{table}[t]\centering
\footnotesize{
\begin{tabular}{|l|r|r|r|r|}\hline
$m_{k,l}$ & $l=1$ & $l=2$ & $l=3$ & $l=4$ \\\hline
 $k=1$ & 4\,536\,960 & 2\,075\,384\,960 & 1\,750\,629\,048\,960 & 1\,951\,117\,108\,140\,160 \\
 $k=2$ & \text{} & 961\,126\,562\,880 & 811\,503\,225\,375\,360 & 904\,721\,970\,681\,455\,680 \\
 $k=3$ & \text{} & \text{} & 685\,189\,180\,065\,298\,560 & 763\,898\,769\,976\,093\,842\,560 \\
 $k=4$ & \text{} & \text{} & \text{} & 851\,650\,443\,220\,977\,804\,680\,320 \\\hline
\end{tabular}}
\caption{\label{tab:MeetingInvariantsX14}Klemm--Pandharipande meeting invariants $m_{k,l}\equiv m_{l,k}$ of the Calabi--Yau fourfold~$X_{1,4}$ up to degree four. For ease of presentation we only list the invariants for $k \le l$. }
\end{table}
%%%%%%%%%%%%%%%%%%%%%

Our findings are consistent with the results presented by Honma and Manabe in ref.~\cite{Honma:2013hma}. There the quantum correlator genus zero invariants $n_{0,d}(\phi_{2,(a)})$ are computed, as for instance also used in ref.~\cite{Klemm:2007in}. With the identification $\phi_{2,(1)}=\sigma_1^2=H_1$ and $\phi_{2,(2)}=5\sigma_2-3\sigma_1^2=H_2$ these invariants are related to the quantum cohomology ring invariants $n^{(1)}_{0,d}$ and $n^{(2)}_{0,d}$ according to
\begin{equation}
\begin{aligned}
n_{0,d}(H_1) &= \int\limits_{X_{1,4}} \sigma_1^2 \cup \left(n^{(1)}_{0,d} \sigma_{1,1}+n^{(2)}_{0,d}\sigma_2\right) = 8 n^{(1)}_{0,d}+12 n^{(2)}_{0,d} \ ,\\
n_{0,d}(H_2) &= \int\limits_{X_{1,4}} \left(5\sigma_2-3\sigma_1^2\right) \cup \left(n^{(1)}_{0,d} \sigma_{1,1}+n^{(2)}_{0,d}\sigma_2\right) = -4 n^{(1)}_{0,d}+4 n^{(2)}_{0,d} \ .
\end{aligned}
\end{equation}
We note that integrality of $n^{(1)}_{0,d}$ and $n^{(2)}_{0,d}$ implies integrality of $n_{0,d}(H_1)$ and $n_{0,d}(H_2)$, while the converse is not true.

Finally, we want to determine the genus one invariants $n_{1,d}$ from the quantity $F_1^{\text{top}}$ specified in eq.~\eqref{eq:F1}. The discriminant locus has two rational factors, namely the large volume divisor $\Delta_\text{LV}=z$ and the discriminant factor $\Delta$ of eq.~\eqref{eq:DisX14}. From the asymptotic behaviour of $F_1^{\text{top}}$ at large volume~\eqref{eq:F1LV} and at the conifold~\eqref{eq:F1Coni} the coefficients $b_1$ and $b_2$ reflecting the holomorphic ambiguity are determined to be
\begin{equation}
1+b_1 =- \frac1{24} \int_X c_3(X) \cup J = \frac{55}{3}\quad \text{and}\quad b_2 = -\frac{1}{24} \ .
\end{equation} 
With the Euler characteristic $\chi = 1\,848$ we thus have
\begin{equation}
F_1^{\text{top}}\,=\, 74\log \Pi_{\mathcal{O}_\text{pt}}+\log \left(\frac1{2\pi i} \frac{\partial z}{\partial t} \right) +\frac{52\, \log z}{3} -\frac{\log (1-2\,816z-65\,536z^2)}{24} \ .
\end{equation}
In the asymptotic large volume limit $z\to 0$ and after reexpressing $z$ in terms of the variable $q$ this expression reduces to
\begin{equation}
\begin{aligned}
F_1^{\text{top}} &= \frac{55 \log (q)}{3}-\frac{8\,720 q}{3}-1\,163\,440 q^2-\frac{8\,709\,831\,680 q^3}{9} + \ldots \\
&=  \frac{55 \log (q)}{3} + \sum_{d=1}^\infty N_{1,d}\, q^d\ .
\end{aligned}
\end{equation}
Hence, we can read of the rational genus one invariants $N_{1,d}$. By the multicovering formula~\eqref{eq:MCg1} these are then translated into the integral genus one invariants, the first few of which are listed in Table~\ref{tab:Genus1X14}. We have checked integrality up to degree $50$, and we observe that $n_{1,1}=n_{1,2}=0$.
\begin{table}\centering
\begin{tabular}{|r|r|r|}
\hline $d$ & $n_{1,d}$ \\\hline
 1 & 0 \\
 2 & 0 \\
 3 & -3\,200 \\
 4 & 370\,151\,480 \\
 5 & 4\,108\,408\,756\,800 \\
 6 & 19\,279\,169\,520\,232\,000 \\
 7 & 66\,081\,794\,099\,798\,279\,680 \\
 8 & 194\,122\,441\,310\,522\,439\,007\,040 \\
 9 & 522\,534\,128\,159\,184\,581\,441\,465\,280 \\
 10 & 1\,332\,480\,344\,031\,795\,460\,733\,665\,780\,608 \\\hline
\end{tabular}
\caption{\label{tab:Genus1X14}Integral genus one Gromov--Witten invariants $n_{1,d}$ of $X_{1,4}$ up to degree $d=10$.}
\end{table}

\subsection{Skew Symmetric Sigma Model Calabi--Yau fourfold $X_{1,17,7}$} \label{sec:Example2}
As our second example we consider the Calabi--Yau fourfold $X_{1,17,7}$ arising as the large volume phase of a certain gauged linear sigma model \cite{Gerhardus:2015sla}. It is the non-complete intersection projective variety
\begin{equation}
   X_{1,17,7} \,=\, \left\{ [x,\omega] \in \mathbb{P}(V \oplus \Lambda^2 V^*) \,\middle|\, \operatorname{rk} \omega \le 2 \,,\, x \in \ker \omega \right\} \cap \mathbb{P}(L) \ ,
\end{equation}
with the vector space $V=\mathbb{C}^7$ and a generic 17~dimensional subspace $L \subset V \oplus \Lambda^2 V^*$. In the following we use the isomorphsim to the incidence correspondence of ref.~\cite{Gerhardus:2015sla} to describe $X_{1,17,7}$ as
\begin{equation}
   X_{1,17,7} \,\simeq\, \left\{ (x,p) \in \mathbb{P}^{16}\times\operatorname{Gr}(2,7) \,\middle|\, G(x,p) = 0  \right\}  \ .
\end{equation}
Here $G(x,p)$ is a generic section of the rank $22$ bundle $\mathcal{B}$
\begin{equation}
    \mathcal{B} \,=\, \frac{\mathcal{O}(1)\otimes\Lambda^2 V^*}{\mathcal{O}(1) \otimes \Lambda^2\mathcal{U}} \oplus 
    \left( \mathcal{O}(1) \otimes \mathcal{U}^* \right) \ ,
\end{equation}
in terms of the hyperplane bundle $\mathcal{O}(1)$ of the projective space $\mathbb{P}^{16}$ and the rank two universal subbundle $\mathcal{U}$ of the Grassmannian $\operatorname{Gr}(2,7)$. In particular, the class $[X_{1,17,7}]$ of the Calabi--Yau fourfold $\iota: X_{1,17,7} \hookrightarrow \mathbb{P}^{16}\times\operatorname{Gr}(2,7)$ becomes the top Chern class of the bundle $\mathcal{B}$, i.e., 
\begin{equation}
    [X_{1,17,7}] \,=\, c_{22}(\mathcal{B}) \ ,
\end{equation}
which is given in terms of the hyperplane class $H$ of $\mathbb{P}^{16}$ and the Schubert classes $\sigma_2$ of $\operatorname{Gr}(2,7)$.\footnote{For a review on Schubert classes see for instance ref.~\cite{MR1288523}.} Then --- for cohomology classes $\iota^*\alpha$ pulled back from the ambient space $\mathbb{P}^{16}\times\operatorname{Gr}(2,7)$ --- we compute the intersection numbers of $X_{1,17,7}$ according to\footnote{For ease of notation, in the following we suppress the pullback for the cohomology class on $X_{1,17,7}$ induced from the ambient space.}
\begin{equation}
   \int_{\mathbb{P}^{16}\times\operatorname{Gr}(2,7)} c_{22}(\mathcal{B}) \cup \alpha \,=\, \int_{X_{1,17,7}} \iota^*\alpha \ .
\end{equation}
Hence, we arrive at the intersection numbers
\begin{equation} \label{eq:SomeIntegralsX1177}
  \int\limits_{X_{1,17,7}} H^4 = 98 \ ,\quad  \int\limits_{X_{1,17,7}} \sigma_2\cup\sigma_2 = 44 \ , \quad  \int\limits_{X_{1,17,7}} \sigma_2\cup H^2 = 65 \  .
\end{equation}  
 Note that on the variety $X_{1,17,7}$ we have in cohomology the equivalences $H \simeq \sigma_1$ (c.f., ref.~\cite{Gerhardus:2015sla}), $16H^3\simeq 49\sigma_3$, and $33H^3 \simeq 98 \sigma_{2,1}$, as well as $11H^4 \simeq 98\sigma_4$, $21 H^4 \simeq 98\sigma_{3,1}$, and $6 H^4\simeq 49\sigma_{2,2}$. As a result we obtain the integral cohomology generators 
\begin{equation}
\begin{aligned}
  H^0(X_{1,17,7},\mathbb{Z}) &=\langle\!\langle 1 \rangle\!\rangle \ ,  
  &H^2(X_{1,17,7},\mathbb{Z})&=\langle\!\langle H \rangle\!\rangle \ , 
  &H^4(X_{1,17,7},\mathbb{Z})&\supset \langle\!\langle H^2,\sigma_{2} \rangle\!\rangle\ , \\[0.2em]
  H^6(X_{1,17,7},\mathbb{Z}) &=\langle\!\langle  \tfrac1{98}H^3  \rangle\!\rangle \ , 
  &H^8(X_{1,17,7},\mathbb{Z})&=\langle\!\langle \tfrac1{98}H^4 \rangle \!\rangle \ .
\end{aligned}
\end{equation}
Similarly as for the previously discussed Calabi--Yau fourfold, the classes $H^2$ and $\sigma_2$ are integral but not necessarily integral generators of $H^4(X_{1,17,7},\mathbb{Z})$. Finally, the total Chern class of the Calabi--Yau fourfold $X_{1,17,7}$ is given by
\begin{equation}
  c(X_{1,17,7}) \,=\, \frac{c(\mathbb{P}^{16}) c(\operatorname{Gr}(2,7))}{c(\mathcal{B})}
   \,=\, 1 + (4 H^2 - 2 \sigma_2) - 328 \frac{H^3}{98} + 672 \frac{H^4}{98} \  ,   
\end{equation}
i.e., the first Chern class vanishes and $\chi=672$ is the Euler characteristic of the Calabi--Yau fourfold $X_{1,17,7}$.

Apart from the canonical B-branes $\mathcal{O}_\text{pt}$, $\mathcal{C}^\bullet[1]$, $\mathcal{E}^\bullet$ and $\mathcal{O}_X$, we construct the 4-branes $\mathcal{S}_1^\bullet$ and $\mathcal{S}_2^\bullet$ associated to the algebraic surfaces $\mathcal{S}_1$ and $\mathcal{S}_2$ of the zero sections of the rank two bundles $\mathcal{O}(1)^{\oplus 2}$ and $\mathcal{U}$ intersected with $X_{1,17,7}$. The Chern characters of the constructed 4-branes is computed by the Grothendieck--Riemann--Roch theorem to be
\begin{equation}\label{eq:ChernCharModel2}
\begin{aligned}
    \operatorname{ch}\,\mathcal{S}^\bullet_1 \,&=\, H^2 - H^3 + \frac{7}{12} H^4 \ , \\
    \operatorname{ch}\,\mathcal{S}^\bullet_2 \,&=\, (H^2 - \sigma_2) + \frac12 ( H^3 - H \sigma_2) + \frac1{12} (H^4 - \sigma_2^2) \ .
\end{aligned}    
\end{equation}
With respect to the B-branes
$
  {\vec{\mathcal{E}}}^\bullet \,=\, \left( \mathcal{E}^\bullet_k \right)_{k=0,\ldots,5} \,=\, 
  \left(  
    \mathcal{O}_\text{pt},\, 
    \mathcal{C}^\bullet[1],\, 
    \mathcal{S}^\bullet_1,\, 
    \mathcal{S}^\bullet_2,\,
    \mathcal{E}^\bullet,\,
    \mathcal{O}_X \right)
$
the asymptotic periods $\vec\Pi^\text{asy} = \left( \Pi_{\mathcal{E}_k^\bullet}^\text{asy}\right)_{k=0,\ldots,5}$ for the Calabi--Yau fourfold $X_{1,17,6}$ become
\begin{equation} \label{eq:IntPeriods2}
    \vec\Pi^{\text{asy}}(t) \,=\,
    \begin{pmatrix} 1 \\
     t  \\
     49 t^2+98 t+\frac{817}{12}  \\
    \frac{33}2 t^2-\frac{33}2 t+\frac{33}{4} \\
     -\frac{49}{3} t^3-\frac{49}2 t^2-\frac{109}{4} t-\frac{229}{24}+\frac{41 i \zeta (3)}{\pi^3}  \\
    \frac{49}{12} t^4+\frac{131}{24}t^2 - \frac{41 i  \zeta (3)}{\pi ^3}t  + \frac{7}{18}
    \end{pmatrix} \ .
\end{equation}
The symmetric intersection pairing is readily computed to be
\begin{equation}
  \chi({\vec{\mathcal{E}}}^\bullet,{\vec{\mathcal{E}}}^\bullet) \,=\,
  \begin{pmatrix}
 0 & 0 & 0 & 0 & 0 & 1 \\
 0 & 0 & 0 & 0 & 1 & 0 \\
 0 & 0 & 98 & 33 & 49 & 79 \\
 0 & 0 & 33 & 12 & -33 & 12 \\
 0 & 1 & 49 & -33 & -30 & -15 \\
 1 & 0 & 79 & 12 & -15 & 2
  \end{pmatrix} \ ,
\end{equation}
which determines the monodromy matrix $M_{t_{\mathcal{O}_X}}$ with eq.~\eqref{eq:STtwist} to be
\begin{equation}
\label{eq:MTox2}
  M_{t_{\mathcal{O}_X}} \,=\, 
  \begin{pmatrix}
    1 & 0 & 0 & 0 & 0 & 0 \\
    0 & 1 & 0 & 0 & 0 & 0 \\
    0 & 0 & 1 & 0 & 0 & 0 \\
    0 & 0 & 0 & 1 & 0 & 0 \\
    0 & 0 & 0 & 0 & 1 & 0 \\
   -1 & 0 & -79 & -12 & 15 & -1
  \end{pmatrix} \ .
\end{equation}

\subsubsection{Picard--Fuchs system}
In ref.~\cite{Gerhardus:2015sla} we have calculated the two sphere partition function of a gauged linear sigma model, which in its large volume phase realizes the non complete intersection fourfold $X_{1,17,7}$. From this calculation the fundamental period has been found as
\begin{equation}
\label{eq:FundPeriod2}
\Pi_0(z) = 1 + 9 z + 469 z^2 + 38\,601 z^3 + 4\,008\,501 z^4 + \ldots \ ,
\end{equation}
where $z = z_{\text{LV}}$ is a coordinate around the large point. This period is annihilated by the order six Picard--Fuchs operator
\begin{equation}
\label{eq:PFModel2}
\begin{aligned}
\mathcal{L}(z) = &+316\,932 (\theta -1) \theta ^5 -98 z\, \theta  \big[700\,453 \theta ^5+1\,335\,058 \theta ^4+1\,609\,080 \theta ^3+879\,285 \theta ^2\\
&+249\,018 \theta +29\,106\big]+962754229z^2\big[ \theta ^6-1\,976\,960\,883 \theta ^5-10\,395\,509\,031 \theta ^4\\
&-14\,991\,662\,969 \theta ^3-10\,456\,423\,600 \theta ^2-3\,667\,629\,910 \theta -521\,151\,456 \big] \\
&+2 z^3 \big[9\,812\,727\,979 \theta ^6+53\,190\,263\,573 \theta ^5+105\,895\,432\,463 \theta ^4\\
&+103\,996\,363\,801 \theta ^3 +54\,017\,188\,106 \theta ^2+14\,078\,111\,747 \theta +1\,415\,445\,066\big]\\
&-2 z^4\big[11\,549\,486\,896 \theta ^6+46\,324\,321\,804 \theta ^5+73\,290\,469\,426 \theta ^4\\
&+60\,074\,870\,026 \theta ^3+27\,353\,847\,169 \theta ^2+6\,669\,746\,719 \theta +696\,036\,075\big]\\
&+174z^5 \big(1\,666\,198 \theta ^6+6\,006\,981 \theta ^5+10\,497\,819 \theta ^4+11\,551\,078 \theta ^3+8\,162\,130 \theta ^2\\
&+3\,331\,047 \theta +588\,537\big]-211\,932 z^6 (\theta +1)^5 (2 \theta +3) \ .
\end{aligned}
\end{equation}
In addition to the singular points at $z=0$ and $z=\infty$ there might be singularities at the zero loci of the polynomial multiplying $\theta^6$ in $\mathcal{L}(z)$,
\begin{equation}
\begin{aligned}
h^{(6)}_{\text{LV}}(z)= -&\left(1 - 188 z - 2\,368 z^2 + 4 z^3\right)\cdot\\
&\left(-316\,932 + 9\,061\,178 z - 9\,747\,741 z^2 +105\,966 z^3\right)\ .
\end{aligned}
\end{equation}
It turns out, however, that at the zeros of the second factor in this polynomial there are six regular solutions. Consequently, these are regular points. On the other hand, the zero loci
\begin{equation}
z_1 \approx-0.084\ , \quad z_2 \approx592.079\ , \quad z_3 \approx0.005 \ ,
\end{equation}
of the first factor, $\Delta(z) = 1 - 188 z - 2\,368 z^2 + 4 z^3$, are indeed singular. The Riemann P-symbol reads
\begin{equation}\label{eq:PSymbolX1177}
\left\{
\begin{array}{ccccc}
0 & \infty & z_1 & z_2 & z_3\\[0.1em]
\hline
0 & 1 & 0 & 0 & 0 \\[0.1em]
0 & 1 & 1 & 1 & 1\\[0.1em]
0 & 1 & 2 & 2 & 2\\[0.1em]
0 & 1 & 3 & 3 & 3\\[0.1em]
0 & 1 & 4 & 4 & 4\\[0.1em]
1 & \frac32 & \frac{3}{2} & \frac{3}{2} & \frac{3}{2}\\
\end{array}
\right\}\ .
\end{equation}

We note that the large volume point $z=0$ again does not have maximally unipotent monodromy. Its structure is, in fact, the same as for the Grassmannian example discussed in the previous section: In addtion to the fundamental period $\Pi_0=\Pi^{(1,0)}_0$ given in eq.~\eqref{eq:FundPeriod2} there is second regular solution,
\begin{equation}\label{eq:RegSol2Model2}
\Pi^{(2,0)}_0 = z\left(1+\frac{6\,125 z}{132}+\frac{1\,524\,635 z^2}{396}+\frac{210\,992\,845 z^3}{528}+\ldots\right)\ .
\end{equation}
The singly logarithmic period, $\Pi_1=\Pi^{(1,1)}_0$, defines the flat coordinate $t$ as in eq.~\eqref{eq:FlatCo} and the period vector $\vec{\Pi}=(\Pi_0,\ldots,\Pi_5)^T$ is as in eq.~\eqref{eq:IntPeriodVector} with the asymptotic limit $\vec{\Pi}^{\text{asy}}$ now given by eq.~\eqref{eq:IntPeriods2}.

By an analytic continuation of $\vec{\Pi}$ to the other four singular points we then again obtain numerical expressions for the monodromy matrices $M_0$, $M_\infty$, $M_{z_1}$, $M_{z_2}$ and $M_{z_3}$ in the large volume basis $\{\Pi_0,\ldots,\Pi_5\}$. Among these matrices only $M_{z_3}$ can possibly agree with $M_{t_{\mathcal{O}_X}}$ in eq.~\eqref{eq:MTox2}. Hence, we compare its last line
\begin{equation}
\left(-1\,,\,0\,,\,-79-\frac{\alpha_2}{99}\,,\,-\frac{4\,675}{392}-\frac{\alpha_3}{99}\,,\,15-\frac{\alpha_4}{99}\,,\,-\frac{4\,675}{4\,704}-\frac{\alpha_5}{99}\right) \ ,
\end{equation}
to the last line of $M_{t_{\mathcal{O}_X}}$ and deduce
\begin{equation}
\alpha_2=\alpha_4=0 \ , \quad \alpha_3 =\frac{2\,871}{392}\ , \quad \alpha_5=\frac{957}{1\,568} \ .
\end{equation}
Inserting these values indeed makes all five monodromy matrices integral and they read
\begin{align}
M_0&=\left(
\begin{array}{*{6}{>{\scriptstyle}c}}
 1 & 1 & 147 & 0 & 0 & 0 \\
 0 & 1 & 98 & 33 & 0 & 0 \\
 0 & 0 & 1 & 0 & -1 & 0 \\
 0 & 0 & 0 & 1 & 0 & 0 \\
 0 & 0 & 0 & 0 & 1 & -1 \\
 0 & 0 & 0 & 0 & 0 & 1 \\
\end{array}
\right) \ , \quad
M_{z_1}=\left(
\begin{array}{*{6}{>{\scriptstyle}c}}
 99 & 49 & 13328 & 588 & -3724 & 343 \\
 196 & 99 & 26656 & 1176 & -7448 & 686 \\
 -2 & -1 & -271 & -12 & 76 & -7 \\
 2 & 1 & 272 & 13 & -76 & 7 \\
 -2 & -1 & -272 & -12 & 77 & -7 \\
 -4 & -2 & -544 & -24 & 152 & -13 \\
\end{array}
\right), \nonumber \\[0.1em]
M_{z_2}&=\left(
\begin{array}{*{6}{>{\scriptstyle}c}}
 4117 & 1568 & 478828 & 23520 & -115248 & 8232 \\
 4809 & 1833 & 559447 & 27480 & -134652 & 9618 \\
 -84 & -32 & -9771 & -480 & 2352 & -168 \\
 0 & 0 & 0 & 1 & 0 & 0 \\
 -168 & -64 & -19544 & -960 & 4705 & -336 \\
 -441 & -168 & -51303 & -2520 & 12348 & -881 \\
\end{array}
\right) \ , \
M_{z_3}=\left(
\begin{array}{*{6}{>{\scriptstyle}c}}
 1 & 0 & 0 & 0 & 0 & 0 \\
 0 & 1 & 0 & 0 & 0 & 0 \\
 0 & 0 & 1 & 0 & 0 & 0 \\
 0 & 0 & 0 & 1 & 0 & 0 \\
 0 & 0 & 0 & 0 & 1 & 0 \\
 -1 & 0 & -79 & -12 & 15 & -1 \\
\end{array}
\right) , \nonumber \\[0.1em]
M_\infty&=\left(
\begin{array}{*{6}{>{\scriptstyle}c}}
 -3975 & -1490 & -459291 & -22890 & 110201 & -7854 \\
 -1757 & -617 & -197897 & -10479 & 46942 & -3318 \\
 67 & 25 & 7728 & 387 & -1853 & 132 \\
 121 & 47 & 14181 & 682 & -3423 & 245 \\
 151 & 57 & 17499 & 867 & -4205 & 300 \\
 198 & 74 & 22862 & 1146 & -5487 & 391 \\
\end{array}
\right)\ .
\end{align}
Note that $M_{z_3} = M_{t_{\mathcal{O}_X}}$ and that the consistency condition $M_\infty M_{z_2} M_{z_3} M_0 M_{z_1} = \mathds{1}$ is fulfilled. While this shows that the $8$-brane $\mathcal{O}_X$ becomes massless at $z_3$, the integral periods
\begin{equation}
\begin{aligned}
\Pi_{\mathcal{B}_{z_1}} &= 49 \Pi_0 + 98 \Pi_1 - \Pi_2+\Pi_3-\Pi_4 - 2\Pi_5 \ , \\
\Pi_{\mathcal{B}_{z_2}} &= -196 \Pi_0 -229 \Pi_1 +4 \Pi_2+8\Pi_4 +21\Pi_5 \ ,
\end{aligned}
\end{equation}
vanish at $z_1$ and $z_2$, respectively. Hence, at each point of $\mathbb{Z}_2$-monodromy there is a massless brane and the monodromies are described by Seidel--Thomas twists. The regular singular point at infinity will be discussed in Section~\ref{sec:PointInf}.

Due to the non-minimal order property of the Calabi--Yau fourfold $X_{1,17,7}$ we can find integral quantum periods, which give rise to flux-induced superpotentials of the form~\eqref{eq:FluxW}. In terms of the flat coordinate $t$ we for instance have
\begin{equation}
\begin{aligned}
  W^{(1)}_\text{flux}(t) \,&=\, \frac{1}{\Pi_0} \left(5\,753 \Pi_0 + 19\,404 \Pi_1 -132 \Pi_2 + 392 \Pi_3  \right) = \frac{11\,484}{4\pi^2} e^{2\pi i t} + \mathcal{O}(e^{4\pi i t})  \ , \\
  W^{(2)}_\text{flux}(t) \,&=\, \frac{1}{\Pi_0} \left(4\,851 \Pi_1 -33 \Pi_2 + 98 \Pi_3  \right) = -\frac{5\,753}{4} + \frac{2\,781}{4\pi^2} e^{2\pi i t} + \mathcal{O}(e^{4\pi i t})  \ , \\
  W^{(3)}_\text{flux}(t) \,&=\, \frac{1}{\Pi_0} \left(5\,753 \Pi_0  -132 \Pi_2 + 392 \Pi_3  \right) = -19\,404\,t + \frac{11\,484}{4\pi^2} e^{2\pi i t} + \mathcal{O}(e^{4\pi i t})  \ .
\end{aligned}
\end{equation}

%%%%%%%%%%%%%%%%%%%%%%%%
\begin{table}[t]\centering
\footnotesize{
\begin{tabular}{|r|r|r|}
\hline $d$ & $n^{(1)}_{0,d}$ & $n^{(2)}_{0,d}$ \\
\hline 
 1 & 0 & 33 \\
 2 & 721 & 170  \\
 3 & 38\,255 & 16\,126  \\
 4 & 3\,042\,676 & 1\,141\,312 \\
 5 & 274\,320\,123 & 100\,955\,257 \\
 6 & 27\,276\,710\,118 & 9\,821\,360\,694 \\
 7 & 2\,897\,092\,850\,989 & 1\,028\,274\,636\,900 \\
 8 & 323\,207\,209\,581\,582 & 113\,458\,193\,073\,000 \\
 9 & 37\,444\,642\,819\,824\,776 & 13\,032\,484\,062\,881\,000 \\
 10 & 4\,469\,922\,540\,366\,355\,762 & 1\,545\,108\,865\,260\,914\,434 \\
 \hline
\end{tabular}}
\caption{\label{tab:Genus0Model2}Genus zero Gromov--Witten invariants $n^{(1)}_{0,d}$ and $n^{(2)}_{0,d}$ of the Calabi--Yau fourfold~$X_{1,17,7}$ associated to $\phi_{2,(1)} = H^2$ and $\phi_{2,(2)} = \sigma_{2}$ up to degree $d=10$.}
\end{table}
%%%%%%%%%%%%%%%%%%%%%%%%%

\subsubsection{Gromov--Witten invariants and quantum cohomology ring}
To determine the Gromov-Witten invariants $n^{(a)}_{0,d}$ of the Calabi--Yau fourfold $X_{1,17,7}$ we insert the intersection numbers \eqref{eq:SomeIntegralsX1177}, the explicit Chern characters~\eqref{eq:ChernCharModel2} as well as the identifications $\phi_{2,(1)}=H^2$ and $\phi_{2,(2)}=\sigma_2$ into eq.~\eqref{eq:Genus0Formula}. This yields the two equations
\begin{equation}
\begin{aligned}
\frac{\partial^2}{\partial t^2}\frac{\Pi_{\mathcal{S}^\bullet_1}}{\Pi_{\mathcal{O}_\text{pt}}} &= 98 + \sum_{d=1}^\infty d^2\frac{q^d}{1-q^d} \left(98n^{(1)}_{0,d}+65n^{(2)}_{0,d}\right)\ , \\
\frac{\partial^2}{\partial t^2}\frac{\Pi_{\mathcal{S}^\bullet_2}}{\Pi_{\mathcal{O}_\text{pt}}} &= 33 + \sum_{d=1}^\infty d^2\frac{q^d}{1-q^d} \left(33n^{(1)}_{0,d}+21n^{(2)}_{0,d}\right)\ ,
\end{aligned}
\end{equation}
from which we find the invariants $n^{(a)}_{0,d}$ and list them up to degree $10$ in Table~\ref{tab:Genus0Model2}. Further, we deduce the associated Klemm--Pandharipande meeting invariants listed in Table~\ref{tab:MeetingModel2}.

%%%%%%%%%%%%%%%%%%%%
\begin{table}[t]\centering
\footnotesize{
\begin{tabular}{|l|r|r|r|r|}\hline
$m_{k,l}$ & $l=1$ & $l=2$ & $l=3$ & $l=4$ \\\hline
 $k=1$ &  60\,784\,240 & 28\,194\,221\,040 & 23\,782\,299\,222\,640 & 26\,506\,970\,805\,517\,040 \\
 $k=2$ &  \text{} & 13\,065\,863\,900\,400 & 11\,031\,985\,902\,832\,240 & 12\,299\,429\,676\,016\,495\,600 \\
 $k=3$ & \text{} & \text{} & 9\,314\,685\,486\,617\,406\,000 & 10\,384\,847\,256\,692\,114\,669\,040 \\
 $k=4$ & \text{} & \text{} & \text{} & 11\,577\,959\,795\,730\,175\,108\,775\,920 \\  \hline
\end{tabular}}
\caption{\label{tab:MeetingModel2}Klemm--Pandharipande meeting invariants $m_{k,l}\equiv m_{l,k}$ of the Calabi--Yau fourfold~$X_{1,17,7}$ up to degree four. For ease of presentation we only list the invariants for $k \le l$. }
\end{table}
%%%%%%%%%%%%%%%%%%%%%

Moreover, we use the quantity $F_1^{\text{top}}$ specified in eq.~\eqref{eq:F1} to determine the genus one invariants $n_{1,d}$. The discriminant locus has two rational factors, these are the large volume divisor $\Delta_{\text{LV}}=z$ and the discriminant factor $\Delta=1-188z-2\,368z^2+4z^3$. The coefficients $b_1$ and $b_2$ reflecting the holomorphic ambiguity are from the asymptotic behavior of $F_1^{\text{top}}$ at large volume~\eqref{eq:F1LV} and at the conifold~\eqref{eq:F1Coni} determined to be
\begin{equation}
1+b_1 = -\frac{1}{24}\int_X c_3(X)\cup J = \frac{328}{24} \quad \text{and} \quad b_2 = -\frac{1}{24}\ .
\end{equation}
With the Euler characteristic $\chi = 672$ we thus find
\begin{equation}
F_1^{\text{top}}\,=\, 25\log \Pi_{\mathcal{O}_\text{pt}}+\log \left(\frac1{2\pi i} \frac{\partial z}{\partial t} \right) +\frac{38\, \log z}{3} -\frac{\log (1-188z-2\,368z^2+4z^3)}{24} \ ,
\end{equation}
which in the large volume limit $z\to 0$ and after reexpressing $z$ in terms of $q$ reduces to
\begin{equation}
\begin{aligned}
F_1^{\text{top}} &= \frac{41 \log (q)}{3}-\frac{473 q}{2}-\frac{13\,949 q^2}{2}-\frac{2\,276\,105 q^3}{6} + \ldots \\
&=  \frac{41 \log (q)}{3} + \sum_{d=1}^\infty N_{1,d}\, q^d\ .
\end{aligned}
\end{equation}
This equation determines the rational genus one invariants $N_{1,d}$, which by the multicovering formula~\eqref{eq:MCg1} encode the integral genus one invariants $n_{1,d}$ listed in Table~\ref{tab:Genus1Model2}. Their integrality has been checked up to degree $50$.

%%%%%%%%%%%%%%%%%%%%%%
\begin{table}\centering
\begin{tabular}{|r|r|r|}
\hline $d$ & $n_{1,d}$ \\\hline
 1 & 0 \\
 2 & 0 \\
 3 & 0 \\
 4 & 0 \\
 5 & 224\,386 \\
 6 & 206\,613\,908 \\
 7 & 83\,707\,955\,196 \\
 8 & 23\,455\,827\,469\,526 \\
 9 & 5\,401\,382\,970\,402\,176 \\
 10 & 1\,107\,021\,477\,254\,814\,128 \\ \hline
\end{tabular}
\caption{\label{tab:Genus1Model2}Integral genus one Gromov--Witten invariants $n_{1,d}$ of $X_{1,17,7}$ up to degree $d=10$.}
\end{table}
%%%%%%%%%%%%%%%%%%%%%%%

\subsubsection{The regular singular point at infinity}\label{sec:PointInf}
From the Riemann P-symbol in eq.~\eqref{eq:PSymbolX1177} we see that the structure of solutions at the singular point $z=\infty$ is similar to that at the large volume point $z=0$. Namely, there are two non-logarithmic solutions, which in terms of $w=z^{-1}$ enjoy the expansions
\begin{equation}
\begin{aligned}
\Pi^{(1,0)}_\infty(w)&=w \,(1 + 21 w + 2\,989 w^2 + 714\,549 w^3 + 217\,515\,501 w^4+\ldots) \ , \\[0.1em]
\Pi^{(2,0)}_\infty(w)&=w^{3/2} \left(1+\frac{10\,085 w}{126}+\frac{782\,127 w^2}{50}+\frac{379\,170\,123\,893 w^3}{88\,200}+\ldots\right)\ .
\end{aligned}
\end{equation}
Moreover, there are four logarithmic solutions $\Pi^{(1,l)}_\infty$ for $l=1,\ldots,4$. As opposed to the second non-logarithmic period at large volume --- $\Pi^{(2,0)}_0$ given in eq.~\eqref{eq:RegSol2Model2} --- the additional solution $\Pi^{(2,0)}_\infty$ has a branch cut arising from the square root of the solution. This already indicates that $z=\infty$ is not large volume limit of a smooth Calabi--Yau fourfold.

Let us now look at the integral period vector $\vec{\widetilde{\Pi}}=(\widetilde{\Pi}_0, \widetilde{\Pi}_1, \widetilde{\Pi}_2, \widetilde{\Pi}_3,\widetilde{\Pi}_4,\widetilde{\Pi}_5)^T$, which is related to the integral period vector $\vec{\Pi}$ by the $SL(5,\mathbb{Z})$ transformation $S$ according to
\begin{equation}
\left(
\begin{array}{c}
\widetilde{\Pi}_0\\\widetilde{\Pi}_1\\ \widetilde{\Pi}_2\\ \widetilde{\Pi}_3 \\ \widetilde{\Pi}_4 \\ \widetilde{\Pi}_5
\end{array}
\right)=
\underbrace{\left(
\begin{array}{cccccc}
 105 & 98 & -2 & -1 & -4 & -8 \\
 -49 & -56 & 1 & 0 & 2 & 5 \\
 -1498 & -1400 & 28 & 14 & 53 & 89 \\
 -648 & -615 & 12 & 6 & 22 & 34 \\
 -330 & -243 & 6 & 5 & 12 & 16 \\
 196 & 229 & -4 & 0 & -8 & -21
\end{array}
\right)}_{S}\cdot
\left(
\begin{array}{c}
\Pi_0\\\Pi_1\\ \Pi_2\\ \Pi_3 \\ \Pi_4 \\ \Pi_5
\end{array}
\right)\ .
\end{equation}
By an analytic continuation of this period vector to $z=\infty$ we find that it corresponds to the following linear combination of solutions $\Pi^{(k,l)}_\infty$:
\begin{equation} \label{eq:Basis2}
\left(
\begin{array}{c}
\widetilde{\Pi}_0\\\widetilde{\Pi}_1\\ \widetilde{\Pi}_2\\ \widetilde{\Pi}_3 \\ \widetilde{\Pi}_4 \\ \widetilde{\Pi}_5
\end{array}
\right)=
\left(
\begin{array}{cccccc}
 1 & 0 & 0 & 0 & 0 & 0 \\
 0 & 1 & 0 & 0 & 0 & 0 \\
 -\frac{29}{2} & \frac{53}{2} & \frac{87}{2} & 0 & 0 & 0 \\
 -\frac{57}{8} & 11 & \frac{87}{4} & 0 & 0 & -\frac{7}{2 \pi ^2} \\
 -\frac{61}{16}+\frac{275 i \zeta (3)}{8 \pi ^3} & 10 & \frac{87}{8} & -\frac{29}{4} & 0 & \frac{21}{4 \pi ^2} \\
 \frac{73}{192} & -\frac{275 i \zeta (3)}{8 \pi ^3} & \frac{29}{8} & 0 & \frac{29}{16} & \frac{7}{8 \pi ^2} 
\end{array}
\right)\cdot
\left(
\begin{array}{c}
\Pi^{(1,0)}_\infty\\\Pi^{(1,1)}_\infty\\ \Pi^{(1,2)}_\infty\\ \Pi^{(1,3)}_\infty \\ \Pi^{(1,4)}_\infty \\ \Pi^{(2,0)}_\infty
\end{array}
\right) 
\end{equation}
Hence, we deduce in the limit $w \to 0$ the asymptotic behavior for $\vec{\widetilde{\Pi}}$ to be
\begin{equation} \label{eq:IntPeriodsInf}
    \vec{\widetilde{\Pi}}^{\text{asy}}(s) \,=\,
    \begin{pmatrix} 1 \\[0.1em]
     s  \\[0.1em]
     \frac{87}{2} s^2+\frac{53}{2} s-\frac{29}{2}  \\[0.1em]
     \frac{87}{4} s^2+11 s-\frac{57}{8}  \\[0.1em]
     -\frac{29}{4} s^3+\frac{87}{8} s^2+10 s-\frac{61}{16}+\frac{275 i \zeta (3)}{8\pi^3}  \\[0.1em]
    \frac{29}{16} s^4+\frac{29}{8}s^2 - \frac{275 i  \zeta (3)}{8\pi ^3}s  + \frac{73}{192}
    \end{pmatrix} \ ,
\end{equation}
in terms of the flat coordinate
\begin{equation}
  s(w) = \frac{1}{2\pi i}  \frac{\widetilde\Pi_1(w)}{\widetilde\Pi_0(w)} \ .
\end{equation}
In the newly defined integral basis $\vec{\widetilde{\Pi}}$ the monodromy matrices at $z=\infty$ and $z_2$ transform into
\begin{equation}
\widetilde{M}_\infty=\left(
\begin{array}{cccccc}
 1 & 1 & 70 & 33 & 6 & 9 \\
 0 & 1 & 87 & 39 & 20 & 9 \\
 0 & 0 & 1 & 1 & -2 & 1 \\
 0 & 0 & 0 & -1 & 3 & -1 \\
 0 & 0 & 0 & 0 & 1 & -1 \\
 0 & 0 & 0 & 0 & 0 & 1
\end{array}
\right)\ ,\quad 
\widetilde{M}_{z_2} =\left(
\begin{array}{cccccc}
 1 & 0 & 0 & 0 & 0 & 0 \\
 0 & 1 & 0 & 0 & 0 & 0 \\
 0 & 0 & 1 & 0 & 0 & 0 \\
 0 & 0 & 0 & 1 & 0 & 0 \\
 0 & 0 & 0 & 0 & 1 & 0 \\
 -1 & 0 & 0 & 0 & 0 & -1
\end{array}
\right) \ ,
\end{equation}
while the intersection pairing becomes 
\begin{equation}
S\, \chi({\vec{\mathcal{E}}}^\bullet,{\vec{\mathcal{E}}}^\bullet)\, S^{\,T}=
\left(
\begin{array}{cccccc}
 0 & 0 & 0 & 0 & 0 & 1 \\
 0 & 0 & 0 & 0 & 1 & 0 \\
 0 & 0 & 174 & 87 & 70 & 0 \\
 0 & 0 & 87 & 44 & 32 & 0 \\
 0 & 1 & 70 & 32 & 32 & 0 \\
 1 & 0 & 0 & 0 & 0 & 2 \\
\end{array}
\right)\ .
\end{equation}

%%%%%%%%%%%%%%%%%%%%%%%%%%%%%%%%%%%%%%%%%%%%%%%%%%%%%%%%%%%%%
\begin{table}\centering
\begin{tabular}{|r|p{3.5cm}|p{3.5cm}|}
\hline $d$ & \hfill$n^{(A)}_{0,d}$& \hfill$n^{(B)}_{0,d}$ \\\hline
$1/2$ &  \hfill--  & \hfill14 \\
  1 &  \hfill7\,569 & \hfill3\,781 \\
 3/2 & \hfill-- & \hfill167 \\
 2 & \hfill735\,324 & \hfill367\,662 \\
 5/2 & \hfill--  & \hfill23\,647 \\
 3 & \hfill129\,395\,187 & \hfill$\frac{258\,790\,207}{4}$ \\
 7/2 & \hfill-- & \hfill$\frac{18\,828\,027}{4}$ \\
 4 & \hfill29\,766\,479\,280 & \hfill14\,883\,239\,640 \\
 9/2 & \hfill-- & \hfill$\frac{9\,280\,303\,369}{8}$ \\
 5 & \hfill7\,978\,989\,505\,959 & \hfill$\frac{15\,957\,978\,988\,271}{4}$ \\
\hline
\end{tabular}
\caption{\label{tab:GWIsInf}Rational genus zero invariants $n^{(A)}_{0,d}$ and $n^{(B)}_{0,d}$ associated to the doubly logarithmic periods $\widetilde{\Pi}_2$ and $\widetilde{\Pi}_3$.}
\end{table}
%%%%%%%%%%%%%%%%%%%%%%%%%%%%%%%%%%%%%%%%%%%%%%%%%%%%%%%%%%%%%

We observe that in terms of the transformed intersection pairing the monodromy matrix $\widetilde{M}_{z_2}$ has the characteristic form of the Seidel--Thomas twist \eqref{eq:STtwist} with respect to the structure sheaf of a geometric target space. However, by the structure of the quantum periods in the vicinity of $w=0$, this target space cannot be a smooth Calabi--Yau fourfold for various reasons. Firstly, as can be seen from eq.~\eqref{eq:Basis2} --- apart from the logarithmic branch cut --- there is also a square root branch cut appearing in one doubly logarithmic, the triply logarithmic and quadruply logarithmic quantum periods. This square root branch cut, however, does not conform with the singularity behavior of quantum volumes of cycles in a large volume phase. Secondly, if the target space were a smooth Calabi--Yau fourfold, the leading asymptotic term $\frac{29}{16}s^4 + \ldots$ in the quadruply logarithmic period would encode the degree $\kappa$ of the Calabi--Yau fourfold according to $\frac{\kappa}{4!}s^4+ \ldots\,$. This yields, however, the non-integral coefficient $\kappa=\frac{87}2$. 

On the other hand, due to the discussed similarities to a large volume phase, it is conceivable that the target space enjoys an interpretation as a singular Calabi--Yau variety --- possibly with a singularity in codimension two, which could explain the square root branch cut starting in one of the doubly logarithmic quantum periods. Having such a geometric picture in our mind, we naively extract an instanton expansion from the doubly logarithmic integral periods $\widetilde{\Pi}_2$ and $\widetilde{\Pi}_3$ according to
\begin{equation}
\begin{aligned}
\widetilde{\Pi}_2&= \frac{87}{2}s^2+\frac{53}{2}s - \frac{29}{2} + \sum_{d=1}^\infty n^{(A)}_{0,d} \operatorname{Li}_2\left(e^{2\pi i s \cdot d}\right)\ , \\
\widetilde{\Pi}_3&= \frac{87}{4}s^2+ 11s - \frac{57}{8} + \sum_{d=1}^\infty n^{(B)}_{0,d/2} \operatorname{Li}_2\left(e^{2\pi i s \cdot d/2}\right)\ .
\end{aligned}
\end{equation}
The leading numbers of this expansion are listed in in Table~\ref{tab:GWIsInf}. Note that the doubly logarithmic solution without the square root branch cut yields a conventional genus zero instanton expansion with integral invariants $n_{0,d}^{(A)}$ for integral $d$. The other doubly logarithmic solution, however, yields instanton invariants $n_{0,d}^{(B)}$ arising also at half instanton degrees, which reflects the square root branch cut behavior of this quantum period. Moreover, the invariants $n_{0,d}^{(B)}$ in general are rational numbers with powers of two in their denominators. It would be interesting to give a geometric interpretation of all the large volume like features, potentially as speculated in terms of a singular Calabi--Yau fourfold variety.

%%%%%%%%%%%%%%%%%%%%%%%%%%%%%%%%%%%
\section{Conclusions} \label{sec:con}
%%%%%%%%%%%%%%%%%%%%%%%%%%%%%%%%%%%
In this work we have studied the Gromov--Witten theory on Calabi--Yau fourfolds, emphasizing the role of non-marginal chiral--anti-chiral operators in the associated quantum chiral rings. We established and demonstrated explicitly that the number of chiral--anti-chiral operators of conformal weight $(2,2)$ --- i.e., operators corresponding to generators of the middle-dimensional cohomology group of the Calabi--Yau fourfold --- yields the number of independent genus zero Gromov--Witten invariants with a single marked point at each degree. We argued that for Calabi--Yau fourfolds with a single K\"ahler modulus such examples arise from the Picard--Fuchs operators of quantum periods with non-minimal order. Namely, the regular singular point associated to the large volume limit is not a regular singular point with maximally unipotent monodromy of the associated Picard--Fuchs operator. Our explicit examples of this phenomenon were constructed from non-complete intersection projective varieties or from complete intersections in non-toric ambient spaces. To deduce their quantum cohomology rings we calculated the integral quantum periods with the help of numerical analytic continuation techniques. Furthermore, we computed the monodromy matrices about all regular singular points in quantum K\"ahler moduli space with respect to the established integral basis. Finally, we determined the Klemm--Pandharipande meeting invariants and the genus one BPS invariants for the analyzed Calabi--Yau fourfolds. The confirmed integrality property of these invariants furnished a non-trivial check on the deduced quantum cohomology rings. As a further check on our results, we independently verified the genus zero Gromov--Witten invariants at degree one entering the quantum cohomology ring with intersection theory methods.

We established that the large volume asymptotics of quantum periods admitted purely instanton generated integral linear combinations. As briefly mentioned, this observation may prove useful in string cosmology for F-term monodromy inflation scenarios \cite{Silverstein:2008sg,Hebecker:2014kva,Marchesano:2014mla,Hebecker:2014eua}. Moreover, such instanton generated quantum periods are interesting from an open-closed string duality point of view \cite{Mayr:2001xk}, which --- in certain geometric situations --- relates closed-string quantum periods of Calabi--Yau fourfolds to open-string quantum periods of Calabi--Yau threefolds with branes \cite{Mayr:2001xk,Alim:2009bx,Grimm:2009ef}. Identifying purely instanton generated open-string quantum periods would hence establish stable brane configurations in Calabi--Yau threefolds at large volume. The absence of perturbative terms in the expansion of open quantum periods would imply that the associated open-closed deformation space were obstructed by closed sphere and open disk instanton effects only. Such setups promise interesting enumerative interpretations in terms of real and Ooguri--Vafa invariants in compact Calabi--Yau geometries \cite{Walcher:2006rs,MR2425184,Jockers:2008pe,Alim:2009rf}.

We would like to point out an observation that the Picard--Fuchs operators of some of our Calabi--Yau fourfold examples --- namely for some of those given as complete intersections in ambient Grassmannian spaces --- exhibit intriguing algebraic properties. That is to say that the fundamental periods factorize into the Hadamard product of two new fundamental periods that are solutions to a Calabi--Yau threefold and elliptic curve Picard--Fuchs differential equations of fourth and first order, respectively.\footnote{We are thankful to Gert Almkvist for pointing and explaining this factorization property.} For instance, the fundamental period~\eqref{eq:FPex1} of the example discussed in Section~\ref{sec:Example1} enjoys the expansion
\begin{equation}
    \Pi_0(z) \,=\, \sum_{n=0}^{+\infty}\sum_{k=0}^n  \binom{2n}{n}^2 \binom{4n}{2n} \binom{n}{k}^2\binom{n+k}n \, z^n \,=\, (\Pi_0^\text{CY3} \star \Pi_0^\text{E})(z) \ ,
\end{equation}
with the fundamental periods 
\begin{equation}
  \Pi_0^\text{CY3}(z)\,=\,\sum_{n=0}^{+\infty}\sum_{k=0}^n  \binom{2n}n \binom{4n}{2n} \binom{n}{k}^2\binom{n+k}n \, z^n  \ , 
  \qquad  \Pi_0^\text{E}(z)\,=\, \sum_{n=0}^{+\infty} \binom{2n}n z^n \ ,
\end{equation}
and the Hadamard product $(f\star g)(z)= \sum_n a_nb_n z^n$ defined in terms of the series expansions $f(z)=\sum_n a_n z^n$ and $g(z)=\sum_n b_n z^n$. In particular, the fundamental period~$\Pi_0^\text{CY3}(z)$ is the solution to the fourth order Picard--Fuchs operator \cite{Almkvist:2005arxiv,vanStraten:2012db}\footnote{Compare with example~AESZ~51 in ref.~\cite{Almkvist:2005arxiv} and the online Calabi--Yau datebase~\cite{vanStraten:2012db}.}
\begin{equation}
  \mathcal{L}^\text{CY3}=\theta^4 - 4 z (4\theta+1)(4\theta+3)(11\theta^2+11\theta+3) -16z^2 (4\theta+1)(4\theta+3)(4\theta+5)(4\theta+7) \ ,
\end{equation}
with maximally unipotent monodromy point at $z=0$. It gives rise to the integral genus zero Gromov--Witten invariants $920$, $50\,520$, $5\,853\,960$, \ldots, c.f.,~ref.~\cite{vanStraten:2012db}. It would be interesting to find a geometric interpretation for these Hadamard factorization of such Calabi--Yau fourfolds, perhaps along the lines of ref.~\cite{Doran:2015qy}.

Finally, let us mention that the non-minimal order property of the analyzed Picard--Fuchs operators for the Calabi--Yau fourfold periods may also exhibit interesting features from a modular form perspective, see, e.g., refs.~\cite{Gannon:2013jua,Doran:2013npa}. At least, we expect that a better understanding of global properties of the quantum K\"ahler moduli space should simplify the required derivation of integral quantum periods.

%%%%%%%%%%%%%%%%%%
\subsection*{Acknowledgments}
%%%%%%%%%%%%%%%%%%
We would like to thank
Gert Almkvist,
Rolf Kappl,
Albrecht Klemm,
Peter Mayr,
Dave Morrison,
Urmi Ninad,
Thorsten Schimannek,
Stephan Stieberger
and
Eva Silverstein
for discussions and correspondences.
A.G. is supported by the graduate school BCGS and the Studienstiftung des deutschen Volkes.

%%%%%%%%%%%%%%%%%%%%%%%%%%%%%%%%%%%
\newpage
\begin{appendix}
%%%%%%%%%%%%%%%%%%%%%%%%%%%%%%%%%%%

%%%%%%%%%%%%%%%%%%%%%%%%%%%%%%%%%%%
\section{Tabulated results of analyzed examples} \label{app:tables}
%%%%%%%%%%%%%%%%%%%%%%%%%%%%%%%%%%%
In this appendix we tabulate the data that specifies the quantum periods and monodromy structure for several Calabi--Yau fourfolds with a order six Picard--Fuchs operator. We also list the leading genus zero Gromov--Witten invariants generating the quantum cohomology rings and the genus one BPS invariants of these Calabi--Yau fourfolds. Two of these examples --- with their tables listed in Appendix~\ref{app:Tab1} and Appendix~\ref{app:Tab2} --- are discussed thoroughly in the main text in Section~\ref{sec:Example1} and Section~\ref{sec:Example2}, respectively. The data of the remaining examples is calculated analogously.
\bigskip

\footnotesize
%\newpage
\subsection{Calabi--Yau fourfold $X_{1,4} \subset \operatorname{Gr}(2,5)$}
\label{app:Tab1}
\begin{tabular}{|L|L|}
\hline
\multicolumn{2}{|l|}{Picard--Fuchs operator:}\\
\multicolumn{2}{|l|}{$\begin{aligned}
\mathcal{L}(z)\,=\,&(\theta -1) \theta ^5-8 z(2 \theta +1) (4 \theta +1) (4 \theta +3) \left(11 \theta^2+11 \theta +3\right) \theta\\ &-64z^2 (2 \theta +1) (2 \theta +3) (4 \theta +1) (4 \theta +3) (4 \theta +5) (4 \theta +7) \end{aligned}$}\\
\hline
\multicolumn{1}{|l|}{Discriminant locus:} & \multicolumn{1}{|l|}{Riemann P-symbol:}\\[0.2em]
\multicolumn{1}{|l|}{$ \Delta(z)=1 - 2\,816 z - 65\,536 z^2$} & \multicolumn{1}{|l|}{\multirow{4}{*}{$\left\{
\begin{array}{cccc}
0 & \infty & z_1 & z_{\mathcal{O}_X}\\[0.1em]
\hline
0 & \frac14 & 0 & 0 \\[0.1em]
0 & \frac12 & 1 & 1 \\[0.1em]
0 & \frac34 & 2 & 2 \\[0.1em]
0 & \frac54 & 3 & 3 \\[0.1em]
0 & \frac32 & 4 & 4 \\[0.1em]
1 & \frac74 & \frac{3}{2} & \frac{3}{2} \\
\end{array}
\right\}$}} \\
&\\
\multicolumn{1}{|l|}{Regular singular points:} & \\
\multicolumn{1}{|l|}{$\begin{aligned} z&=0\\z&=\infty\\z&=z_{\mathcal{O}_X}(=z_2)\approx 3.5\cdot 10^{-4}\\z&=z_1 \approx- 0.043\end{aligned}$} & \\
\hline
\multicolumn{1}{|l|}{Intersection pairing:} & \multicolumn{1}{|l|}{Large volume asymptotics:} \\[0.2em]
\multicolumn{1}{|l|}{$\chi = \left(
\begin{array}{*{6}{>{\scriptstyle}c}}
 0 & 0 & 0 & 0 & 0 & 1 \\
 0 & 0 & 0 & 0 & 1 & 0 \\
 0 & 0 & 20 & 8 & 10 & 24 \\
 0 & 0 & 8 & 4 & -8 & 6 \\
 0 & 1 & 10 & -8 & -14 & -7 \\
 1 & 0 & 24 & 6 & -7 & 2 \\
\end{array}
\right)$} & 
\multicolumn{1}{|l|}{$ \vec\Pi^{\text{asy}}(t) = \left(
\begin{array}{*{1}{>{\scriptstyle}c}}
1\\
t\\
10t^2+20t+\frac{107}{6}\\
4t^2-4t+\frac{7}{2}\\
-\frac{10}{3}t^3-5t^2-\frac{19}{2}t-\frac{47}{12}+\frac{55 i \zeta(3)}{\pi^3}\\
\frac{5}{6}t^4+\frac{37}{12}t^2-\frac{55 i \zeta(3)}{\pi^3}t+\frac{7}{144}
\end{array}\right)$}\\
\hline
\multicolumn{2}{|l|}{Monodromy matrices:}\\[0.2em]
\multicolumn{2}{|l|}{$
\begin{aligned}
M_0&=\left(
\begin{array}{*{6}{>{\scriptstyle}c}}
 1 & 1 & 30 & 0 & 0 & 0 \\
 0 & 1 & 20 & 8 & 0 & 0 \\
 0 & 0 & 1 & 0 & -1 & 0 \\
 0 & 0 & 0 & 1 & 0 & 0 \\
 0 & 0 & 0 & 0 & 1 & -1 \\
 0 & 0 & 0 & 0 & 0 & 1 \\
\end{array}
\right)  \quad
M_\infty =\left(
\begin{array}{*{6}{>{\scriptstyle}c}}
 -19 & -11 & -670 & -72 & 270 & -40 \\
 -40 & -19 & -1340 & -168 & 540 & -80 \\
 2 & 1 & 67 & 8 & -27 & 4 \\
 -3 & -1 & -90 & -13 & 35 & -5 \\
 2 & 1 & 66 & 8 & -27 & 4 \\
 5 & 2 & 156 & 22 & -63 & 9 \\
\end{array}
\right)  \\[0.2em]
%%%%%%%%%%%%%%
M_{z_1}&=\left(
\begin{array}{*{6}{>{\scriptstyle}c}}
 21 & 10 & 700 & 80 & -300 & 50 \\
 40 & 21 & 1400 & 160 & -600 & 100 \\
 -2 & -1 & -69 & -8 & 30 & -5 \\
 2 & 1 & 70 & 9 & -30 & 5 \\
 -2 & -1 & -70 & -8 & 31 & -5 \\
 -4 & -2 & -140 & -16 & 60 & -9 \\
\end{array}
\right)  \quad
M_{z_{\mathcal{O}_X}} = \left(
\begin{array}{*{6}{>{\scriptstyle}c}}
 1 & 0 & 0 & 0 & 0 & 0 \\
 0 & 1 & 0 & 0 & 0 & 0 \\
 0 & 0 & 1 & 0 & 0 & 0 \\
 0 & 0 & 0 & 1 & 0 & 0 \\
 0 & 0 & 0 & 0 & 1 & 0 \\
 -1 & 0 & -24 & -6 & 7 & -1 \\
\end{array}
\right)
\end{aligned}
$}\\
\hline
\end{tabular}

\newpage
\begin{tabular}{|R|R|}\hline
\multicolumn{2}{|l|}{Generators of cohomology ring:}\\[0.2em]
\multicolumn{2}{|l|}{
$\begin{aligned}
\left\langle\!\left\langle 1 \,;\,\sigma_1\,;\,\sigma_{1,1},\sigma_2\,;\,\frac{\sigma_3}{4}\,;\,\frac{\sigma_{2,2}}{4}\right\rangle\!\right\rangle
\,\subset\, H^{0,0}(X)\oplus \ldots \oplus H^{4,4}(X)
\end{aligned}$}\\[8pt]
\hline
\multicolumn{2}{|l|}{Total Chern character:}\\[0.2em]
\multicolumn{2}{|l|}{
$\begin{aligned}
 c(X) \,=\, 1 + (8 \sigma_{1,1} + 7 \sigma_2) -440 \frac{\sigma_3}{4} + 1\,848 \frac{\sigma_{2,2}}{4}
\end{aligned}$}\\[8pt]
\hline
\multicolumn{2}{|l|}{Intersection numbers:}\\[0.2em]
\multicolumn{2}{|l|}{
$\begin{aligned}
\sigma_{1,1}.\sigma_{1,1}= 4 \ , \quad \sigma_{1,1}.\sigma_2 = 4 \ , \quad \sigma_2.\sigma_2=8
\end{aligned}$}\\\hline
\multicolumn{2}{|l|}{Zeros of integral quantum periods:}\\[0.2em]
\multicolumn{2}{|l|}{
$\begin{aligned}
\Pi_5 &= 0 \quad \text{at } z_{\mathcal{O}_X} \ , \\
\Pi_{\mathcal{B}_{z_1}}&=10 \Pi_0 + 20 \Pi_1 - \Pi_2 + \Pi_3 - \Pi_4 - 2\Pi_5 = 0 \quad \text{at } z_1
\end{aligned}$}\\[8pt]
 \hline
\multicolumn{2}{|l|}{Genus zero Gromov--Witten invariants $n^{(1)}_{0,d}$ (left) and $n^{(2)}_{0,d}$ (right)}\\\hline
400 & 520 \\
208\,240 & 226\,480 \\
175\,466\,480 & 191\,464\,760 \\
196\,084\,534\,160 & 213\,155\,450\,240 \\
255\,402\,582\,828\,400 & 277\,092\,686\,601\,400 \\
367\,048\,595\,782\,193\,680 & 397\,700\,706\,634\,553\,680 \\
564\,810\,585\,071\,858\,496\,880 & 611\,416\,342\,763\,726\,567\,800 \\
913\,929\,133\,261\,543\,393\,001\,760 & 988\,670\,017\,271\,687\,389\,572\,480 \\
1\,536\,929\,129\,164\,031\,410\,293\,358\,720 & 1\,661\,748\,145\,541\,449\,358\,296\,013\,440 \\
 2\,664\,576\,223\,763\,330\,924\,317\,069\,072\,400 & 2\,879\,777\,881\,450\,393\,936\,532\,565\,976\,400 \\ \hline
\multicolumn{2}{|l|}{Genus one Gromov--Witten invariants $n_{1,d}$}\\\hline
\multicolumn{2}{|r|}{0} \\
\multicolumn{2}{|r|}{0} \\
\multicolumn{2}{|r|}{-3\,200} \\
\multicolumn{2}{|r|}{370\,151\,480} \\
\multicolumn{2}{|r|}{4\,108\,408\,756\,800} \\
\multicolumn{2}{|r|}{19\,279\,169\,520\,232\,000} \\
\multicolumn{2}{|r|}{66\,081\,794\,099\,798\,279\,680} \\
\multicolumn{2}{|r|}{194\,122\,441\,310\,522\,439\,007\,040} \\
\multicolumn{2}{|r|}{522\,534\,128\,159\,184\,581\,441\,465\,280} \\
\multicolumn{2}{|r|}{1\,332\,480\,344\,031\,795\,460\,733\,665\,780\,608} \\\hline
\end{tabular}

\newpage
\subsection{Calabi--Yau fourfold $X_{2,3} \subset \operatorname{Gr}(2,5)$}
\begin{tabular}{|L|L|}
\hline
\multicolumn{2}{|l|}{Picard--Fuchs operator:}\\
\multicolumn{2}{|l|}{$\begin{aligned}
\mathcal{L}(z)\,=\,&(\theta -1) \theta ^5-6 z (2 \theta +1) (3 \theta +1) (3 \theta +2) \left(11 \theta ^2+11 \theta +3\right) \theta \\
&-36z^2 (2 \theta +1) (2 \theta +3) (3 \theta +1) (3 \theta +2) (3 \theta +4) (3 \theta +5)\end{aligned}$}\\
\hline
\multicolumn{1}{|l|}{Discriminant locus:} & \multicolumn{1}{|l|}{Riemann P-symbol:}\\[0.2em]
\multicolumn{1}{|l|}{$ \Delta(z)= 1 - 1\,188 z - 11\,664 z^2$}& \multicolumn{1}{|l|}{\multirow{4}{*}{$\left\{
\begin{array}{cccc}
0 & \infty & z_1 & z_{\mathcal{O}_X}\\[0.1em]
\hline
0 & \frac13 & 0 & 0 \\[0.1em]
0 & \frac12 & 1 & 1 \\[0.1em]
0 & \frac23 & 2 & 2 \\[0.1em]
0 & \frac43 & 3 & 3 \\[0.1em]
0 & \frac32 & 4 & 4 \\[0.1em]
1 & \frac53 & \frac{3}{2} & \frac{3}{2} \\
\end{array}
\right\}$}} \\
&\\
\multicolumn{1}{|l|}{Regular singular points:} & \\
\multicolumn{1}{|l|}{$\begin{aligned} z&=0\\z&=\infty\\z&=z_{\mathcal{O}_X}\approx 8.3\cdot 10^{-4}\\z&=z_1 \approx -0.10 \end{aligned}$} & \\
\hline
\multicolumn{1}{|l|}{Intersection pairing:} & \multicolumn{1}{|l|}{Large volume asymptotics:} \\[0.2em]
\multicolumn{1}{|l|}{$\chi = \left(
\begin{array}{*{6}{>{\scriptstyle}c}}
 0 & 0 & 0 & 0 & 0 & 1 \\
 0 & 0 & 0 & 0 & 1 & 0 \\
 0 & 0 & 30 & 12 & 15 & 26 \\
 0 & 0 & 12 & 6 & -12 & 5 \\
 0 & 1 & 15 & -12 & -16 & -8 \\
 1 & 0 & 26 & 5 & -8 & 2 \\
\end{array}
\right)$} & 
\multicolumn{1}{|l|}{$ \vec\Pi^{\text{asy}}(t) = \left(
\begin{array}{*{1}{>{\scriptstyle}c}}
1\\
t\\
15t^2+30t+\frac{77}{4}\\
6t^2-6t+\frac{9}{4}\\
-5t^3-\frac{15}{2}t^2-\frac{47}{4}t-\frac{37}{8}+\frac{45 i \zeta(3)}{\pi^3}\\
\frac{5}{4}t^4+\frac{27}{8}t^2-\frac{45 i \zeta(3)}{\pi^3}t+\frac{23}{96}
\end{array}\right)$}\\
\hline
\multicolumn{2}{|l|}{Monodromy matrices:}\\[0.2em]
\multicolumn{2}{|l|}{$
\begin{aligned}
M_0&=\left(
\begin{array}{*{6}{>{\scriptstyle}c}}
 1 & 1 & 45 & 0 & -5 & 0 \\
 0 & 1 & 30 & 12 & 0 & 0 \\
 0 & 0 & 1 & 0 & -1 & 0 \\
 0 & 0 & 0 & 1 & 0 & 0 \\
 0 & 0 & 0 & 0 & 1 & -1 \\
 0 & 0 & 0 & 0 & 0 & 1 \\
\end{array}
\right)  \quad
M_\infty =\left(
\begin{array}{*{6}{>{\scriptstyle}c}}
 -25 & -13 & -1015 & -46 & 390 & -50 \\
 -60 & -29 & -2400 & -132 & 930 & -120 \\
 2 & 1 & 80 & 4 & -31 & 4 \\
 -3 & -1 & -105 & -8 & 40 & -5 \\
 2 & 1 & 79 & 4 & -31 & 4 \\
 5 & 2 & 184 & 13 & -72 & 9 \\
\end{array}
\right)  \\[0.2em]
%%%%%%%%%%%%%%
M_{z_1}&=\left(
\begin{array}{*{6}{>{\scriptstyle}c}}
 25 & 12 & 1020 & 48 & -420 & 60 \\
 60 & 31 & 2550 & 120 & -1050 & 150 \\
 -2 & -1 & -84 & -4 & 35 & -5 \\
 2 & 1 & 85 & 5 & -35 & 5 \\
 -2 & -1 & -85 & -4 & 36 & -5 \\
 -4 & -2 & -170 & -8 & 70 & -9 \\
\end{array}
\right)  \quad
M_{z_{\mathcal{O}_X}} = \left(
\begin{array}{*{6}{>{\scriptstyle}c}}
 1 & 0 & 0 & 0 & 0 & 0 \\
 0 & 1 & 0 & 0 & 0 & 0 \\
 0 & 0 & 1 & 0 & 0 & 0 \\
 0 & 0 & 0 & 1 & 0 & 0 \\
 0 & 0 & 0 & 0 & 1 & 0 \\
 -1 & 0 & -26 & -5 & 8 & -1 \\
\end{array}
\right)
\end{aligned}
$}\\ \hline
\end{tabular}

\newpage
\begin{tabular}{|R|R|}
\hline
\multicolumn{2}{|l|}{Generators of cohomology ring:}\\[0.2em]
\multicolumn{2}{|l|}{
$\begin{aligned}
\left(1\,;\,\sigma_1\,;\,\sigma_{1,1},\sigma_2\,;\,\frac{\sigma_3}{6}\,;\,\frac{\sigma_{2,2}}{6}\right)\,\in\, H^{0,0}(X)\oplus \ldots \oplus H^{4,4}(X)
\end{aligned}$}\\
\hline
\multicolumn{2}{|l|}{Total Chern character:}\\[0.2em]
\multicolumn{2}{|l|}{
$\begin{aligned}
 c(X) \,=\, 1 + (6 \sigma_{1,1} + 5 \sigma_2) -360 \frac{\sigma _3}{6} + 1188 \frac{\sigma _{2,2}}{6}
\end{aligned}$}\\
\hline
\multicolumn{2}{|l|}{Intersection numbers:}\\[0.2em]
\multicolumn{2}{|l|}{
$\begin{aligned}
\sigma_{1,1}.\sigma_{1,1}= 6 \ , \quad \sigma_{1,1}.\sigma_2 = 6 \ , \quad \sigma_2.\sigma_2=12
\end{aligned}$}\\\hline
\multicolumn{2}{|l|}{Zeros of integral quantum periods:}\\[0.2em]
\multicolumn{2}{|l|}{
$\begin{aligned}
\Pi_5 &= 0 \quad \text{at } z_{\mathcal{O}_X} \ , \\
\Pi_{\mathcal{B}_{z_1}}&=12 \Pi_0 + 30 \Pi_1 - \Pi_2 + \Pi_3 - \Pi_4 - 2\Pi_5 = 0 \quad \text{at } z_1
\end{aligned}$}\\[8pt]
 \hline
\multicolumn{2}{|l|}{Genus zero Gromov--Witten invariants $n^{(1)}_{0,d}$ (left) and $n^{(2)}_{0,d}$ (right)}\\\hline
 150 & 210 \\
 34\,635 & 38\,175 \\
 12\,266\,460 & 13\,599\,540 \\
 5\,755\,894\,980 & 6\,352\,627\,620 \\
 3\,144\,906\,174\,450 & 3\,462\,780\,142\,950 \\
 1\,895\,113\,546\,937\,010 & 2\,083\,385\,152\,900\,350 \\
 1\,222\,482\,269\,477\,448\,870 & 1\,342\,443\,529\,699\,952\,610 \\
 829\,123\,506\,499\,521\,864\,000 & 909\,737\,222\,891\,667\,295\,200 \\
 584\,369\,804\,499\,128\,982\,030\,870 & 640\,780\,961\,536\,667\,529\,927\,090 \\
 424\,582\,414\,793\,779\,873\,760\,931\,825 & 465\,334\,861\,886\,835\,590\,355\,227\,325 \\ \hline
\multicolumn{2}{|l|}{Genus one Gromov--Witten invariants $n_{1,d}$}\\\hline
\multicolumn{2}{|r|}{0} \\
\multicolumn{2}{|r|}{0} \\
\multicolumn{2}{|r|}{-40} \\
\multicolumn{2}{|r|}{6\,629\,085} \\
\multicolumn{2}{|r|}{33\,762\,865\,500} \\
\multicolumn{2}{|r|}{72\,983\,984\,748\,600} \\
\multicolumn{2}{|r|}{111\,703\,298\,516\,011\,620} \\
\multicolumn{2}{|r|}{143\,677\,197\,771\,963\,884\,280} \\
\multicolumn{2}{|r|}{167\,307\,680\,280\,218\,203\,241\,460} \\
\multicolumn{2}{|r|}{183\,135\,579\,515\,334\,103\,668\,439\,662} \\\hline
\end{tabular}

\newpage
\subsection{Calabi--Yau fourfold $X_{1^3,3} \subset \operatorname{Gr}(2,6)$}
\begin{tabular}{|L|L|}
\hline
\multicolumn{2}{|l|}{Picard--Fuchs operator:}\\
\multicolumn{2}{|l|}{$\begin{aligned}
\mathcal{L}(z)\,=\,& (\theta -1) \theta ^5-3 z(2 \theta +1) (3 \theta +1) (3 \theta +2) \left(13 \theta ^2+13 \theta +4\right)
   \theta
\\&-27 z^2 (3 \theta +1) (3 \theta +2)^2 (3 \theta +4)^2 (3 \theta +5) z^2
\end{aligned}$}\\
\hline
\multicolumn{1}{|l|}{Discriminant locus:} & \multicolumn{1}{|l|}{Riemann P-symbol:}\\[0.2em]
\multicolumn{1}{|l|}{$ \Delta(z)= 1 - 702 z - 19\,683 z^2$}& \multicolumn{1}{|l|}{\multirow{4}{*}{$\left\{
\begin{array}{cccc}
0 & \infty & z_1 & z_{\mathcal{O}_X}\\[0.1em]
\hline
0 & \frac13 & 0 & 0 \\[0.1em]
0 & \frac23 & 1 & 1 \\[0.1em]
0 & \frac23 & 2 & 2 \\[0.1em]
0 & \frac43 & 3 & 3 \\[0.1em]
0 & \frac43 & 4 & 4 \\[0.1em]
1 & \frac53 & \frac{3}{2} & \frac{3}{2} \\
\end{array}
\right\}$}} \\
&\\
\multicolumn{1}{|l|}{Regular singular points:} & \\
\multicolumn{1}{|l|}{$\begin{aligned} z&=0\\z&=\infty\\z&=z_{\mathcal{O}_X}=729^{-1}\\z&=z_1 = -27^{-1} \end{aligned}$} & \\
\hline
\multicolumn{1}{|l|}{Intersection pairing:} & \multicolumn{1}{|l|}{Large volume asymptotics:} \\[0.2em]
\multicolumn{1}{|l|}{$\chi = \left(
\begin{array}{*{6}{>{\scriptstyle}c}}
 0 & 0 & 0 & 0 & 0 & 1 \\
 0 & 0 & 0 & 0 & 1 & 0 \\
 0 & 0 & 42 & 15 & 21 & 41 \\
 0 & 0 & 15 & 6 & -15 & 8 \\
 0 & 1 & 21 & -15 & -20 & -10 \\
 1 & 0 & 41 & 8 & -10 & 2 \\
\end{array}
\right)$} & 
\multicolumn{1}{|l|}{$ \vec\Pi^{\text{asy}}(t) = \left(
\begin{array}{*{1}{>{\scriptstyle}c}}
1\\
t\\
21t^2+42t+\frac{131}{4}\\
\frac{15}{2}t^2-\frac{15}{2}t+5\\
-7t^3-\frac{21}{2}t^2-\frac{61}{4}t-\frac{47}{8}+\frac{213 i \zeta(3)}{4\pi^3}\\
\frac{7}{4}t^4+\frac{33}{8}t^2-\frac{213 i \zeta(3)}{4\pi^3}t+\frac{3}{16}
\end{array}\right)$}\\
\hline
\multicolumn{2}{|l|}{Monodromy matrices:}\\[0.2em]
\multicolumn{2}{|l|}{$
\begin{aligned}
M_0&=\left(
\begin{array}{*{6}{>{\scriptstyle}c}}
 1 & 1 & 63 & 0 & 0 & 0 \\
 0 & 1 & 42 & 15 & 0 & 0 \\
 0 & 0 & 1 & 0 & -1 & 0 \\
 0 & 0 & 0 & 1 & 0 & 0 \\
 0 & 0 & 0 & 0 & 1 & -1 \\
 0 & 0 & 0 & 0 & 0 & 1 \\
\end{array}
\right)  \quad
M_\infty =\left(
\begin{array}{*{6}{>{\scriptstyle}c}}
 -62 & -22 & -3360 & -300 & 1029 & -105 \\
 -126 & -41 & -6720 & -645 & 2058 & -210 \\
 3 & 1 & 160 & 15 & -49 & 5 \\
 -4 & -1 & -200 & -22 & 60 & -6 \\
 3 & 1 & 159 & 15 & -49 & 5 \\
 7 & 2 & 359 & 38 & -110 & 11 \\
\end{array}
\right)  \\[0.2em]
%%%%%%%%%%%%%%
M_{z_1}&=\left(
\begin{array}{*{6}{>{\scriptstyle}c}}
 43 & 21 & 2730 & 210 & -966 & 126 \\
 84 & 43 & 5460 & 420 & -1932 & 252 \\
 -2 & -1 & -129 & -10 & 46 & -6 \\
 2 & 1 & 130 & 11 & -46 & 6 \\
 -2 & -1 & -130 & -10 & 47 & -6 \\
 -4 & -2 & -260 & -20 & 92 & -11 \\
\end{array}
\right)  \quad
M_{z_{\mathcal{O}_X}} = \left(
\begin{array}{*{6}{>{\scriptstyle}c}}
 1 & 0 & 0 & 0 & 0 & 0 \\
 0 & 1 & 0 & 0 & 0 & 0 \\
 0 & 0 & 1 & 0 & 0 & 0 \\
 0 & 0 & 0 & 1 & 0 & 0 \\
 0 & 0 & 0 & 0 & 1 & 0 \\
 -1 & 0 & -41 & -8 & 10 & -1 \\
\end{array}
\right)
\end{aligned}
$}\\
\hline
\end{tabular}

\newpage
\begin{tabular}{|R|R|}\hline
\multicolumn{2}{|l|}{Generators of cohomology ring:}\\[0.2em]
\multicolumn{2}{|l|}{
$\begin{aligned}
\left(1\,;\,\sigma_1\,;\,\sigma_{1,1},\sigma_2\,;\,\frac{\sigma_3}{12}\,;\,\frac{\sigma_{2,2}}{6}\right)\,\in\, H^{0,0}(X)\oplus \ldots \oplus H^{4,4}(X)
\end{aligned}$}\\
\hline
\multicolumn{2}{|l|}{Total Chern character:}\\[0.2em]
\multicolumn{2}{|l|}{
$\begin{aligned}
 c(X) \,=\, 1 + (6 \sigma_{1,1} + 4 \sigma_2) - 426 \frac{\sigma _3}{12} + 1368 \frac{\sigma _{2,2}}{6}
\end{aligned}$}\\
\hline
\multicolumn{2}{|l|}{Intersection numbers:}\\[0.2em]
\multicolumn{2}{|l|}{
$\begin{aligned}
\sigma_{1,1}.\sigma_{1,1}= 6 \ , \quad \sigma_{1,1}.\sigma_2 = 9 \ , \quad \sigma_2.\sigma_2=18
\end{aligned}$}\\\hline
\multicolumn{2}{|l|}{Zeros of integral quantum periods:}\\[0.2em]
\multicolumn{2}{|l|}{
$\begin{aligned}
\Pi_5 &= 0 \quad \text{at } z_{\mathcal{O}_X} \ , \\
\Pi_{\mathcal{B}_{z_1}}&=21 \Pi_0 + 42 \Pi_1 - \Pi_2 + \Pi_3 - \Pi_4 - 2\Pi_5 = 0 \quad \text{at } z_1
\end{aligned}$}\\[8pt]
\hline
\multicolumn{2}{|l|}{Genus zero Gromov--Witten invariants $n^{(1)}_{0,d}$ (left) and $n^{(2)}_{0,d}$ (right)}\\\hline
 45 & 129 \\
 11\,169 & 13\,731 \\
 2\,334\,015 & 2\,977\,203 \\
 670\,339\,377 & 843\,149\,973 \\
 222\,531\,477\,228 & 278\,449\,436\,724 \\
 81\,416\,926\,226\,097 & 101\,484\,761\,783\,937 \\
 31\,861\,797\,197\,835\,564 & 39\,609\,507\,515\,035\,620 \\
 13\,104\,024\,227\,969\,549\,085 & 16\,258\,171\,900\,604\,949\,897 \\
 5\,598\,901\,286\,610\,753\,390\,696 & 6\,935\,937\,444\,307\,917\,236\,520 \\
 2\,465\,575\,949\,291\,932\,283\,056\,560 & 3\,050\,652\,167\,218\,394\,830\,016\,340 \\\hline
\multicolumn{2}{|l|}{Genus one Gromov--Witten invariants $n_{1,d}$}\\\hline
\multicolumn{2}{|r|}{0} \\
 \multicolumn{2}{|r|}{0} \\
\multicolumn{2}{|r|}{ 20} \\
 \multicolumn{2}{|r|}{117\,369} \\
 \multicolumn{2}{|r|}{1\,111\,542\,426} \\
\multicolumn{2}{|r|}{ 2\,030\,821\,680\,744} \\
\multicolumn{2}{|r|}{ 2\,190\,254\,867\,538\,498} \\
\multicolumn{2}{|r|}{ 1\,859\,490\,547\,470\,080\,793} \\
\multicolumn{2}{|r|}{ 1\,386\,159\,363\,843\,011\,650\,458} \\
\multicolumn{2}{|r|}{ 955\,211\,114\,503\,390\,944\,999\,069} \\\hline
\end{tabular}

\newpage
\subsection{Calabi--Yau fourfold $X_{1^2,2^2} \subset \operatorname{Gr}(2,6)$}
\begin{tabular}{|L|L|}
\hline
\multicolumn{2}{|l|}{Picard--Fuchs operator:}\\
\multicolumn{2}{|l|}{$\begin{aligned}
\mathcal{L}(z)\,=\,&(\theta -1) \theta ^5 -4z (2 \theta +1)^3 \left(13 \theta ^2+13 \theta +4\right) \theta -48z^2 (2 \theta +1)^2 (2 \theta +3)^2 (3 \theta +2) (3 \theta +4)
\end{aligned}$}\\
\hline
\multicolumn{1}{|l|}{Discriminant locus:} & \multicolumn{1}{|l|}{Riemann P-symbol:}\\[0.2em]
\multicolumn{1}{|l|}{$ \Delta(z)=1 - 416 z - 6\,912 z^2$}& \multicolumn{1}{|l|}{\multirow{4}{*}{$\left\{
\begin{array}{cccc}
0 & \infty & z_1 & z_{\mathcal{O}_X}\\[0.1em]
\hline
0 & \frac12 & 0 & 0 \\[0.1em]
0 & \frac12 & 1 & 1 \\[0.1em]
0 & \frac23 & 2 & 2 \\[0.1em]
0 & \frac43 & 3 & 3 \\[0.1em]
0 & \frac32 & 4 & 4 \\[0.1em]
1 & \frac32 & \frac{3}{2} & \frac{3}{2} \\
\end{array}
\right\}$}} \\
&\\
\multicolumn{1}{|l|}{Regular singular points:} & \\
\multicolumn{1}{|l|}{$\begin{aligned} z&=0\\z&=\infty\\z&=z_{\mathcal{O}_X}=432^{-1}\\z&=z_1 = -16^{-1} \end{aligned}$} & \\
\hline
\multicolumn{1}{|l|}{Intersection pairing:} & \multicolumn{1}{|l|}{Large volume asymptotics:} \\[0.2em]
\multicolumn{1}{|l|}{$\chi = \left(
\begin{array}{*{6}{>{\scriptstyle}c}}
 0 & 0 & 0 & 0 & 0 & 1 \\
 0 & 0 & 0 & 0 & 1 & 0 \\
 0 & 0 & 56 & 20 & 28 & 50 \\
 0 & 0 & 20 & 8 & -20 & 9 \\
 0 & 1 & 28 & -20 & -22 & -11 \\
 1 & 0 & 50 & 9 & -11 & 2 \\
\end{array}
\right)$} & 
\multicolumn{1}{|l|}{$ \vec\Pi^{\text{asy}}(t) = \left(
\begin{array}{*{1}{>{\scriptstyle}c}}
1\\
t\\
28t^2+56t+\frac{124}{3}\\
10t^2-10t+\frac{35}{6}\\
-\frac{28}{3}t^3-14t^2-18t-\frac{20}{3}+\frac{43 i \zeta(3)}{\pi^3}\\
\frac{7}{3}t^4+\frac{13}{3}t^2-\frac{43 i \zeta(3)}{\pi^3}t+\frac{47}{144}
\end{array}\right)$}\\
\hline
\multicolumn{2}{|l|}{Monodromy matrices:}\\[0.2em]
\multicolumn{2}{|l|}{$
\begin{aligned}
M_0&=\left(
\begin{array}{*{6}{>{\scriptstyle}c}}
 1 & 1 & 84 & 0 & 0 & 0 \\
 0 & 1 & 56 & 20 & 0 & 0 \\
 0 & 0 & 1 & 0 & -1 & 0 \\
 0 & 0 & 0 & 1 & 0 & 0 \\
 0 & 0 & 0 & 0 & 1 & -1 \\
 0 & 0 & 0 & 0 & 0 & 1 \\
\end{array}
\right)  \quad
M_\infty =\left(
\begin{array}{*{6}{>{\scriptstyle}c}}
 -83 & -29 & -5572 & -400 & 1512 & -140 \\
 -168 & -55 & -11144 & -860 & 3024 & -280 \\
 3 & 1 & 199 & 15 & -54 & 5 \\
 -4 & -1 & -248 & -23 & 66 & -6 \\
 3 & 1 & 198 & 15 & -54 & 5 \\
 7 & 2 & 446 & 39 & -121 & 11 \\
\end{array}
\right)  \\[0.2em]
%%%%%%%%%%%%%%
M_{z_1}&=\left(
\begin{array}{*{6}{>{\scriptstyle}c}}
 57 & 28 & 4592 & 280 & -1456 & 168 \\
 112 & 57 & 9184 & 560 & -2912 & 336 \\
 -2 & -1 & -163 & -10 & 52 & -6 \\
 2 & 1 & 164 & 11 & -52 & 6 \\
 -2 & -1 & -164 & -10 & 53 & -6 \\
 -4 & -2 & -328 & -20 & 104 & -11 \\
\end{array}
\right)  \quad
M_{z_{\mathcal{O}_X}} = \left(
\begin{array}{*{6}{>{\scriptstyle}c}}
 1 & 0 & 0 & 0 & 0 & 0 \\
 0 & 1 & 0 & 0 & 0 & 0 \\
 0 & 0 & 1 & 0 & 0 & 0 \\
 0 & 0 & 0 & 1 & 0 & 0 \\
 0 & 0 & 0 & 0 & 1 & 0 \\
 -1 & 0 & -50 & -9 & 11 & -1 \\
\end{array}
\right)
\end{aligned}
$}\\
\hline
\end{tabular}

\newpage
\begin{tabular}{|R|R|}\hline
\multicolumn{2}{|l|}{Generators of cohomology ring:}\\[0.2em]
\multicolumn{2}{|l|}{
$\begin{aligned}
\left(1\,;\,\sigma_1\,;\,\sigma_{1,1},\sigma_2\,;\,\frac{\sigma_3}{16}\,;\,\frac{\sigma_{2,2}}{8}\right)\,\in\, H^{0,0}(X)\oplus \ldots \oplus H^{4,4}(X)
\end{aligned}$}\\
\hline
\multicolumn{2}{|l|}{Total Chern character:}\\[0.2em]
\multicolumn{2}{|l|}{
$\begin{aligned}
 c(X) \,=\, 1 + (5 \sigma_{1,1} + 3 \sigma_2) -344 \frac{\sigma _3}{16} + 888 \frac{\sigma _{2,2}}{8}
\end{aligned}$}\\
\hline
\multicolumn{2}{|l|}{Intersection numbers:}\\[0.2em]
\multicolumn{2}{|l|}{
$\begin{aligned}
\sigma_{1,1}.\sigma_{1,1}= 8 \ , \quad \sigma_{1,1}.\sigma_2 = 12 \ , \quad \sigma_2.\sigma_2=24
\end{aligned}$}\\\hline
\multicolumn{2}{|l|}{Zeros of integral quantum periods:}\\[0.2em]
\multicolumn{2}{|l|}{
$\begin{aligned}
\Pi_5 &= 0 \quad \text{at } z_{\mathcal{O}_X} \ , \\
\Pi_{\mathcal{B}_{z_1}}&=28 \Pi_0 + 56 \Pi_1 - \Pi_2 + \Pi_3 - \Pi_4 - 2\Pi_5 = 0 \quad \text{at } z_1
\end{aligned}$}\\[8pt]
 \hline
\multicolumn{2}{|l|}{Genus zero Gromov--Witten invariants $n^{(1)}_{0,d}$ (left) and $n^{(2)}_{0,d}$ (right)}\\\hline
 20 & 76 \\
 3\,710 & 4\,662 \\
 456\,996 & 601\,308 \\
 77\,744\,208 & 100\,674\,808 \\
 15\,262\,779\,768 & 19\,647\,842\,856 \\
 3\,300\,982\,396\,086 & 4\,230\,686\,882\,622 \\
 763\,420\,513\,970\,084 & 975\,446\,610\,603\,036 \\
 185\,520\,589\,035\,937\,760 & 236\,505\,646\,336\,207\,216 \\
 46\,831\,421\,841\,938\,832\,444 & 59\,596\,808\,422\,526\,994\,692 \\
 12\,183\,382\,927\,032\,659\,991\,892 & 15\,482\,698\,161\,874\,509\,215\,956 \\ \hline
\multicolumn{2}{|l|}{Genus one Gromov--Witten invariants $n_{1,d}$}\\\hline
\multicolumn{2}{|r|}{0} \\
\multicolumn{2}{|r|}{0} \\
\multicolumn{2}{|r|}{0} \\
\multicolumn{2}{|r|}{17\,898} \\
\multicolumn{2}{|r|}{60\,657\,824} \\
\multicolumn{2}{|r|}{65\,864\,201\,248} \\
\multicolumn{2}{|r|}{43\,546\,640\,994\,304} \\
\multicolumn{2}{|r|}{22\,541\,684\,709\,460\,560} \\
\multicolumn{2}{|r|}{10\,173\,360\,305\,632\,854\,080} \\
\multicolumn{2}{|r|}{4\,221\,177\,321\,952\,488\,663\,680} \\\hline
\end{tabular}

\newpage
\subsection{Calabi--Yau fourfold $X_{1^5,2} \subset \operatorname{Gr}(2,7)$}
\begin{tabular}{|L|L|}
\hline
\multicolumn{2}{|l|}{Picard--Fuchs operator:}\\
\multicolumn{2}{|l|}{$\begin{aligned}
\mathcal{L}(z)\,=\,&+9 (\theta -1) \theta ^5-6 z\theta  \left(310 \theta ^5+919 \theta ^4+884 \theta ^3+476 \theta ^2+132 \theta +15\right)\\
&-4z^2 \left(21311 \theta ^6+78951 \theta ^5+154395 \theta ^4+180544 \theta ^3+121086 \theta ^2+42546 \theta +6048\right)\\
&-8 z^3(2 \theta +1) \left(57561 \theta ^5+249372 \theta ^4+412273 \theta ^3+310581 \theta^2+104388 \theta +11691\right)\\
&-16z^4 (2 \theta +1) (2 \theta +3) \left(10501 \theta ^4+20138 \theta ^3+13096 \theta ^2+2676 \theta -154\right)\\
&+1184z^5 (\theta +1)^3 (2 \theta+1) (2 \theta +3) (2 \theta +5)
\end{aligned}$}\\
\hline
\multicolumn{1}{|l|}{Discriminant locus:} & \multicolumn{1}{|l|}{Riemann P-symbol:}\\[0.2em]
\multicolumn{1}{|l|}{$ \Delta(z)=1 - 228 z - 4\,624 z^2 + 64 z^3$} & \multicolumn{1}{|l|}{\multirow{4}{*}{$\left\{
\begin{array}{ccccc}
0 & \infty & z_1 & z_2 & z_{\mathcal{O}_X}\\[0.1em]
\hline
0 & \frac12 & 0 & 0 & 0 \\[0.1em]
0 & 1 & 1 & 1 & 1 \\[0.1em]
0 & 1 & 2 & 2 & 2\\[0.1em]
0 & 1 & 3 & 3 & 3\\[0.1em]
0 & \frac32 & 4 & 4 & 4\\[0.1em]
1 & \frac52 & \frac{3}{2} & \frac{3}{2} &\frac{3}{2} \\
\end{array}
\right\}$}} \\
&\\
\multicolumn{1}{|l|}{Regular singular points:} & \\
\multicolumn{1}{|l|}{$\begin{aligned} z&=0\\z&=\infty\\z&= z_{\mathcal{O}_X}\approx 0.004\\z&= z_1 \approx -0.053 \\z&=z_2\approx 72.3 \end{aligned}$} & \\
\hline
\multicolumn{1}{|l|}{Intersection pairing:} & \multicolumn{1}{|l|}{Large volume asymptotics:} \\[0.2em]
\multicolumn{1}{|l|}{$\chi = \left(
\begin{array}{*{6}{>{\scriptstyle}c}}
 0 & 0 & 0 & 0 & 0 & 1 \\
 0 & 0 & 0 & 0 & 1 & 0 \\
 0 & 0 & 84 & 28 & 42 & 70 \\
 0 & 0 & 28 & 10 & -28 & 11 \\
 0 & 1 & 42 & -28 & -28 & -14 \\
 1 & 0 & 70 & 11 & -14 & 2 \\
\end{array}
\right)$} & 
\multicolumn{1}{|l|}{$ \vec\Pi^{\text{asy}}(t) = \left(
\begin{array}{*{1}{>{\scriptstyle}c}}
1\\
t\\
42t^2+84t+\frac{119}{2}\\
14t^2-14t+\frac{89}{12}\\
-14t^3-21t^2-\frac{49}{2}t-\frac{35}{4}+\frac{91 i \zeta(3)}{2\pi^3}\\
\frac{7}{2}t^4+\frac{21}{4}t^2-\frac{91 i \zeta(3)}{2\pi^3}t+\frac{65}{192}
\end{array}\right)$}\\
\hline
\multicolumn{2}{|l|}{Monodromy matrices:}\\[0.2em]
\multicolumn{2}{|l|}{$
\begin{aligned}
M_0&=\left(
\begin{array}{*{6}{>{\scriptstyle}c}}
 1 & 1 & 126 & 0 & 0 & 0 \\
 0 & 1 & 84 & 28 & 0 & 0 \\
 0 & 0 & 1 & 0 & -1 & 0 \\
 0 & 0 & 0 & 1 & 0 & 0 \\
 0 & 0 & 0 & 0 & 1 & -1 \\
 0 & 0 & 0 & 0 & 0 & 1 \\
\end{array}
\right)  \quad
M_\infty =\left(
\begin{array}{*{6}{>{\scriptstyle}c}}
 -533 & -43 & -36946 & -3626 & 6902 & -252 \\
 -1008 & -83 & -70252 & -6916 & 13244 & -504 \\
 13 & 1 & 897 & 89 & -167 & 6 \\
 -14 & -1 & -966 & -99 & 182 & -7 \\
 16 & 1 & 1078 & 110 & -195 & 6 \\
 42 & 2 & 2772 & 294 & -490 & 13 \\
\end{array}
\right)  \\[0.2em]
%%%%%%%%%%%%%%
M_{z_1}&=\left(
\begin{array}{*{6}{>{\scriptstyle}c}}
 85 & 42 & 9996 & 504 & -2940 & 294 \\
 168 & 85 & 19992 & 1008 & -5880 & 588 \\
 -2 & -1 & -237 & -12 & 70 & -7 \\
 2 & 1 & 238 & 13 & -70 & 7 \\
 -2 & -1 & -238 & -12 & 71 & -7 \\
 -4 & -2 & -476 & -24 & 140 & -13 \\
\end{array}
\right)  \quad
M_{z_2}=\left(
\begin{array}{*{6}{>{\scriptstyle}c}}
 97 & 0 & 5824 & 672 & -896 & 0 \\
 420 & 1 & 25480 & 2940 & -3920 & 0 \\
 -3 & 0 & -181 & -21 & 28 & 0 \\
 12 & 0 & 728 & 85 & -112 & 0 \\
 0 & 0 & 0 & 0 & 1 & 0 \\
 -9 & 0 & -546 & -63 & 84 & 1 \\
\end{array}
\right) \\[0.2em]
%%%%%%%%%%%%%%
M_{z_{\mathcal{O}_X}} &= \left(
\begin{array}{*{6}{>{\scriptstyle}c}}
 1 & 0 & 0 & 0 & 0 & 0 \\
 0 & 1 & 0 & 0 & 0 & 0 \\
 0 & 0 & 1 & 0 & 0 & 0 \\
 0 & 0 & 0 & 1 & 0 & 0 \\
 0 & 0 & 0 & 0 & 1 & 0 \\
 -1 & 0 & -70 & -11 & 14 & -1 \\
\end{array}
\right)
\end{aligned}
$}\\
\hline
\end{tabular}

\newpage
\begin{tabular}{|R|R|}\hline
\multicolumn{2}{|l|}{Generators of cohomology ring:}\\[0.2em]
\multicolumn{2}{|l|}{
$\begin{aligned}
\left(1\,;\,\sigma_1\,;\,\sigma_{1,1},\sigma_2\,;\,\frac{\sigma_3}{28}\,;\,\frac{\sigma_{2,2}}{10}\right)\,\in\, H^{0,0}(X)\oplus \ldots \oplus H^{4,4}(X)
\end{aligned}$}\\
\hline
\multicolumn{2}{|l|}{Total Chern character:}\\[0.2em]
\multicolumn{2}{|l|}{
$\begin{aligned}
 c(X) \,=\, 1 + (5 \sigma_{1,1} + 2 \sigma_2) -364 \frac{\sigma _3}{28} + 846 \frac{\sigma _{2,2}}{10}
\end{aligned}$}\\
\hline
\multicolumn{2}{|l|}{Intersection numbers:}\\[0.2em]
\multicolumn{2}{|l|}{
$\begin{aligned}
\sigma_{1,1}.\sigma_{1,1}= 10 \ , \quad \sigma_{1,1}.\sigma_2 = 18 \ , \quad \sigma_2.\sigma_2=38
\end{aligned}$}\\\hline
\multicolumn{2}{|l|}{Zeros of integral quantum periods:}\\[0.2em]
\multicolumn{2}{|l|}{
$\begin{aligned}
\Pi_5 &= 0 \quad \text{at } z_{\mathcal{O}_X} \ , \\
\Pi_{\mathcal{B}_{z_1}}&=42 \Pi_0 + 84 \Pi_1 - \Pi_2 + \Pi_3 - \Pi_4 - 2\Pi_5 = 0 \quad \text{at } z_1\ , \\
\Pi_{\mathcal{B}_{z_2}}&=32 \Pi_0 + 140 \Pi_1 - \Pi_2 + 4\Pi_3 -3\Pi_5 = 0 \quad \text{at } z_2
\end{aligned}$}\\[17pt]
 \hline
\multicolumn{2}{|l|}{Genus zero Gromov--Witten invariants $n^{(1)}_{0,d}$ (left) and $n^{(2)}_{0,d}$ (right)}\\\hline 
 -10 & 46 \\
 1\,009 & 1\,499 \\
 66\,436 & 111\,012 \\
 6\,611\,218 & 10\,644\,996 \\
 744\,513\,554 & 1\,186\,881\,242 \\
 92\,436\,371\,702 & 146\,004\,322\,222 \\
 12\,248\,099\,597\,230 & 19\,229\,229\,169\,542 \\
 1\,704\,064\,096\,112\,480 & 2\,663\,089\,251\,024\,164 \\
 246\,133\,929\,404\,316\,702 & 383\,301\,240\,195\,065\,542 \\
 36\,625\,042\,233\,637\,069\,635 & 56\,876\,037\,388\,681\,122\,041 \\ \hline
\multicolumn{2}{|l|}{Genus one Gromov--Witten invariants $n_{1,d}$}\\\hline
\multicolumn{2}{|r|}{0} \\
\multicolumn{2}{|r|}{0} \\
\multicolumn{2}{|r|}{0} \\
\multicolumn{2}{|r|}{175} \\
\multicolumn{2}{|r|}{1\,251\,544} \\
\multicolumn{2}{|r|}{1\,106\,013\,132} \\
\multicolumn{2}{|r|}{502\,633\,629\,368} \\
\multicolumn{2}{|r|}{165\,747\,820\,001\,414} \\
\multicolumn{2}{|r|}{458\,876\,986\,698\,030\,32} \\
\multicolumn{2}{|r|}{11\,434\,511\,768\,888\,583\,676} \\\hline
\end{tabular}

\newpage
\subsection{Calabi--Yau fourfold $X_{1^8} \subset \operatorname{Gr}(2,8)$}
\begin{tabular}{|L|L|}
\hline
\multicolumn{2}{|l|}{Picard--Fuchs operator:}\\
\multicolumn{2}{|l|}{$\begin{aligned}
\mathcal{L}(z)\,=\,&+121 (\theta -1) \theta ^5-22z\, \theta  \left(438 \theta ^5+2094 \theta ^4+1710 \theta ^3+950 \theta ^2+275 \theta +33\right)\\
&+z^2\big(-839313 \theta ^6-2471661 \theta^5-4037556 \theta ^4-4497304 \theta ^3-3093948 \theta ^2-1158740 \theta\\
&-180048\big)-2z^3 \big(5746754 \theta ^6+26470666 \theta ^5+51184224 \theta ^4+50480470 \theta^3+26295335 \theta ^2\\
&+6684843 \theta +604098\big)-4z^4 \big(4081884 \theta ^6+14894484 \theta ^5+18825903 \theta ^4 +7472030 \theta ^3\\
&-3698839 \theta ^2-4099839\theta-993618\big)+56 z^5\big(29592 \theta ^6+255960 \theta ^5+806448 \theta ^4+1272787\theta^3\\
&+1088403 \theta ^2+483431 \theta +87609\big)+1568 z^6(\theta +1)^3 (2 \theta+3) (4 \theta +3) (4 \theta +5)
\end{aligned}$}\\
\hline
\multicolumn{1}{|l|}{Discriminant locus:} & \multicolumn{1}{|l|}{Riemann P-symbol:}\\[0.2em]
\multicolumn{1}{|l|}{$ \Delta(z)=(1 + 16 z) (1 - 136 z + 16 z^2)$} & \multicolumn{1}{|l|}{\multirow{4}{*}{$\left\{
\begin{array}{ccccc}
0 & \infty & z_1 & z_2 & z_{\mathcal{O}_X}\\[0.1em]
\hline
0 & \frac34 & 0 & 0 & 0 \\[0.1em]
0 & 1 & 1 & 1 & 1 \\[0.1em]
0 & 1 & 2 & 2 & 2\\[0.1em]
0 & 1 & 3 & 3 & 3\\[0.1em]
0 & \frac54 & 4 & 4 & 4\\[0.1em]
1 & \frac32 & \frac{3}{2} & \frac{3}{2} &\frac{3}{2} \\
\end{array}
\right\}$}} \\
&\\
\multicolumn{1}{|l|}{Regular singular points:} & \\
\multicolumn{1}{|l|}{$\begin{aligned} z&=0\\z&=\infty\\z&= z_{\mathcal{O}_X}\approx 0.007\\z&= z_1 =-16^{-1} \\z&=z_2\approx 8.5 \end{aligned}$} & \\
\hline
\multicolumn{1}{|l|}{Intersection pairing:} & \multicolumn{1}{|l|}{Large volume asymptotics:} \\[0.2em]
\multicolumn{1}{|l|}{$\chi = \left(
\begin{array}{*{6}{>{\scriptstyle}c}}
 0 & 0 & 0 & 0 & 0 & 1 \\
 0 & 0 & 0 & 0 & 1 & 0 \\
 0 & 0 & 132 & 42 & 66 & 102 \\
 0 & 0 & 42 & 14 & -42 & 14 \\
 0 & 1 & 66 & -42 & -36 & -18 \\
 1 & 0 & 102 & 14 & -18 & 2 \\
\end{array}
\right)$} & 
\multicolumn{1}{|l|}{$ \vec\Pi^{\text{asy}}(t) = \left(
\begin{array}{*{1}{>{\scriptstyle}c}}
1\\
t\\
66t^2+132t+\frac{179}{2}\\
21t^2-21t+\frac{119}{12}\\
-22t^3-33t^2-\frac{69}{2}t-\frac{47}{4}+\frac{42 i \zeta(3)}{\pi^3}\\
\frac{11}{2}t^4+\frac{25}{4}t^2-\frac{42 i \zeta(3)}{\pi^3}t+\frac{115}{288}
\end{array}\right)$}\\
\hline
\multicolumn{2}{|l|}{Monodromy matrices:}\\[0.3em]
\multicolumn{2}{|l|}{$
\begin{aligned}
M_0&=\left(
\begin{array}{*{6}{>{\scriptstyle}c}}
 1 & 1 & 198 & 0 & 0 & 0 \\
 0 & 1 & 132 & 42 & 0 & 0 \\
 0 & 0 & 1 & 0 & -1 & 0 \\
 0 & 0 & 0 & 1 & 0 & 0 \\
 0 & 0 & 0 & 0 & 1 & -1 \\
 0 & 0 & 0 & 0 & 0 & 1 \\
\end{array}
\right)  \, \,
M_\infty =\left(
\begin{array}{*{6}{>{\scriptstyle}c}}
 -1109 & -67 & -111054 & -10010 & 17610 & -462 \\
 -2100 & -131 & -211308 & -19026 & 33780 & -924 \\
 17 & 1 & 1699 & 154 & -269 & 7 \\
 -18 & -1 & -1800 & -167 & 288 & -8 \\
 20 & 1 & 1968 & 182 & -305 & 7 \\
 50 & 2 & 4848 & 462 & -738 & 15 \\
\end{array}
\right)  \\[0.3em]
%%%%%%%%%%%%%%
M_{z_1}&=\left(
\begin{array}{*{6}{>{\scriptstyle}c}}
 133 & 66 & 23760 & 924 & -6336 & 528 \\
 264 & 133 & 47520 & 1848 & -12672 & 1056 \\
 -2 & -1 & -359 & -14 & 96 & -8 \\
 2 & 1 & 360 & 15 & -96 & 8 \\
 -2 & -1 & -360 & -14 & 97 & -8 \\
 -4 & -2 & -720 & -28 & 192 & -15 \\
\end{array}
\right)  \, \,
M_{z_2}=\left(
\begin{array}{*{6}{>{\scriptstyle}c}}
 157 & 0 & 14040 & 1456 & -1872 & 0 \\
 648 & 1 & 58320 & 6048 & -7776 & 0 \\
 -3 & 0 & -269 & -28 & 36 & 0 \\
 12 & 0 & 1080 & 113 & -144 & 0 \\
 0 & 0 & 0 & 0 & 1 & 0 \\
 -9 & 0 & -810 & -84 & 108 & 1 \\
\end{array}
\right) \\[0.3em]
%%%%%%%%%%%%%%
M_{z_{\mathcal{O}_X}} &= \left(
\begin{array}{*{6}{>{\scriptstyle}c}}
 1 & 0 & 0 & 0 & 0 & 0 \\
 0 & 1 & 0 & 0 & 0 & 0 \\
 0 & 0 & 1 & 0 & 0 & 0 \\
 0 & 0 & 0 & 1 & 0 & 0 \\
 0 & 0 & 0 & 0 & 1 & 0 \\
 -1 & 0 & -102 & -14 & 18 & -1 \\
\end{array}
\right)
\end{aligned}
$}\\
\hline
\end{tabular}

\newpage
\begin{tabular}{|R|R|}\hline
\multicolumn{2}{|l|}{Generators of cohomology ring:}\\[0.2em]
\multicolumn{2}{|l|}{
$\begin{aligned}
\left(1\,;\,\sigma_1\,;\,\sigma_{1,1},\sigma_2\,;\,\frac{\sigma_3}{48}\,;\,\frac{\sigma_{2,2}}{14}\right)\,\in\, H^{0,0}(X)\oplus \ldots \oplus H^{4,4}(X)
\end{aligned}$}\\
\hline
\multicolumn{2}{|l|}{Total Chern character:}\\[0.2em]
\multicolumn{2}{|l|}{
$\begin{aligned}
 c(X) \,=\, 1 + (5 \sigma_{1,1} + \sigma_2) -336 \frac{\sigma _3}{48} + 636 \frac{\sigma _{2,2}}{14}
\end{aligned}$}\\
\hline
\multicolumn{2}{|l|}{Intersection numbers:}\\[0.2em]
\multicolumn{2}{|l|}{
$\begin{aligned}
\sigma_{1,1}.\sigma_{1,1}= 14 \ , \quad \sigma_{1,1}.\sigma_2 = 28 \ , \quad \sigma_2.\sigma_2=62
\end{aligned}$}\\\hline
\multicolumn{2}{|l|}{Zeros of integral quantum periods:}\\[0.2em]
\multicolumn{2}{|l|}{
$\begin{aligned}
\Pi_5 &= 0 \quad \text{at } z_{\mathcal{O}_X} \ , \\
\Pi_{\mathcal{B}_{z_1}}&=66 \Pi_0 + 132 \Pi_1 - \Pi_2 + \Pi_3 - \Pi_4 - 2\Pi_5 = 0 \quad \text{at } z_1\ , \\
\Pi_{\mathcal{B}_{z_2}}&=52 \Pi_0 + 216 \Pi_1 - \Pi_2 + 4\Pi_3 -3\Pi_5 = 0 \quad \text{at } z_2
\end{aligned}$}\\[17pt]
 \hline
\multicolumn{2}{|l|}{Genus zero Gromov--Witten invariants $n^{(1)}_{0,d}$ (left) and $n^{(2)}_{0,d}$ (right)}\\\hline
 -20 & 28 \\
 222 & 462 \\
 7\,564 & 18\,732 \\
 433\,184 & 999\,488 \\
 27\,132\,712 & 61\,606\,888 \\
 1\,883\,975\,,918 & 4\,190\,840\,486 \\
 138\,861\,570\,764 & 305\,141\,892\,524 \\
 10\,734\,197\,390\,880 & 23\,363\,298\,862\,176 \\
 860\,337\,105\,561\,204 & 1\,859\,026\,775\,810\,036 \\
 70\,983\,785\,067\,825\,508 & 152\,499\,803\,765\,006\,068 \\ \hline
\multicolumn{2}{|l|}{Genus one Gromov--Witten invariants $n_{1,d}$}\\\hline
\multicolumn{2}{|r|}{0} \\
\multicolumn{2}{|r|}{0} \\
\multicolumn{2}{|r|}{0} \\
\multicolumn{2}{|r|}{0} \\
\multicolumn{2}{|r|}{24\,528} \\
\multicolumn{2}{|r|}{14\,591\,360} \\
\multicolumn{2}{|r|}{4\,331\,039\,424} \\
\multicolumn{2}{|r|}{882\,540\,559\,446} \\
\multicolumn{2}{|r|}{145\,991\,147\,911\,616} \\
\multicolumn{2}{|r|}{21\,275\,702\,877\,573\,816} \\\hline
\end{tabular}

\newpage
\subsection{Skew Symmetric Sigma Model Calabi--Yau fourfold $X_{1,17,7}$}
\label{app:Tab2}
\begin{tabular}{|L|L|}
\hline
\multicolumn{2}{|l|}{Picard--Fuchs operator:}\\
\multicolumn{2}{|l|}{{$\begin{aligned}
\mathcal{L}(z) \,=\, &+316\,932 (\theta -1) \theta ^5 -98 z\, \theta  \big(700\,453 \theta ^5+1\,335\,058 \theta ^4+1\,609\,080 \theta ^3+879\,285 \theta ^2 +249\,018 \theta\\
&+29\,106\big)+962754229z^2\big( \theta ^6-1\,976\,960\,883 \theta ^5-10\,395\,509\,031 \theta ^4-14\,991\,662\,969 \theta ^3\\
&-10\,456\,423\,600 \theta ^2-3\,667\,629\,910 \theta -521\,151\,456 \big) +2 z^3 \big(9\,812\,727\,979 \theta ^6\\
&+53\,190\,263\,573 \theta ^5+105\,895\,432\,463 \theta ^4+103\,996\,363\,801 \theta ^3 +54\,017\,188\,106 \theta ^2\\
&+14\,078\,111\,747 \theta +1\,415\,445\,066\big)-2 z^4\big(11\,549\,486\,896 \theta ^6+46\,324\,321\,804 \theta ^5\\
&+73\,290\,469\,426 \theta ^4+60\,074\,870\,026 \theta ^3+27\,353\,847\,169 \theta ^2+6\,669\,746\,719 \theta+696\,036\,075\big)\\
&+174z^5 \big(1\,666\,198 \theta ^6+6\,006\,981 \theta ^5+10\,497\,819 \theta ^4+11\,551\,078 \theta ^3+8\,162\,130 \theta ^2\\
&+3\,331\,047 \theta +588\,537\big)-211\,932 z^6 (\theta +1)^5 (2 \theta +3)
\end{aligned}$}}\\
\hline
\multicolumn{1}{|l|}{Discriminant locus:} & \multicolumn{1}{|l|}{Riemann P-symbol:}\\[0.2em]
\multicolumn{1}{|l|}{$ \Delta(z)=(1 - 188 z - 2368 z^2 + 4 z^3)$} & \multicolumn{1}{|l|}{\multirow{4}{*}{$\left\{
\begin{array}{ccccc}
0 & \infty & z_1 & z_2 & z_{\mathcal{O}_X}\\[0.1em]
\hline
0 & 1 & 0 & 0 & 0 \\[0.1em]
0 & 1 & 1 & 1 & 1 \\[0.1em]
0 & 1 & 2 & 2 & 2\\[0.1em]
0 & 1 & 3 & 3 & 3\\[0.1em]
0 & 1 & 4 & 4 & 4\\[0.1em]
1 & \frac32 & \frac{3}{2} & \frac{3}{2} &\frac{3}{2} \\
\end{array}
\right\}$}} \\[0.3em]
&\\
\multicolumn{1}{|l|}{Regular singular points:} & \\
\multicolumn{1}{|l|}{$\begin{aligned} z&=0\\z&=\infty\\z&= z_{\mathcal{O}_X}(=z_3)\approx 0.005\\z&= z_1 \approx-0.084 \\z&=z_2\approx 592 \end{aligned}$} & \\
\hline
\multicolumn{1}{|l|}{Intersection pairing:} & \multicolumn{1}{|l|}{Large volume asymptotics:} \\[0.2em]
\multicolumn{1}{|l|}{$\chi = \left(
\begin{array}{*{6}{>{\scriptstyle}c}}
 0 & 0 & 0 & 0 & 0 & 1 \\
 0 & 0 & 0 & 0 & 1 & 0 \\
 0 & 0 & 98 & 33 & 49 & 79 \\
 0 & 0 & 33 & 12 & -33 & 12 \\
 0 & 1 & 49 & -33 & -30 & -15 \\
 1 & 0 & 79 & 12 & -15 & 2 \\
\end{array}
\right)$} & 
\multicolumn{1}{|l|}{$ \vec\Pi^{\text{asy}}(t) = \left(
\begin{array}{*{1}{>{\scriptstyle}c}}
1\\
t\\
49t^2+98t+\frac{817}{12}\\
\frac{33}{2}t^2-\frac{33}{2}t+\frac{33}{4}\\
-\frac{49}{3}t^3-\frac{49}{2}t^2-\frac{109}{4}t-\frac{229}{24}+\frac{41 i \zeta(3)}{\pi^3}\\
\frac{49}{12}t^4+\frac{131}{24}t^2-\frac{41 i \zeta(3)}{\pi^3}t+\frac{7}{18}
\end{array}\right)$}\\
\hline
\multicolumn{2}{|l|}{Monodromy matrices:}\\[0.2em]
\multicolumn{2}{|l|}{\scriptsize
$\begin{aligned}
M_0&=\left(
\begin{array}{*{6}{>{\scriptstyle}c}}
 1 & 1 & 147 & 0 & 0 & 0 \\
 0 & 1 & 98 & 33 & 0 & 0 \\
 0 & 0 & 1 & 0 & -1 & 0 \\
 0 & 0 & 0 & 1 & 0 & 0 \\
 0 & 0 & 0 & 0 & 1 & -1 \\
 0 & 0 & 0 & 0 & 0 & 1 \\
\end{array}
\right)  \quad
M_{z_1}=\left(
\begin{array}{*{6}{>{\scriptstyle}c}}
 99 & 49 & 13328 & 588 & -3724 & 343 \\
 196 & 99 & 26656 & 1176 & -7448 & 686 \\
 -2 & -1 & -271 & -12 & 76 & -7 \\
 2 & 1 & 272 & 13 & -76 & 7 \\
 -2 & -1 & -272 & -12 & 77 & -7 \\
 -4 & -2 & -544 & -24 & 152 & -13 \\
\end{array}
\right) \\[0.2em]
%%%%%%%%%%%%%%
M_{z_2}&=\left(
\begin{array}{*{6}{>{\scriptstyle}c}}
 4117 & 1568 & 478828 & 23520 & -115248 & 8232 \\
 4809 & 1833 & 559447 & 27480 & -134652 & 9618 \\
 -84 & -32 & -9771 & -480 & 2352 & -168 \\
 0 & 0 & 0 & 1 & 0 & 0 \\
 -168 & -64 & -19544 & -960 & 4705 & -336 \\
 -441 & -168 & -51303 & -2520 & 12348 & -881 \\
\end{array}
\right)\quad
M_{z_{\mathcal{O}_X}} = \left(
\begin{array}{*{6}{>{\scriptstyle}c}}
 1 & 0 & 0 & 0 & 0 & 0 \\
 0 & 1 & 0 & 0 & 0 & 0 \\
 0 & 0 & 1 & 0 & 0 & 0 \\
 0 & 0 & 0 & 1 & 0 & 0 \\
 0 & 0 & 0 & 0 & 1 & 0 \\
 -1 & 0 & -79 & -12 & 15 & -1 \\
\end{array}
\right)\\[0.2em]
%%%%%%%%%%%%%%
M_\infty &=\left(
\begin{array}{*{6}{>{\scriptstyle}c}}
 -3975 & -1490 & -459291 & -22890 & 110201 & -7854 \\
 -1757 & -617 & -197897 & -10479 & 46942 & -3318 \\
 67 & 25 & 7728 & 387 & -1853 & 132 \\
 121 & 47 & 14181 & 682 & -3423 & 245 \\
 151 & 57 & 17499 & 867 & -4205 & 300 \\
 198 & 74 & 22862 & 1146 & -5487 & 391 \\
\end{array}
\right)  \\[0.2em]
\end{aligned}
$}\\
\hline
\end{tabular}

\newpage
\footnotesize
\begin{tabular}{|R|R|}\hline
\multicolumn{2}{|l|}{Generators of cohomology ring:}\\[0.2em]
\multicolumn{2}{|l|}{
$\begin{aligned}
\left(1\,;\,H\,;\,\sigma_{2},H^2\,;\,\frac{H^3}{98}\,;\,\frac{H^4}{98}\right)\,\in\, H^{0,0}(X)\oplus \ldots \oplus H^{4,4}(X)
\end{aligned}$}\\
\hline
\multicolumn{2}{|l|}{Total Chern character:}\\[0.2em]
\multicolumn{2}{|l|}{
$\begin{aligned}
 c(X) \,=\,  1 + (4 H^2 - 2 \sigma_2) -328\frac{H^3}{98} + 672\frac{H^4}{98}
\end{aligned}$}\\
\hline
\multicolumn{2}{|l|}{Intersection numbers:}\\[0.2em]
\multicolumn{2}{|l|}{
$\begin{aligned}
\sigma_{2}.\sigma_{2}= 44 \ , \quad \sigma_{2}.H^2 = 65 \ , \quad H^2.H^2=98
\end{aligned}$}\\\hline
\multicolumn{2}{|l|}{Zeros of integral quantum periods:}\\[0.2em]
\multicolumn{2}{|l|}{
$\begin{aligned}
\Pi_5 &= 0 \quad \text{at } z_{\mathcal{O}_X} \ , \\
\Pi_{\mathcal{B}_{z_1}}&=49 \Pi_0 + 98 \Pi_1 - \Pi_2 + \Pi_3 - \Pi_4 - 2\Pi_5 = 0 \quad \text{at } z_1\ , \\
\Pi_{\mathcal{B}_{z_2}}&=-196 \Pi_0 -229 \Pi_1 +4 \Pi_2 + 8\Pi_4 +21\Pi_5 = 0 \quad \text{at } z_2
\end{aligned}$}\\[17pt]
 \hline
\multicolumn{2}{|l|}{Genus zero Gromov--Witten invariants $n^{(1)}_{0,d}$ (left) and $n^{(2)}_{0,d}$ (right)}\\\hline 
 0 & 33 \\
 721 & 170 \\
 38\,255  & 16\,126  \\
 3\,042\,676 & 1\,141\,312 \\
 274\,320\,123 & 100\,955\,257  \\
 27\,276\,710\,118 & 9\,821\,360\,694 \\
 2\,897\,092\,850\,989 & 1\,028\,274\,636\,900 \\
 323\,207\,209\,581\,582 & 113\,458\,193\,073\,000 \\
 37\,444\,642\,819\,824\,776 & 13\,032\,484\,062\,881\,000  \\
 4\,469\,922\,540\,366\,355\,762 & 1\,545\,108\,865\,260\,914\,434 \\ \hline
\multicolumn{2}{|l|}{Genus one Gromov--Witten invariants $n_{1,d}$}\\\hline
\multicolumn{2}{|r|}{0} \\
\multicolumn{2}{|r|}{0} \\
\multicolumn{2}{|r|}{0} \\
\multicolumn{2}{|r|}{0} \\
\multicolumn{2}{|r|}{224\,386} \\
\multicolumn{2}{|r|}{206\,613\,908} \\
\multicolumn{2}{|r|}{83\,707\,955\,196} \\
\multicolumn{2}{|r|}{23\,455\,827\,469\,526} \\
\multicolumn{2}{|r|}{5\,401\,382\,970\,402\,176} \\
\multicolumn{2}{|r|}{1\,107\,021\,477\,254\,814\,128} \\\hline
\end{tabular}
\normalsize

%%%%%%%%%%%%%%%%%%%%%%%%%%%%%%%%%%%
\newpage
\section{Lines on Calabi--Yau fourfolds} \label{app:InterTheory}
%%%%%%%%%%%%%%%%%%%%%%%%%%%%%%%%%%%

To verify the computed integral quantum periods and the deduced quantum cohomology ring, we here enumerate the number of lines with a marked point located on a codimension two algebraic cycle in the studied Calabi--Yau fourfolds, which arise as complete intersections in Grassmannian spaces $\operatorname{Gr}(2,n)$ for various choices of $n$. Note that the presented derivation generalizes to other complete intersection varieties embedded into general Grassmannians $\operatorname{Gr}(k,n)$ as well, and this appendix is rather independent from the main text.

The moduli space $\mathcal{M}_1$ of lines with a marked point in the ambient Grassmannian variety $\operatorname{Gr}(2,n)$ is the flag variety $\operatorname{Fl}(1,2,3,n)$, whose points are the flags $V_1 \subset V_2 \subset V_3 \subset V_n$ of complex vector spaces with $V_\ell \simeq \mathbb{C}^\ell$. For such a flag the two-dimensional quotient vector space $V_3/V_1$ describes the projective line $\mathbb{P}(V_3/V_1)$. The points in $\mathbb{P}(V_3/V_1)$ are the one-dimensional subvector spaces $\Lambda_1 \subset V_3/V_1$, which canonically define two planes $V_1 \oplus \Lambda_1$ to be identified with points in the Grassmannian variety $\operatorname{Gr}(2,n)$. Furthermore, the subvector space $\Lambda_1 = V_2/V_1$ corresponds to the marked point on the projective line that maps to the two plane $V_1\oplus V_2/V_1 \simeq V_2$ in $\operatorname{Gr}(2,n)$. It defines the evaluation map of the marked point
\begin{equation}
  \operatorname{ev}_1: \mathcal{M}_1  
   \to \operatorname{Gr}(2,n) \ ,\
   V_1 \subset V_2 \subset V_3 \subset V_n \mapsto V_2 \ .
\end{equation}  

We realize the flag variety $\mathcal{M}_1\simeq\operatorname{Fl}(1,2,3,n)$ in terms of the nested fibrations of projective spaces \cite{MR658304}:
\begin{equation} \label{eq:bundle}
\begin{CD}
   \mathcal{U}_1 \oplus \mathcal{Q}_1 @. \mathcal{U}_2 \oplus \mathcal{Q}_2 @.  \mathcal{U}_3 \oplus \mathcal{Q}_3\\
   @VVV @VVV @VVV\\ 
   \mathbb{P}(V_n) @<{\pi_1}<< \mathbb{P}(\mathcal{Q}_1)@<{\pi_2}<< \mathbb{P}(\mathcal{Q}_2)
\end{CD} 
\end{equation}
Here, $\mathcal{U}_1$, $\mathcal{U}_2$ and $\mathcal{U}_3$ are the universal line bundles of the (fibered) projective spaces, whereas $\mathcal{Q}_1$, $\mathcal{Q}_2$ and $\mathcal{Q}_3$ are their respective quotient bundles of dimension $(n-1)$, $(n-2)$ and $(n-3)$, i.e., 
\begin{equation} \label{eq:relbundles}
   \mathcal{U}_1 \oplus \mathcal{Q}_1 = V_n \ , \qquad
   \mathcal{U}_2 \oplus \mathcal{Q}_2 = \pi_1^*\mathcal{Q}_1 \ , \qquad
   \mathcal{U}_3 \oplus \mathcal{Q}_3 = \pi_2^*\mathcal{Q}_2 \ .
\end{equation}

Let $\mathcal{M}_2$ be the moduli space of lines with two marked points in $\operatorname{Gr}(2,n)$ given by the fibration
\begin{equation}
\begin{CD}
  \mathbb{P}(\pi_2^*\mathcal{U}_2\oplus\mathcal{U}_3)@>>>\mathcal{M}_2 \\
  @. @VfVV \\
  @. \mathcal{M}_1 \\
\end{CD} \ .
\end{equation}
The projection $f$ to the base $\mathcal{M}_1$ is the forgetful map that removes the second marked point, whereas its evaluation map reads
\begin{equation}
  \operatorname{ev}_2: \mathcal{M}_2 \to \operatorname{Gr}(2,n) \ , 
  (\Lambda_1, V_1 \subset V_2 \subset V_3 \subset V_n) \mapsto V_1 \oplus \Lambda_1 \ ,
\end{equation}
in terms of the one-dimensional vector space $\Lambda_1$ for the points of the projective fibers and the flag $V_1 \subset V_2 \subset V_3 \subset V_n$ for the base point in $\mathcal{M}_1$.

The cohomology of the flag variety $\mathcal{M}_1$ --- as given in the nested fibration~\eqref{eq:bundle} --- becomes \cite{MR658304} 
\begin{equation} \label{eq:Hring}
     H^*(\mathcal{M}_1,\mathbb{Q}) \,=\, \mathbb{Q}[H_1,H_2,H_3,\xi^{(1)},\ldots,\xi^{(n-3)}] / \mathcal{I} \ .
\end{equation}
The generators of the cohomology ring arise from the Chern classes of the bundles over the nested firbation~\eqref{eq:bundle} as
\begin{equation}
\begin{aligned}
  &H_1 \,=\,- \pi_2^*\pi_1^* c_1(\mathcal{U}_1) \ , \quad
  H_2 \,=\, -\pi_2^* c_1(\mathcal{U}_2) \ , \quad
  H_3 \,=\, - c_1(\mathcal{U}_3) \ , \\[1ex]
  &\xi^{(\ell)}_3 \,=\, c_\ell(\mathcal{Q}_3) \ , \quad \ell=1,\ldots,n-3 \ ,
\end{aligned}  
\end{equation}  
where $H_1$, $H_2$, and $H_3$ are the hyperplane classes of the (fibered) projective spaces and $\xi^{(\ell)}_3$ the Chern classes of the quotient bundle over the last fibered projective space $\mathbb{P}(\mathcal{Q}_2)$. The ideal $\mathcal{I}$ is generated by the homogenous terms (with respect to the form degree of the generators) in the expression 
\begin{equation}
  1-(1- H_1)(1-H_2)(1-H_3)(1+\xi^{(1)}_3+\ldots+ \xi^{(n-3)}_3) \ .
\end{equation}
Note that the relations in the ideal $\mathcal{I}$ determine the cohomology classes $\xi^{(\ell)}_3$ in terms of the hyperplane class generators $H_1$, $H_2$, and $H_3$. Furthermore, the total Chern class of the quotient bundles $\pi_2^*\pi_1^*\mathcal{Q}_1$ and $\pi_2^*\mathcal{Q}_2$ read
\begin{equation}
\begin{aligned}
  \pi_2^*\pi_1^*c(\mathcal{Q}_1) &= 1 + \xi^{(1)}_1 + \ldots + \xi^{(n-1)}_1 = \frac{1}{1-H_1} 
  \ \in\, H^*(\mathcal{M}_1)\ ,  \\
  \pi_2^*c(\mathcal{Q}_2) &= 1 + \xi^{(2)}_2 + \ldots + \xi^{(n-2)}_2 = \frac{1}{(1-H_1)(1-H_2)} 
  \  \in\, H^*(\mathcal{M}_1) \ . \nonumber
\end{aligned}   
\end{equation}

Now we want to enumerate the number of lines on Calabi--Yau fourfolds, which for our class of examples are given as complete intersections $X_{k_1,\ldots,k_\alpha}$ (with $k_i \ge 1$) embedded in the Grassmannian spaces $\operatorname{Gr}(2,n)$ as the zero locus of a generic section in $\mathcal{O}(k_1 \sigma_1)\oplus \ldots\oplus \mathcal{O}(k_\alpha \sigma_1)$. Since $\dim_\mathbb{C} \operatorname{Gr}(2,n) = 2(n-2)$ and $c_1(\operatorname{Gr}(2,n) ) = n\,\sigma_1$, we obtain four-dimensional Calabi--Yau varieties in $\operatorname{Gr}(2,n)$ only for
\begin{equation} \label{eq:cond}
    \alpha \,=\, 2(n-4) \ , \qquad n = k_1 + \ldots + k_\alpha \ .
\end{equation}

In the next step, we impose the complete intersection constraints on the level of the moduli space $\mathcal{M}_1$. We observe that the line bundles $\mathcal{O}(k_i \sigma_1)$ induce on $\mathcal{M}_1$ the vector bundles 
\begin{equation}
  \mathcal{B}(k) \,=\, f_*\operatorname{ev}_2^*\mathcal{O}(k \sigma_1) \ .
\end{equation}
These bundles are explicitly determined to be 
\begin{equation}
  \mathcal{B}(k) \,=\, \operatorname{Sym}^{k}\left[ \pi_1^*\pi_2^*\mathcal{U}_1 \otimes \left( \pi_2^*\mathcal{U}_2\oplus\mathcal{U}_3\right)  \right ] \ ,
\end{equation}  
in terms of symmetrized tensor products of the rank two bundle $\pi_1^*\pi_2^*\mathcal{U}_1 \otimes \left( \pi_2^*\mathcal{U}_2\oplus\mathcal{U}_3\right)$ on $\mathcal{M}_1$. By construction the zeros of induced sections on $\mathcal{B}(k_i)$ describe the loci in $\operatorname{Gr}(2,n)$, where the entire projective line of $\mathcal{M}_1$ vanishes. Thus the zero locus of the induced section of the bundle $\mathcal{B}(k_1)\oplus\ldots\oplus\mathcal{B}_\alpha$ on $\mathcal{M}_1$ describes the moduli space of lines with a single marked point of the Calabi--Yau variety $X_{k_1,\ldots,k_\alpha}$. 

To enumerate genus zero Gromov--Witten invariants at degree one on $X_{k_1,\ldots,k_\alpha}$, it remains to  restrict the marked point on the lines to one of the codimension two (pulled-back) Schubert classes $\sigma_{1,1}$ or $\sigma_{2}$ in $X_{k_1,\ldots,k_\alpha}$. On the moduli space $\mathcal{M}_1$ these classes become $\operatorname{ev}_1^*\sigma_{1,1}$ and $\operatorname{ev}_1^*\sigma_{2}$, respectively. Note that the quotient bundle $\mathcal{Q}_{\operatorname{Gr}(2,n)}$ of $\operatorname{Gr}(2,n)$ pulls back to $\operatorname{ev}_1^*\mathcal{Q}_{\operatorname{Gr}(2,n)} \simeq \pi_2^*\mathcal{Q}_2$, which --- due to $c(\mathcal{Q}_{\operatorname{Gr}(2,n)}) = 1 + \sigma_1 + \sigma_2 + \ldots$ --- implies together with eqs.~\eqref{eq:relbundles} and \eqref{eq:Hring} that $\operatorname{ev}_1^*\sigma_{1} = H_1 + H_2$ and $\operatorname{ev}_1^*\sigma_{2}=H_1^2+H_2^2 + H_1 H_2$. Thus, with $\sigma_1^2 = \sigma_2 + \sigma_{1,1}$ we find
\begin{equation}
    \operatorname{ev}_1^*\sigma_{1,1} \,=\, H_1 H_2 \ , \qquad
    \operatorname{ev}_1^*\sigma_{2} \,=\,H_1^2+H_2^2 + H_1 H_2 \ .
\end{equation}

%%%%%%%%%%%%%%%%%%%%%%
\begin{table}\centering
\begin{tabular}{|c|r|r|}
\hline Calabi--Yau fourfold & $N(\sigma_{1,1})$ &$N(\sigma_{2})$ \\\hline
$X_{1,4} \subset \operatorname{Gr}(2,5)$ & 3\,680 & 5\,760 \\
$X_{2,3} \subset \operatorname{Gr}(2,5)$ & 2\,160 & 3\,420 \\
$X_{1^3,3} \subset \operatorname{Gr}(2,6)$ & 1\,431 & 2\,727 \\
$X_{1^2,2^2} \subset \operatorname{Gr}(2,6)$ & 1\,072 & 2\,064 \\
$X_{1^5,2} \subset \operatorname{Gr}(2,7)$ & 728 & 1\,568 \\ 
$X_{1^8} \subset \operatorname{Gr}(2,8)$ & 504 & 1\,176 \\ \hline
\end{tabular}
\caption{The table enumerates lines with marked points on the codimension two (pulled-back) Schubert cycles $\sigma_1$ and $\sigma_2$ for the listed Calabi--Yau fourfolds embedded as complete intersections in Grassmannians. These number are calculated from the derived intersection formula~\eqref{eq:numlines}, and the results correctly relate with eq.~\eqref{eq:Nton11} to the genus zero Gromov--Witten invariants at degree one tabulated in Appendix~\ref{app:tables}.}\label{tab:NumberLines}
\end{table}
%%%%%%%%%%%%%%%%%%%%%%%

With all the necessary ingredients assembled, we now count the number of lines with its marked point restricted to a codimension two Schubert cycle in $X_{k_1,\ldots,k_n}$ according to
\begin{equation} \label{eq:numlines}
\begin{aligned}
  N(\sigma_{1,1}) \,&=\, \int_{\mathcal{M}_1} 
  c_\text{top}(\mathcal{B}(k_1)) \cup \ldots \cup c_\text{top}(\mathcal{B}(k_\alpha)) \cup
   \operatorname{ev}_1^*\sigma_{1,1} \ , \\
   N(\sigma_{2}) \,&=\, \int_{\mathcal{M}_1} 
  c_\text{top}(\mathcal{B}(k_1)) \cup \ldots \cup c_\text{top}(\mathcal{B}(k_\alpha)) \cup
   \operatorname{ev}_1^*\sigma_{2} \ .
\end{aligned}   
\end{equation}   
Here $c_\text{top}$ denotes the top Chern class of the bundles $\mathcal{B}(k_i)$, which by construction have rank $k_i+1$. Thus --- imposing the Calabi--Yau fourfold conditions~\eqref{eq:cond} --- the integrand becomes an element of $H^{(3n-6,3n-6)}(\mathcal{M}_1)$, which indeed represents a top form on $\mathcal{M}_1$ because $\dim_\mathbb{C} \mathcal{M}_1=3n-6$. The numbers of lines $N(\sigma_{1,1})$ and $N(\sigma_{2})$ obtained in this way compare to the genus zero Gromov--Witten invariants $n_{0,1}^{(1)}$ and $n_{0,1}^{(2)}$ of the quantum cohomology ring as (c.f., Section~\ref{sec:Example1} and Appendix~\ref{app:tables})
\begin{equation} \label{eq:Nton11}
\begin{pmatrix}
    N(\sigma_{1,1}) \\N(\sigma_{2})
\end{pmatrix} \,=\,
\begin{pmatrix}
  \sigma_{1,1}.\sigma_{1,1} & \sigma_{1,1}.\sigma_{2}\\
  \sigma_{1,1}.\sigma_{2} & \sigma_{2}.\sigma_{2}
\end{pmatrix}      
\begin{pmatrix}
    n_{0,1}^{(1)} \\ n_{0,1}^{(2)}
\end{pmatrix}\ ,
\end{equation}
in terms of the intersection pairings of the Schubert cycles $\sigma_{1,1}$ and $\sigma_2$ on the Calabi--Yau fourfold $X_{k_1,\ldots,k_\alpha}$.

In this work we explicitly analyze the Calabi--Yau fourfolds $X_{1,4}$, $X_{2,3}$, $X_{1^3,3}$, $X_{1^2,2^2}$, $X_{1^5,2}$, and $X_{1^8}$ (with the obvious notation for repeated indices and the corresponding embedding space $\operatorname{Gr}(2,n)$ determined through Calabi--Yau fourfold conditions~\eqref{eq:cond}). For these Calabi--Yau fourfolds we explicitly count the number of lines according to eq.~\eqref{eq:numlines} as listed in Table~\ref{tab:NumberLines}. For all our examples we find agreement with the genus zero Gromov--Witten invariants at degree one listed in Appendix~\ref{app:tables}. This furnishes another non-trivial check on the deduced linear combinations for the integral doubly logarithmic quantum periods at large volume.

%%%%%%%%%%%%%%%%%%%%%%%%%%%%%%%%%%%
\end{appendix}
%%%%%%%%%%%%%%%%%%%%%%%%%%%%%%%%%%%

%%%%%%%%%%%%%%%%%%%%%%%%%%%%%%%%%%%
\newpage
\ifx\undefined\bysame
\newcommand{\bysame}{\leavevmode\hbox to3em{\hrulefill}\,}
\fi

%%%%%%%%%%%%%%%%%%%%%%%%%%%%%%%%%%%

\end{document}